\documentclass[preprint,12pt]{elsarticle}

\usepackage{amssymb}
\usepackage{amsmath}

\usepackage{tabularx}
\usepackage{algorithm}
\usepackage{booktabs}
\usepackage{algpseudocode}
\usepackage{subcaption}

\usepackage{xcolor}
\usepackage{booktabs}
\usepackage{multirow}
\usepackage{siunitx}
\usepackage{graphicx} 

\journal{Chaos, Solitons \& Fractals}
\begin{document}

\begin{frontmatter}
	
	
	
	\title{PPO-ACT: Proximal Policy Optimization with Adversarial Curriculum Transfer for Spatial Public Goods Games}
	
	
	\author[1,2]{Zhaoqilin Yang}
	
	
	\ead{zqlyang@gzu.edu.cn}
	
	\tnotetext[1]{https://github.com/geek12138/PPO-ACT}
	
	\affiliation[1]{organization={State Key Laboratory of Public Big Data, College of Computer Science and Technology},
		addressline={Guizhou University}, 
		city={Guiyang},
		postcode={550025}, 
		state={Guizhou},
		country={China}}

	\affiliation[2]{organization={Institute of Cryptography and Data Security},
		addressline={Guizhou University}, 
		city={Guiyang},
		postcode={550025}, 
		state={Guizhou},
		country={China}}

		\author[3]{Chanchan Li}		

		\ead{ccli@gzu.edu.cn}
		
		\affiliation[3]{organization={State Key Laboratory of Public Big Data, College of Mathematics and Statistics},
			addressline={Guizhou University}, 
			city={Guiyang},
			postcode={550025}, 
			state={Guizhou},
			country={China}}
		
		\author[4]{Xin Wang}  
		
		\ead{xinwang2@bjtu.edu.cn}
		
		\affiliation[4]{organization={School of Mathematics and Statistics},
			addressline={Beijing Jiaotong University}, 
			city={Beijing},
			postcode={100044}, 
			state={Beijing},
			country={China}}
		
		\author[5,2]{Youliang Tian\corref{cor1}}
		\ead{yltian@gzu.edu.cn}
		\cortext[cor1]{Corresponding author}
		
			\affiliation[5]{organization={State Key Laboratory of Public Big Data, College of Big Data and Information Engineering},
			addressline={Guizhou University}, 
			city={Guiyang},
			postcode={550025}, 
			state={Guizhou},
			country={China}}

	\begin{abstract}
		This study investigates cooperation evolution mechanisms in the spatial public goods game. A novel deep reinforcement learning framework, Proximal Policy Optimization with Adversarial Curriculum Transfer (PPO-ACT), is proposed to model agent strategy optimization in dynamic environments. Traditional evolutionary game models often exhibit limitations in modeling long-term decision-making processes. Imitation-based rules (e.g., Fermi) lack strategic foresight, while tabular methods (e.g., Q-learning) fail to capture spatial-temporal correlations. Deep reinforcement learning effectively addresses these limitation by bridging policy gradient methods with evolutionary game theory. Our study pioneers the application of proximal policy optimization's continuous strategy optimization capability to public goods games through a two-stage adversarial curriculum transfer training paradigm. The experimental results show that PPO-ACT performs better in critical enhancement factor regimes. Compared to conventional standard proximal policy optimization methods, Q-learning and Fermi update rules, achieve earlier cooperation phase transitions and maintain stable cooperative equilibria. This framework exhibits better robustness when handling challenging scenarios like all-defector initial conditions. Systematic comparisons reveal the unique advantage of policy gradient methods in population-scale cooperation, i.e., achieving spatiotemporal payoff coordination through value function propagation. Our work provides a new computational framework for studying cooperation emergence in complex systems, algorithmically validating the punishment promotes cooperation hypothesis while offering methodological insights for multi-agent system strategy design.
	\end{abstract}
	
	
	
	
\end{frontmatter}

	
	
	\section{Introduction}
	\label{sec1}
	Cooperation is a fundamental mechanism for sustaining and developing human societies and implies the foundational significance throughout civilizational evolution \cite{dawes_1988_anomalies,perc_2016_phase,perc_2017_statistical}. Collective behavior has driven human progress throughout history and continues to do so today. In early societies, cooperative efforts were central to hunting and agriculture. Modern civilization coordinates production and global cooperation. These patterns enhance environmental adaptation while accelerating knowledge growth and technological advancement.	Contemporary global challenges such as climate change and public health crises further underscore the urgency of advancing cooperation mechanism research. Establishing sustainable cooperative paradigms amidst complex tensions between individual interests and collective welfare has become a central focus of interdisciplinary studies \cite{pennisi_2005_did,kennedy_2005_don}. Evolutionary game theory provides a systematic theoretical framework for studying the dynamic evolution of cooperative behavior \cite{nowak_1992_evolutionary,hauert_2005_game,szabo_2007_evolutionary}. Among diverse applications, the public goods game (PGG) has become a cornerstone model for analyzing multi-agent cooperation dilemmas due to its precise characterization of tensions between individual and collective interests \cite{nowak_1993_spatial,macy_2002_learning,wang_2015_universal}. 
	It is worth noting that Duh et al. \cite{DUH2021110720} investigated public goods games on random geometric graphs in hyperbolic spaces.
	This model reveals how cooperative behavior evolves in populations and provides key insights for solving real-world collective action problems.
	
	The PGG reveals the core paradox in human cooperation, i.e., individuals create collective benefits by contributing to shared resources, yet face the dilemma of mismatched personal costs and returns. This game structure perfectly captures social dilemmas.  Individual rational choices (defection) conflict with collective optimal solutions (cooperation). This tension inevitably leads to systemic free-riding. Spatially structured PGG creates unique evolutionary dynamics through networked interactions. Local cooperation can spread globally via specific topological structures. To address this classical challenge, academia has developed systematic solution frameworks, primarily including: 1) Positive incentive mechanisms: enhancing cooperation's attractiveness through rewards \cite{chen_2015_first,dos_2015_evolution,okada_2015_effect} and reputation systems \cite{tang_2024_cooperative,quan_2020_information,shen_2022_high}; 2) Negative constraint mechanisms: suppressing defection spread via punishment \cite{helbing_2010_punish,chen_2014_probabilistic,chen_2015_competition,liu_2018_synergy} and exclusion \cite{liu_2017_competitions,szolnoki_2017_alliance}; 3) Institutional design mechanisms: restructuring payoff matrices through taxation \cite{griffin_2017_cyclic,wang_2021_tax,lee_2024_supporting} and heterogeneous investment rules \cite{cao_2010_evolutionary}. These mechanisms have been rigorously validated theoretically. And their design principles closely match real-world cooperation policies.
	
	Traditional evolutionary game theory primarily relies on classical frameworks such as Fermi update rules \cite{szabo_1998_evolutionary} and replicator dynamics \cite{schuster_1983_replicator}. The classical frameworks effectively model immediate payoff effects and network influences in strategy diffusion. While demonstrating these capabilities, the approaches exhibit limitations in simulating key aspects of human decision-making complexity. A notable gap exists in modeling adaptive learning processes from historical experience and insufficient representation of strategic planning for long-term benefits. This gap has driven researchers toward the reinforcement learning framework \cite{izquierdo_2007_transient,lipowski_2009_statistical}. Reinforcement learning \cite{sutton_1998_reinforcement} creates a closed-loop learning system based on state-action-reward cycles. This allows agents to continuously improve their strategies in changing environments. It effectively handles long-term planning through value propagation while optimizing strategies with experience replay. Moreover, it dynamically maintains the balance between exploration and exploitation. Recent advances in reinforcement learning have enabled more sophisticated modeling of emergent cooperation. Phan et al. \cite{phan_2024_emergent} proposed a mutual acknowledgment token exchange that implements emergent cooperation via a reinforcement learning-based token exchange, achieving high fault tolerance in multi-agent systems. This paradigm shift enables researchers to more accurately simulate real decision-makers trade-off processes between short-term benefits and long-term returns when analyzing cooperation strategy evolution \cite{jia_2021_local,wang_2022_levy,song_2022_reinforcement}.
	
	The Q-learning algorithm is widely used in evolutionary game theory because of its theoretical simplicity and practical effectiveness \cite{watkins_1992_q,hasselt_2010_double}. It allows agents to adapt decisions based on historical experience and current environment states by building a Q-table to store state-action values \cite{han_2021_evolutionary,shi_2022_analysis}. Notably, when applied to spatial PGG, Q-learning's unique value iteration mechanism can maintain stable cooperative equilibrium even under free-riding incentives \cite{szolnoki_2009_topology,szolnoki_2010_impact,szabo_2013_coexistence}. Recent advances have further expanded Q-learning's application dimensions. For instance, Yan et al. \cite{yan_2024_periodic} innovatively combined periodic strategy updates with punishment mechanisms to establish an autonomous decision-making agent model. Similarly, Shen et al. \cite{shen_2024_learning} combined Q-learning with Fermi update rules. These advancements demonstrate how Q-learning provides new theoretical insights into the interaction between learning and imitation in spatial public goods games. 
	
	While Q-learning's tabular approach remains effective for small discrete action spaces, its reliance on value iteration becomes computationally prohibitive as strategy dimensions increase. In contrast, the Proximal Policy Optimization (PPO) algorithm \cite{John_2017_arxiv} addresses these limitations through neural network-based policy parameterization, which not only circumvents Q-table memory constraints but also enables direct optimization of stochastic strategies through gradient ascent. Moreover, PPO's clipped objective function inherently stabilizes policy updates, a critical advantage when coordinating multiple agents in high-dimensional discrete environments. Its reliable training stability and computational efficiency establish it as the benchmark for policy gradient methods. This dual-network architecture allows PPO to optimize policy parameters while ensuring training stability, leading to a more effective balance between exploration and exploitation. Unlike value-based methods that suffer from estimation bias in value functions, PPO directly optimizes policies to avoid suboptimal convergence. These characteristics make PPO particularly suitable for multi-agent scenarios requiring long-term policy planning. 
	Contemporary advances \cite{Liu_2025_TPAMI, RAO_2025_kbs, ZHU_2025_kbs} in multi-agent reinforcement learning have yielded substantial theoretical and empirical evidence supporting its efficacy in complex cooperative task domains. Building on these foundations,
	Sun et al. \cite{sun_2024_intuitionistic} innovatively integrated multi-attribute decision theory with PPO to address slow convergence in intelligent wargame training and agents' low success rates with specific rules. Yu et al. \cite{yu_2022_surprising} demonstrated PPO's strong performance in cooperative multi-agent settings, achieving competitive or superior results in both final returns and sample efficiency. However, integrating modern reinforcement learning algorithms like PPO with evolutionary game theory still faces significant challenges. Current research has yet to fully uncover the diffusion dynamics of policy gradient methods in structured populations. The interaction effects between network topology and distributed learning processes remain insufficiently explored. These open questions provide promising directions for future research.

	We propose Proximal Policy Optimization with Adversarial Curriculum Transfer (PPO-ACT) for PGG. Our study pioneers the application of PPO in evolutionary game theory. The framework addresses cooperation evolution in spatial PGG through a novel two-stage training paradigm that builds upon curriculum learning \cite{bengio_2009_curriculum} while introducing adversarial dynamics. Curriculum learning is used to enhance tasks based on reinforcement learning, such as robot navigation \cite{YIN2024123202}, unmanned aerial vehicles collaborative work \cite{SEONG2024124379}.
	It employs a dual-network architecture where the policy and value networks share underlying feature extraction layers. The framework's two-stage curriculum learning design significantly enhances adaptability in complex environments. In Phase 1, cooperative foundations are established in high-reward conditions. In Phase 2, knowledge transfers to low-reward scenarios. This staged approach enables cooperators to better resist defectors under resource scarcity. Systematic simulation experiments validate the effectiveness of PPO-ACT. Results show this framework outperforms Q-learning and traditional evolutionary game methods by significantly enhancing cooperative behavior equilibrium levels and effectively suppressing free-riding diffusion. It shows strong adaptability to diverse initial conditions and changes in the environment. The Adversarial Curriculum Transfer (ACT) process facilitates quicker convergence by biasing the population toward cooperative strategies, outperforming random initialization approaches. Our research offers novel theoretical perspectives for understanding cooperation evolution mechanisms in social dilemmas through the lens of curriculum-based reinforcement learning. The methodological innovations also provide new analytical tools applicable to related studies in economics, ecology, and sociology.
	
	The paper is structured as follows. Section \ref{sec:model} introduces the proposed model and elaborates the strategy update rules.  Section \ref{sec:exp} describes the simulation experiments and provides result analysis. In Section \ref{sec:con}, the conclusion summarizes the findings.
	
	\vspace{-3mm}
	\section{Model}
	\label{sec:model}
	
	Consider a spatial PGG model defined on an $N=L\times{L}$ regular grid with periodic boundary conditions and von Neumann neighborhood ($k=4$). Each grid cell represents an agent, where agents form teams with their $k=4$ nearest neighbors for gameplay. Consider an agent set $\mathcal{\hat{A}} = \{\hat{a}_1,...,\hat{a}_N\}$ where each agent participates in $G=5$ overlapping game groups. Each group $\mathcal{G}_i$ is centered around an agent $i \in \mathcal{\hat{A}}$, forming a local interaction neighborhood for evolutionary game dynamics. Define the strategy space $\mathcal{S}=\{C,D\}$, where $C$ denotes the cooperation strategy and $D$ denotes defection.  The cooperation strategy contributes 1 unit to the public pool while the defection strategy contributes nothing. For any game group $g\in\mathcal{G}_x$, let $N_C^g$ denote the count of cooperators in the group, where each group consists of $k+1$ members. The resulting individual payoff function for the group is given by:
	\begin{equation}
		\Pi(s_i^g) = 
		\begin{cases}
			\frac{rN_C^g}{k+1} - 1, & s_i^g = C \\
			\frac{rN_C^g}{k+1}, & s_i^g = D
		\end{cases},
	\end{equation}
	where, $\Pi(s_i^g)$ denotes the immediate payoff of agent $i$ in group $g$, $N_C^g = \sum_{j \in g} \mathbb{I}(s_j^g=C)$ is the number of cooperators in group $g$. $r>1$ is the enhancement factor. $s_i^g \in \{C,D\}$ denotes the strategy adopted by agent $i$ in group $g$. The cumulative payoff of agent $i$ is the sum of payoffs from all game groups it participates in:
	\begin{equation}
		\Pi_i = \sum_{g\in\mathcal{G}_i} \Pi(s_i^g).
	\end{equation}
	
	We integrate the deep reinforcement learning algorithm PPO with evolutionary game theory. This integration establishes a new dynamical model for studying cooperation evolution in spatial PGG. Our work pioneers the application of the PPO algorithm in spatial evolutionary game research. Building upon this foundation, we further incorporate curriculum learning \cite{bengio_2009_curriculum} with PPO to develop PPO-ACT, which enhances cooperative strategy adaptation through staged environmental challenges. ACT implements a two-stage curriculum learning scheme that transfers the PPO from high-reward to low-reward conditions. The following sections will provide detailed explanations of the key concepts in this model.
	
	\subsection{PPO}
	
	The PPO framework employs an Actor-Critic architecture as its core component, where the PPO algorithm \cite{John_2017_arxiv} directly optimizes parameterized policy functions through policy gradient methods. The PPO algorithm demonstrates fundamental improvements over traditional policy gradient methods through its clipped objective function. This clipping mechanism explicitly constrains policy update magnitudes to prevent training instability caused by excessive policy oscillations. The approach simultaneously enhances both algorithmic stability and sample efficiency, addressing key limitations of conventional methods. The PPO in spatial PGG employs an Actor-Critic architecture where the Actor-network generates policy distributions, and the Critic network evaluates state values. This dual-network design allows agents to balance immediate rewards against long-term evolutionary outcomes when choosing between cooperation and defection strategies. This integrated modeling framework extends beyond conventional evolutionary game theory by introducing new analytical dimensions. It establishes a computationally efficient yet theoretically rigorous approach for investigating cooperation dynamics in complex social dilemmas. Our PPO-ACT framework effectively captures the complex interactions between agent-level policy learning and population-level behavioral patterns using policy gradient methods. It also achieves endogenous expression of social norms in the policy optimization process. Furthermore, it offers new theoretical perspectives for understanding the generation and maintenance mechanisms of cooperative behaviors in the real world. The PPO objective function comprises three key components: the clipped policy objective, value function objective, and entropy regularization term.
	
	The clipped policy objective function is given by:
	\begin{equation}
		L^{CLIP}(\theta) = \mathbb{E}_t \left[ \min \left( r_t(\theta) \cdot A_t, \text{clip}(r_t(\theta), 1-\varepsilon, 1+\varepsilon) \cdot A_t \right) \right],
	\end{equation}
	where $\varepsilon$ is the clipping parameter that constrains the magnitude of policy updates. $\mathbb{E}_t$ denotes the conditional expectation operator at timestep $t$. The operator $\text{clip}(r_t(\theta), 1-\varepsilon, 1+\varepsilon)$ restricts $r_t(\theta)$ within $[1-\varepsilon, 1+\varepsilon]$. Here $r_t(\theta)$ denotes the policy update ratio:
	\begin{equation}
		r_t(\theta) = \frac{\pi_\theta(a_t|s_t)}{\pi_{\theta_{\text{old}}}(a_t|s_t)},
	\end{equation}
	where $\pi_\theta(a_t|s_t)$ is the current policy representing the probability of selecting action $a_t$ at state $s_t$ in timestep $t$. $\pi_{\theta_{\text{old}}}(a_t|s_t)$ denotes the old policy. $\theta$ represents the trainable parameter set of the Policy Network. In the PGG, the agent's action $ a_t $ is a binary contribution choice $ \hat{c} \in \{0,1\} $, where $ \hat{c} = 1 $ denotes cooperation and $ \hat{c} = 0 $ denotes defection. The state $ s_t $ includes both the agent's historical contribution record and the neighbors' contribution information. The advantage function $A_t$ measures the relative benefit of taking action $a_t$ at state $s_t$, computed via generalized advantage estimation with the formula:
	\begin{equation}
		A_t= \sum_{l=0}^{\infty} (\gamma \lambda)^l \psi_{t+l},
	\end{equation}
	where $\psi_t = r_t + \gamma V(s_{t+1}) - V(s_t)$ is the temporal difference error, representing the difference between the current reward at timestep $t$ and value function predictions. $r_t$ denotes the immediate reward obtained at timestep $t$. $\gamma\in[0, 1)$ is the discount factor that determines the importance of future rewards. $\lambda\in[0,1)$ controls the weighting of future advantage estimations. $V(s_t)$ is the state-value function representing the expected cumulative reward at state $s_t$, and $V(s_{t+1})$ is the state-value function for the next state $s_{t+1}$.
	
	The value function objective is given by:
	\begin{equation}
		L^{VF}(\theta) = \mathbb{E}_t \left[ \left( V_\theta(s_t) - V_t^{\text{target}} \right)^2 \right],
	\end{equation}
	where $V_\theta(s_t)$ is the state-value function parameterized by $\theta$, representing the expected cumulative reward at state $s_t$. $V_t^{\text{target}}$ is the target value based on actual returns.
	
	The entropy regularization term $L^{ENT}(\theta)$ promotes policy exploration:
	\begin{equation}
		L^{ENT}(\theta) = \mathbb{E}_t \left[ -\pi_\theta(a_t|s_t) \log \pi_\theta(a_t|s_t) \right].
	\end{equation}
	
	The final PPO objective function is:
	\vspace{-1mm}
	\begin{equation}
		L^{PPO}(\theta) = L^{CLIP}(\theta) - \delta \cdot L^{VF}(\theta) + \rho \cdot L^{ENT}(\theta),
		\vspace{-1mm}
	\end{equation}
	where $\theta$ is the weight hyperparameter for the value function objective, balancing the importance between policy optimization and value function fitting. $\rho$ is the weight hyperparameter for the entropy regularization term, controlling the degree of policy exploration. This objective function is maximized with respect to $\theta$ during policy updates.
	
\begin{figure}
		\centering
		\includegraphics[width=0.75\linewidth]{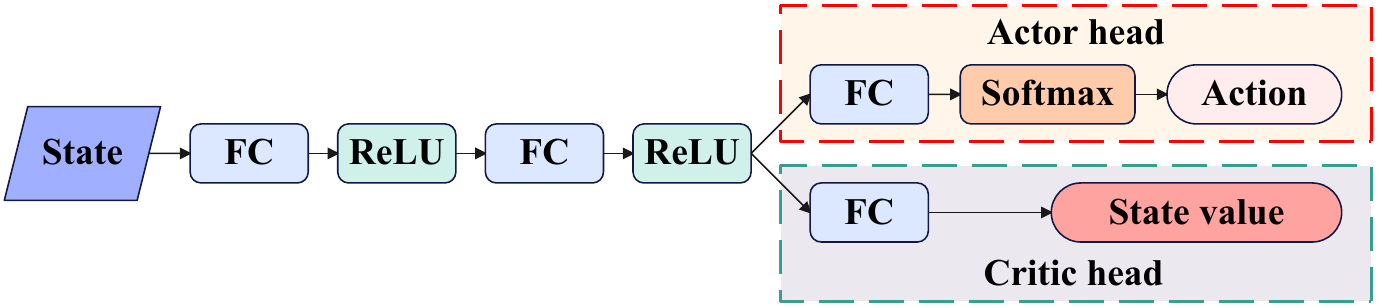}
		\caption{The PPO-ACT framework employs an Actor-Critic architecture where a shared feature extractor processes environmental states, enabling simultaneous optimization of both policy distribution $\pi(a|s)$ through the Actor pathway and state-value estimation $V(s)$ via the Critic pathway, with parameter sharing between these components significantly enhancing spatial correlation learning in PGG.}
		\label{fig:AC}
		\vspace{-5mm}
	\end{figure}

	In PPO, the policy loss $\pi_\theta(a_t|s_t)$ (Actor part) and value loss $V_\theta(s_t)$ (Critic part) are computed by the network shown in Figure \ref{fig:AC}. The network takes the current state as input and shares a feedforward neural network layer for state feature extraction. This feedforward network consists of two fully connected layers with ReLU \cite{Glorot_2011_relu} activation functions. The shared layer takes the input state $\mathbf{x}$ and outputs the hidden representation $\mathbf{h}$:
	\vspace{-1mm}
	\begin{equation}
		\mathbf{h} = \text{ReLU}(\mathbf{W}_2 \cdot \text{ReLU}(\mathbf{W}_1 \cdot x + \mathbf{b}_1) + \mathbf{b}_2),
		\vspace{-1mm}
	\end{equation}
	where $ \mathbf{W}_1, \mathbf{W}_2 $ are weight matrices and $ \mathbf{b}_1, \mathbf{b}_2 $ are bias terms.
	
	The Actor network consists of a shared layer, fully connected layers, and softmax. The Actor part takes the hidden representation $\mathbf{h}$ as input and outputs the action probability distribution $ \pi(a_t|s_t) $. The Actor head maps the hidden representation $ \mathbf{h} $ to action logits $ \mathbf{a} \in \mathbb{R}^{2} $, corresponding to two possible actions (e.g., cooperate or defect):
	\vspace{-1mm}
	\begin{equation}
		\mathbf{a} = \mathbf{W}_{\text{actor}} \cdot \mathbf{h} + \mathbf{b}_{\text{actor}},
		\vspace{-1mm}
	\end{equation}
	where $ \mathbf{W}_{\text{actor}}$ and $ \mathbf{b}_{\text{actor}} $ are the weight and bias of the Actor head.
	
	The action probabilities $\mathbf{p} \in \mathbb{R}^{2}$ are obtained via the softmax function:
	\vspace{-1mm}
	\begin{equation}
		\pi(a|s) = \text{softmax}(\mathbf{a}).
		\vspace{-1mm}
	\end{equation}
	
	The Critic network consists of shared and fully connected layers. The Critic head takes the hidden representation $\mathbf{h}$ as input and outputs the state value $V(s)$:
	\vspace{-1mm}
	\begin{equation}
		V(s) = \mathbf{W}_{critic} \cdot \mathbf{h} + \mathbf{b}_{critic},
		\vspace{-1mm}
	\end{equation}
	where $\mathbf{W}_{\text{critic}}$ and $\mathbf{b}_{\text{critic}}$ are the weight matrix and bias term of the Critic head, respectively.
	
	\vspace{-1mm}
	\subsection{PPO-ACT}
	PPO-ACT enables dynamic adjustment between cooperation and defection behaviors in spatial PGG through a synergistic combination of policy optimization and curriculum learning. The framework integrates two core mechanisms: PPO performs gradient-based policy optimization, and ACT facilitates strategy transfer across varying game conditions. An agent's state $s_t$ includes its current strategy (cooperation or defection), neighbors' strategy distribution, and public pool contribution status. The agent's action $a_t$ is strategy selection, either cooperation ($C$) or defection ($D$). The reward function $R_t$ is defined as the agent's payoff in the game:
	\begin{equation}
		R_t = \Pi_i,
	\end{equation}
	where $\Pi_i$ is the cumulative payoff of agent $i$, calculated by:
	\begin{equation}
		\Pi_i = \sum^G_{g=1} \Pi_i^g,
	\end{equation}
	with $\Pi_i^g$ being agent $i$'s payoff in group $g$.
	
	The complete algorithmic procedure for each iteration is presented in the following pseudocode:

	\vspace{-2.5mm}	
	\begin{algorithm}[H]
		\caption{PPO-ACT Framework for Spatial PGG}
		\scalebox{0.76}{%
			\begin{minipage}{1.1\linewidth} 
		\begin{algorithmic}[1]
			\State \textbf{Initialize:}
			\State \quad Enhancement factor $r_1$, $r_2$
			\State \quad Train epochs $T_1$, $T_2$
			\State \quad Policy $\pi_\theta$, value function $V_\phi$ with shared features

			\State \textbf{Phase 1: Cooperative Policy Initialization}
			\State Initial enhancement factor $r_1$
			\For{$t = 1$ \textbf{to} $T_1$}
			\For{each agent $i$}
			\State Select action $a_t$ according to current policy $\pi_\theta$
			\State Execute PGG, compute payoff $\Pi_i$ and reward $\mathcal{R}_t$
			\EndFor
			\State Standard PPO updates:
			\State \quad Compute advantage function $A_t$ and target value $V_t^{\text{target}}$
			\State \quad Compute PPO objective function $L^{PPO}(\theta)$
			\State \quad Update policy network $\pi_\theta$ and value network $V_\theta$ via gradient descent
			\EndFor

			\State \textbf{Phase 2: Adversarial Curriculum Transfer}
			\State Initial enhancement factor $r_2$
			\For{$t = 1$ \textbf{to} $T_2$}
			\For{each agent $i$}
			\State Select action $a_t$ according to current policy $\pi_\theta$
			\State Execute PGG, compute payoff $\Pi_i$ and reward $\mathcal{R}_t$
			\EndFor
			\State Standard PPO updates:
			\State \quad Compute advantage function $A_t$ and target value $V_t^{\text{target}}$
			\State \quad Compute PPO objective function $L^{PPO}(\theta)$
			\State \quad Update policy network $\pi_\theta$ and value network $V_\theta$ via gradient descent
			\EndFor
		\end{algorithmic}
	\end{minipage}%
} 
	\end{algorithm}
	
	\section{Experimental results}
	\label{sec:exp}
	\subsection{Experimental setup}
	
	Table \ref{table:para} shows the default experimental parameter settings. The model parameters are optimized using the Adam optimizer \cite{Diederik_2015_ICLR} with an initial learning rate of $\alpha$. We employ PyTorch's StepLR learning rate scheduler to enhance training stability and convergence. This scheduler multiplies the learning rate by 0.5 every 1000 iterations during training. This scheduling strategy ensures that the learning rate gradually decreases as training progresses, enabling the model to fine-tune its parameters more effectively. Initial parameters of $r=5.0$, $\alpha=0.001$, $\gamma=0.99$, and $\rho=0.01$ are employed during cooperative policy initialization training to bolster agent exploratory behavior. In the experiments, defection behavior is represented by 0 (black) in the grid plots, while cooperation behavior is represented by 1 (white).
	\begin{table}[h]
		\centering
		\caption{Default Experimental Parameters}
		\begin{tabular}{ccc}
			\toprule
			Parameter & Value & Description \\
			\midrule
			$L$ & $200$ & Side length of initialized grid \\
			$\alpha$ & $0.01$ & Initial learning rate \\
			$\varepsilon$ & 0.2 & Clipping parameter for policy updates \\
			$\gamma$ & 0.96 & Discount factor for future rewards \\
			$\lambda$ & 0.95 & Weight for future advantage estimation \\
			$\delta$ & 0.5 & Weight for value function loss \\
			$\rho$ & 0.001 & Weight for entropy regularization \\
			$r_1$ & 5.0 & Enhancement factor of Phase 1 \\
			$r_2$ & 4.0 & Enhancement factor of Phase 2 \\
			$T_1$ & 1000 & Train epochs of Phase 1 \\
			$T_2$ & 9000 & Train epochs of Phase 2 \\
			\bottomrule
		\end{tabular}
		\label{table:para}
	\end{table}

	\subsection{PPO-ACT with half-and-half initialization}
	\label{exp_hh}
	
\begin{figure*}[htbp!]
		\begin{minipage}{0.45\linewidth}
			\begin{minipage}{\linewidth}
				\centering
				\includegraphics[width=\linewidth]{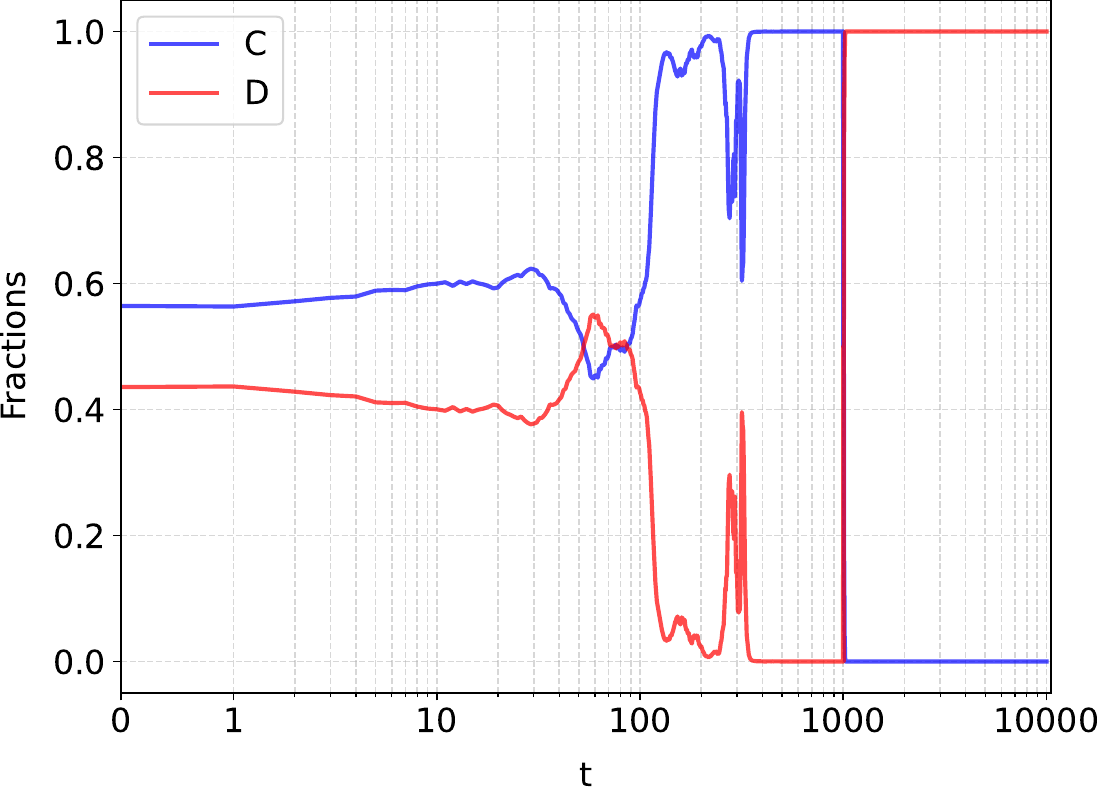}\\
			\end{minipage}
			\vspace{1mm}
			\\
			\begin{minipage}{0.188\linewidth}
				\centering
				\fbox{\includegraphics[width=\linewidth]{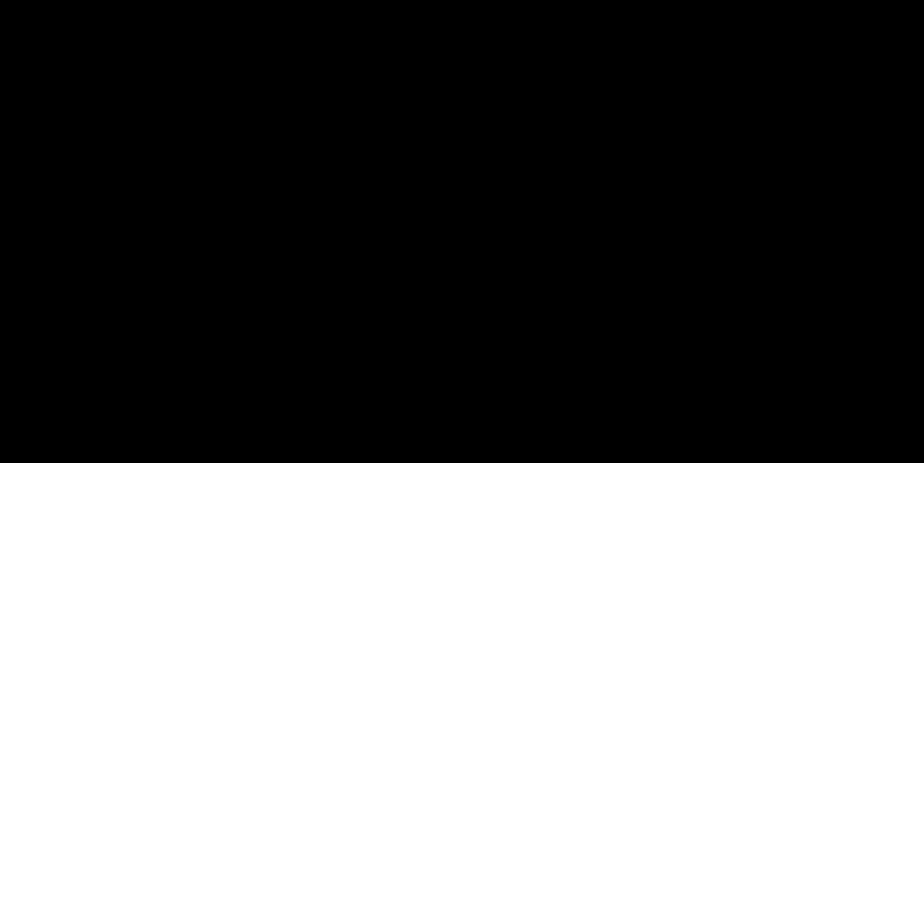}}\\
				\vspace{-2mm}
				{\footnotesize t=0}
			\end{minipage}
			\begin{minipage}{0.188\linewidth}
				\centering
				\fbox{\includegraphics[width=\linewidth]{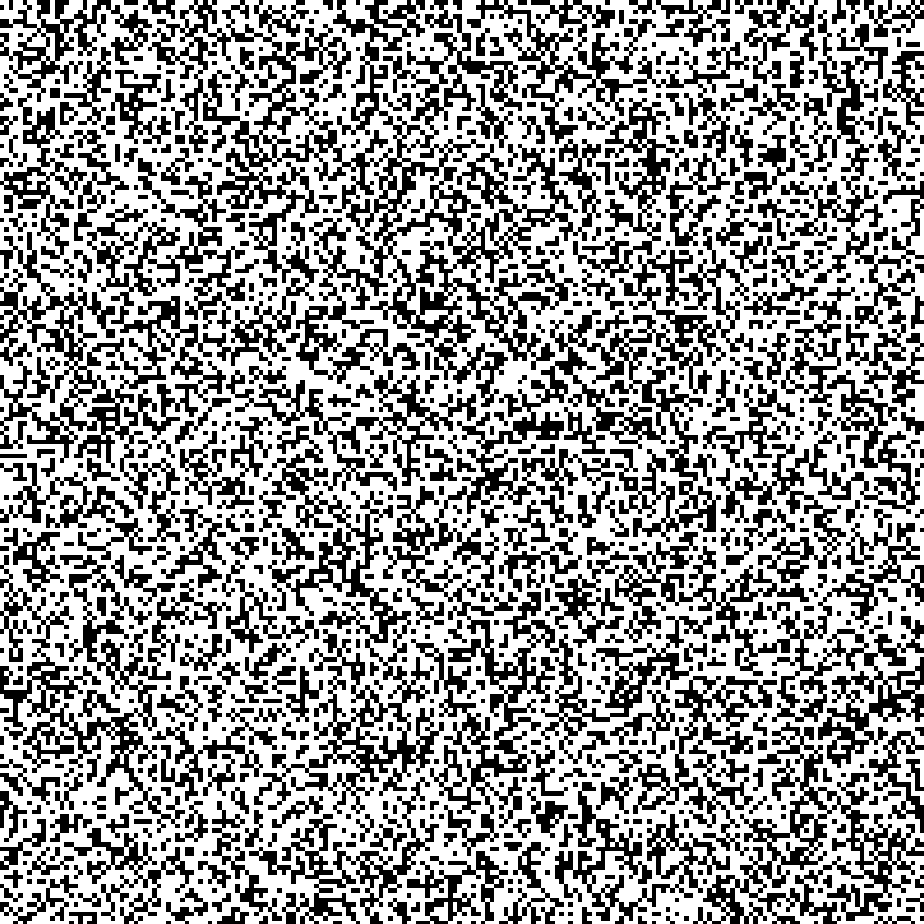}}\\
				\vspace{-2mm}
				{\footnotesize t=10}
			\end{minipage}
			\begin{minipage}{0.188\linewidth}
				\centering
				\fbox{\includegraphics[width=\linewidth]{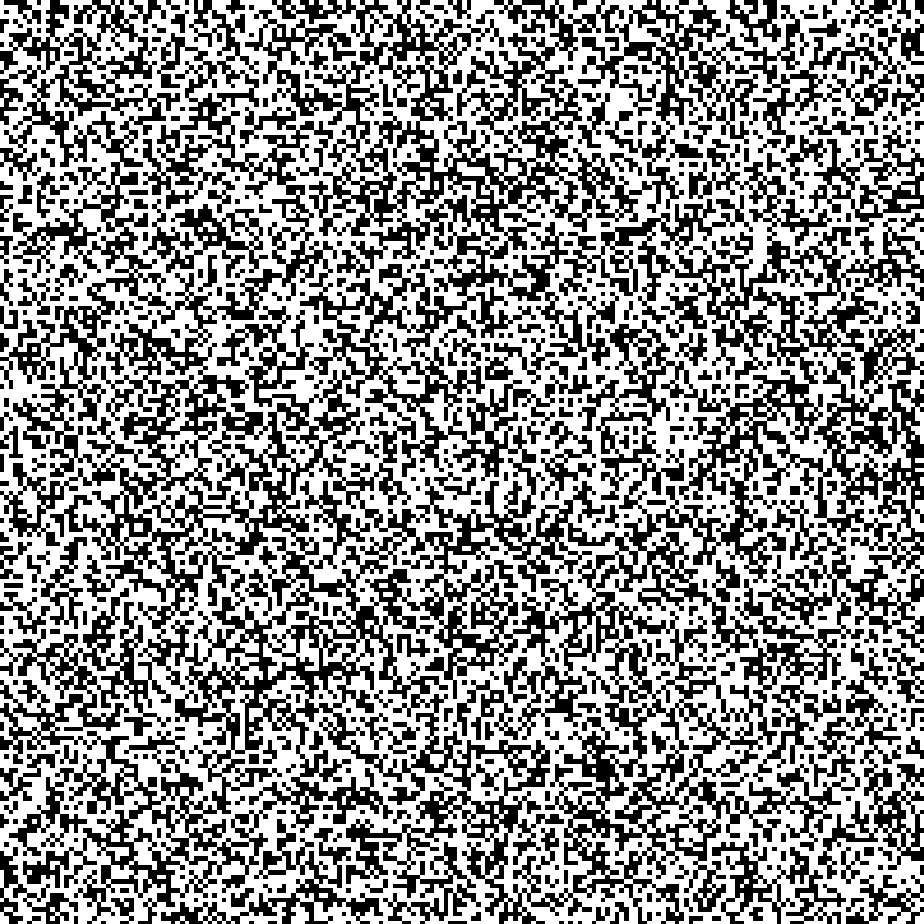}}\\
				\vspace{-2mm}
				{\footnotesize t=100}
			\end{minipage}
			\begin{minipage}{0.188\linewidth}
				\centering
				\fbox{\includegraphics[width=\linewidth]{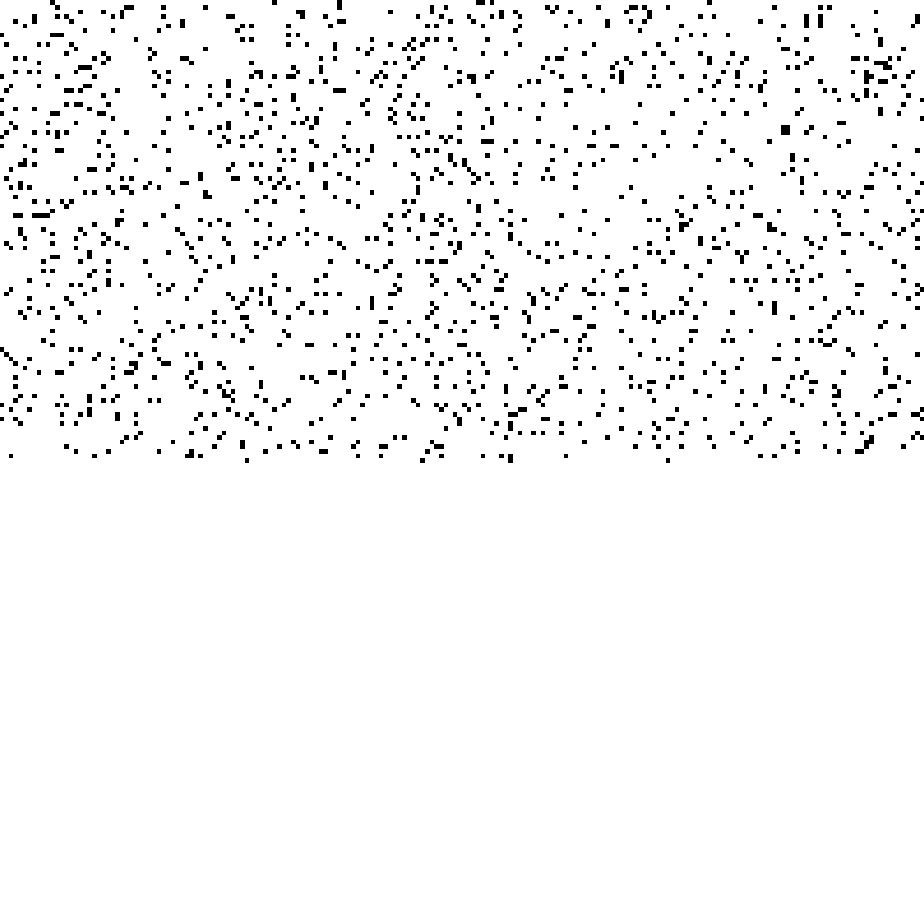}}\\
				\vspace{-2mm}
				{\footnotesize t=1000}
			\end{minipage}
			\begin{minipage}{0.188\linewidth}
				\centering
				\fbox{\includegraphics[width=\linewidth]{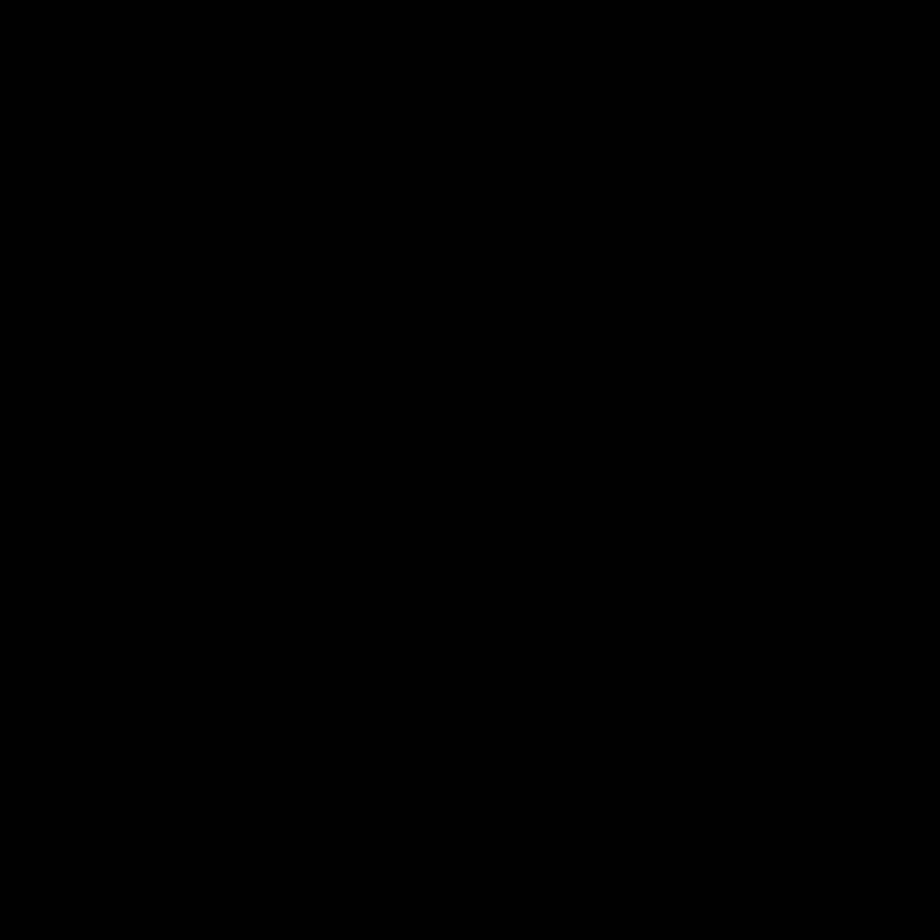}}\\
				\vspace{-2mm}
				{\footnotesize t=10000}
			\end{minipage}
			\vspace{-2mm}
			\caption*{\footnotesize (a) r=3.0 (Phase 1+2)}
		\end{minipage}
		\hfill
		\begin{minipage}{0.45\linewidth}
			\begin{minipage}{\linewidth}
				\centering
				\includegraphics[width=\linewidth]{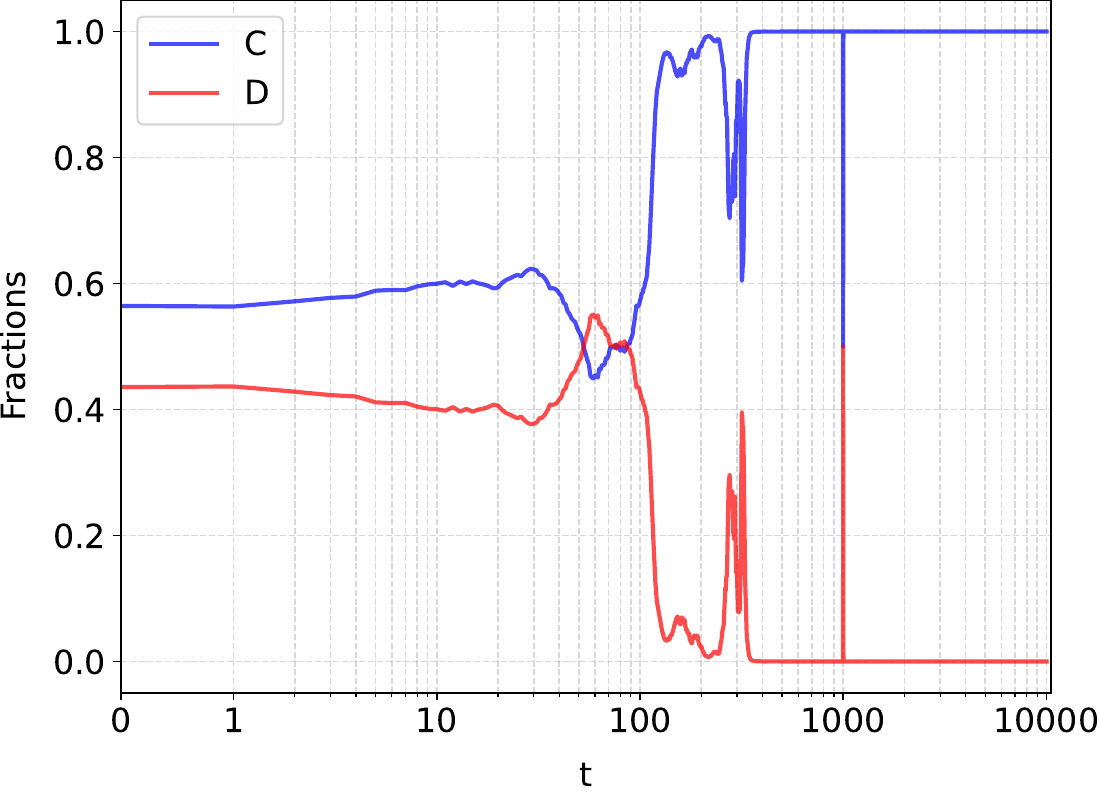} \\
			\end{minipage}
			\vspace{1mm}
			\\
			\begin{minipage}{0.188\linewidth}
				\centering
				\fbox{\includegraphics[width=\linewidth]{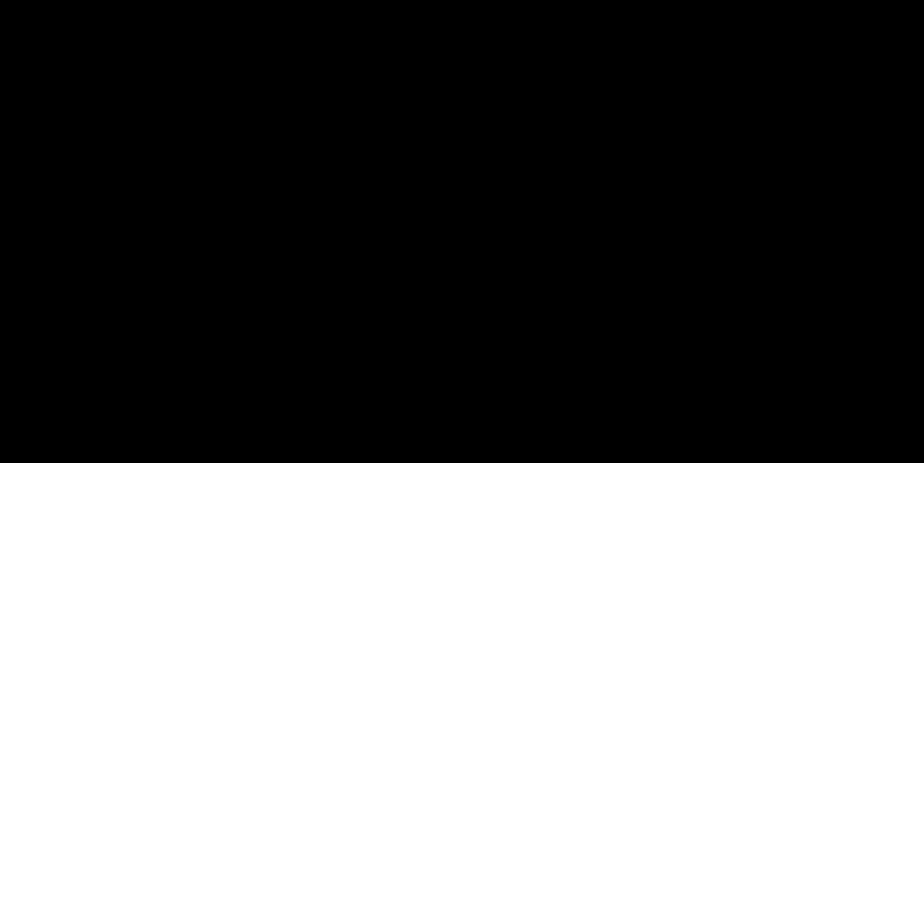}}\\
				\vspace{-2mm}
				{\footnotesize t=0}
			\end{minipage}
			\begin{minipage}{0.188\linewidth}
				\centering
				\fbox{\includegraphics[width=\linewidth]{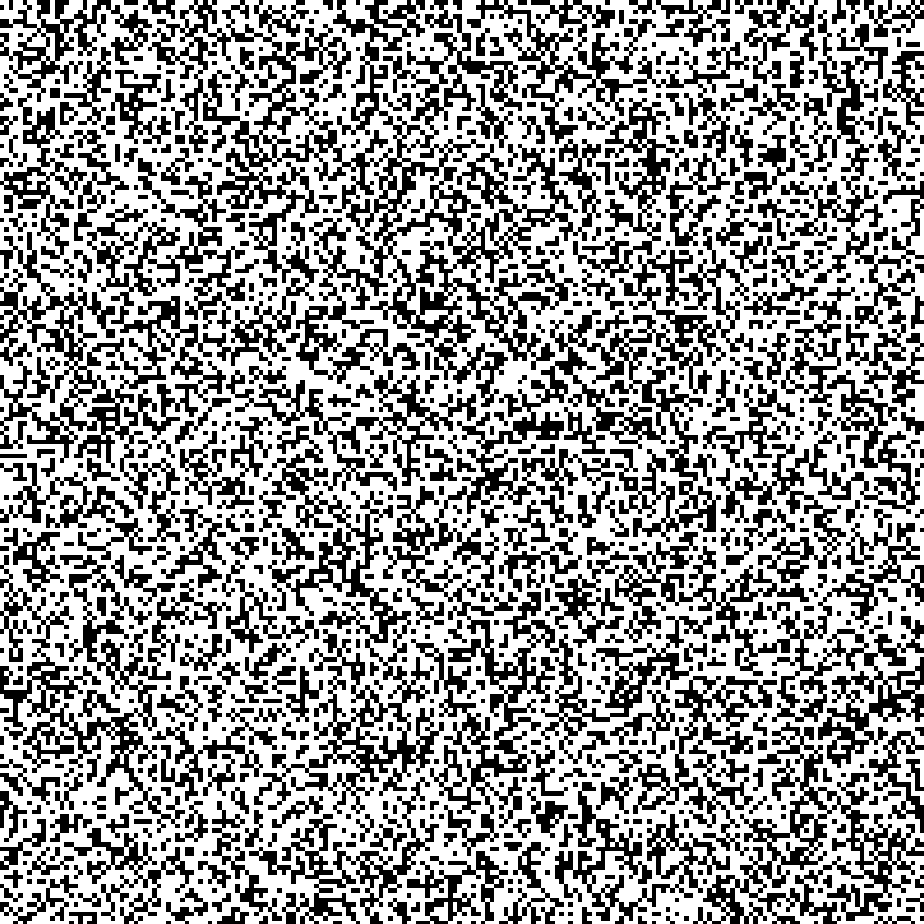}}\\
				\vspace{-2mm}
				{\footnotesize t=10}
			\end{minipage}
			\begin{minipage}{0.188\linewidth}
				\centering
				\fbox{\includegraphics[width=\linewidth]{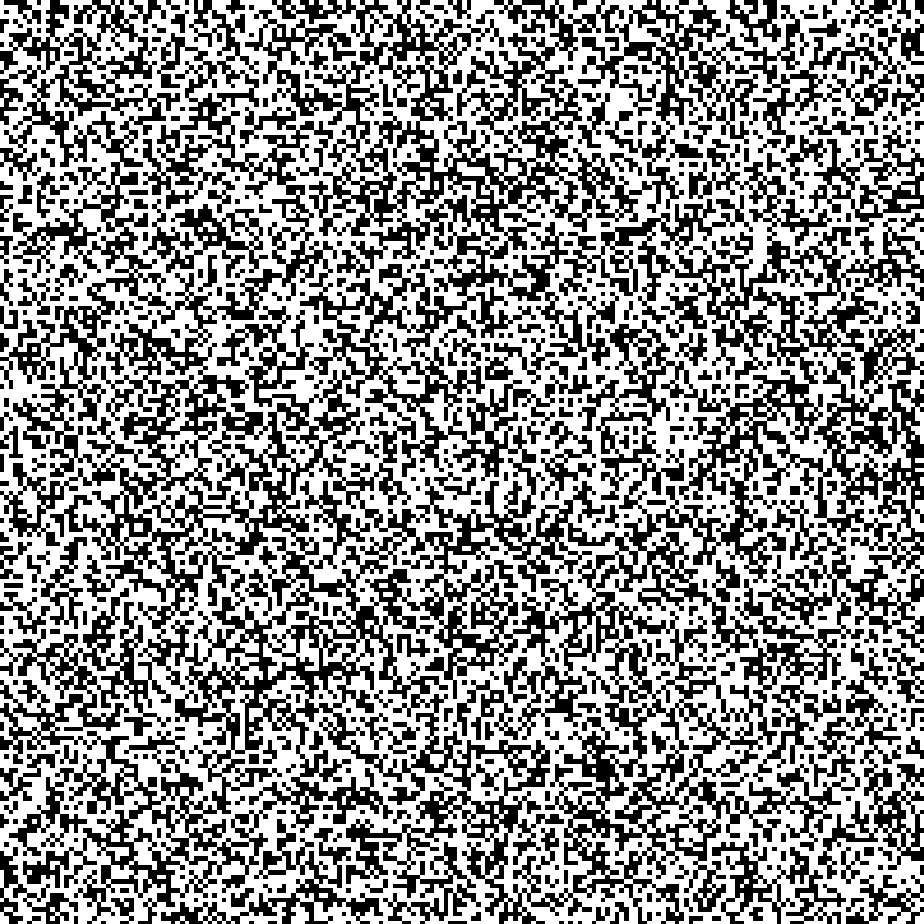}}\\
				\vspace{-2mm}
				{\footnotesize t=100}
			\end{minipage}
			\begin{minipage}{0.188\linewidth}
				\centering
				\fbox{\includegraphics[width=\linewidth]{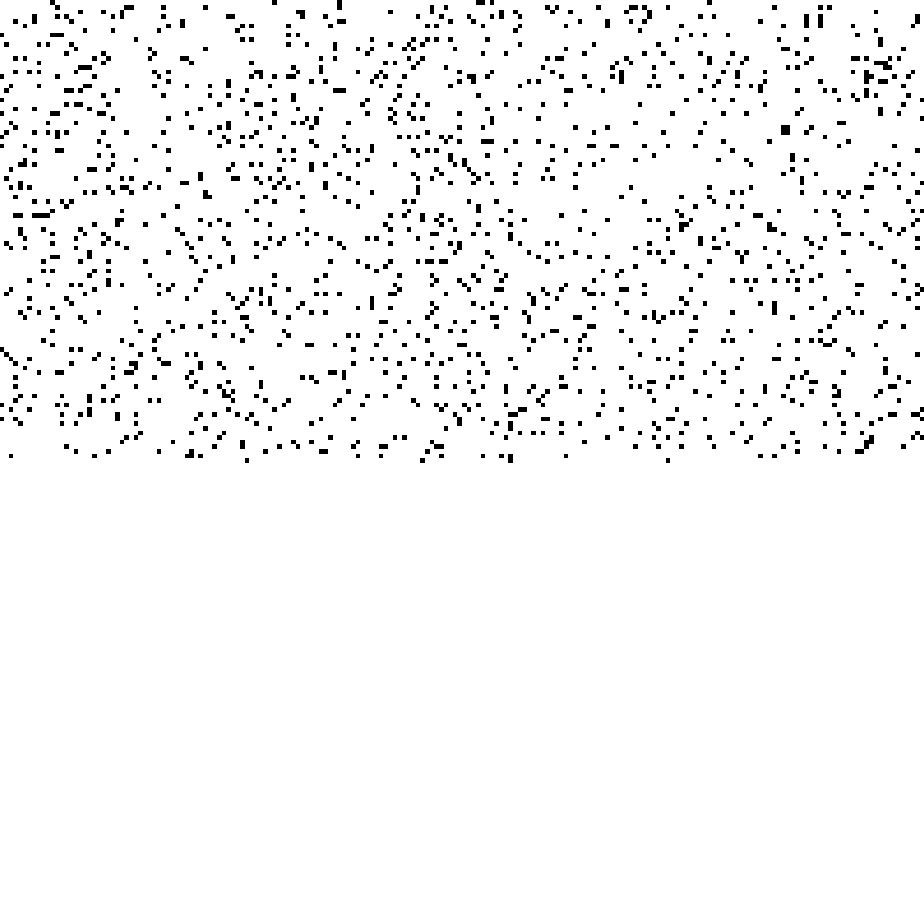}}\\
				\vspace{-2mm}
				{\footnotesize t=1000}
			\end{minipage}
			\begin{minipage}{0.188\linewidth}
				\centering
				\fbox{\includegraphics[width=\linewidth]{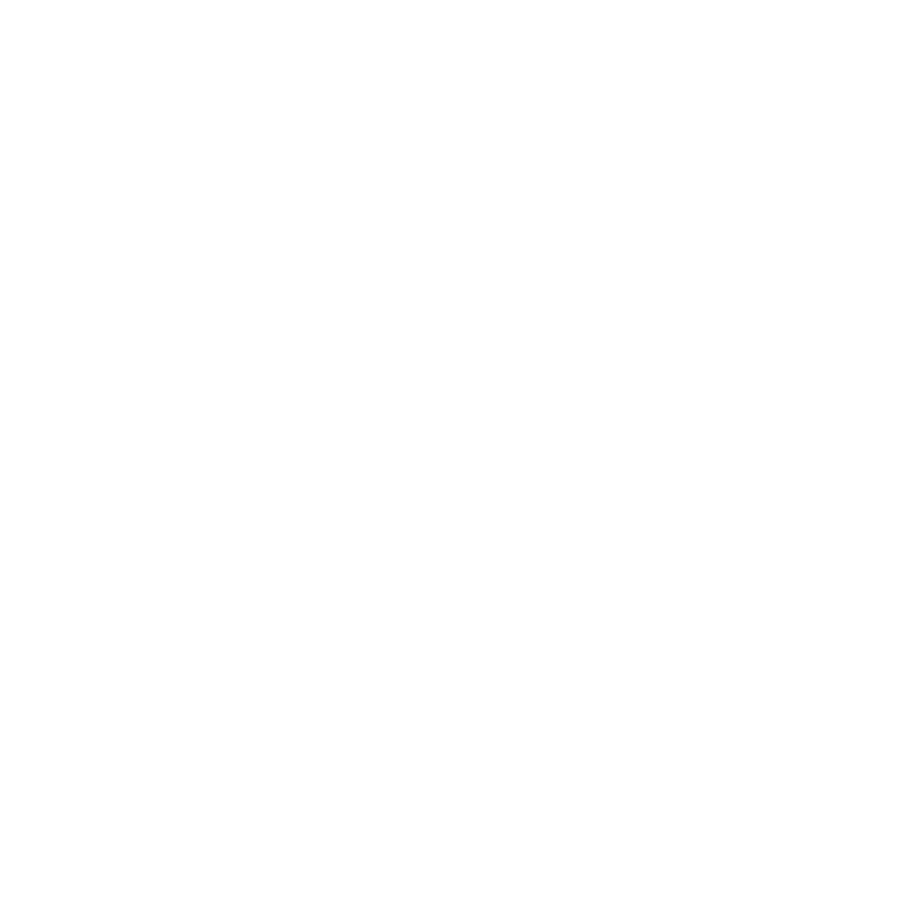}}\\
				\vspace{-2mm}
				{\footnotesize t=10000}
			\end{minipage}
			\vspace{-2mm}
			\caption*{\footnotesize (b) r=4.0 (Phase 1+2)}
		\end{minipage}
		\\[3mm]
		\begin{minipage}{0.45\linewidth}
			\begin{minipage}{\linewidth}
				\centering
				\includegraphics[width=\linewidth]{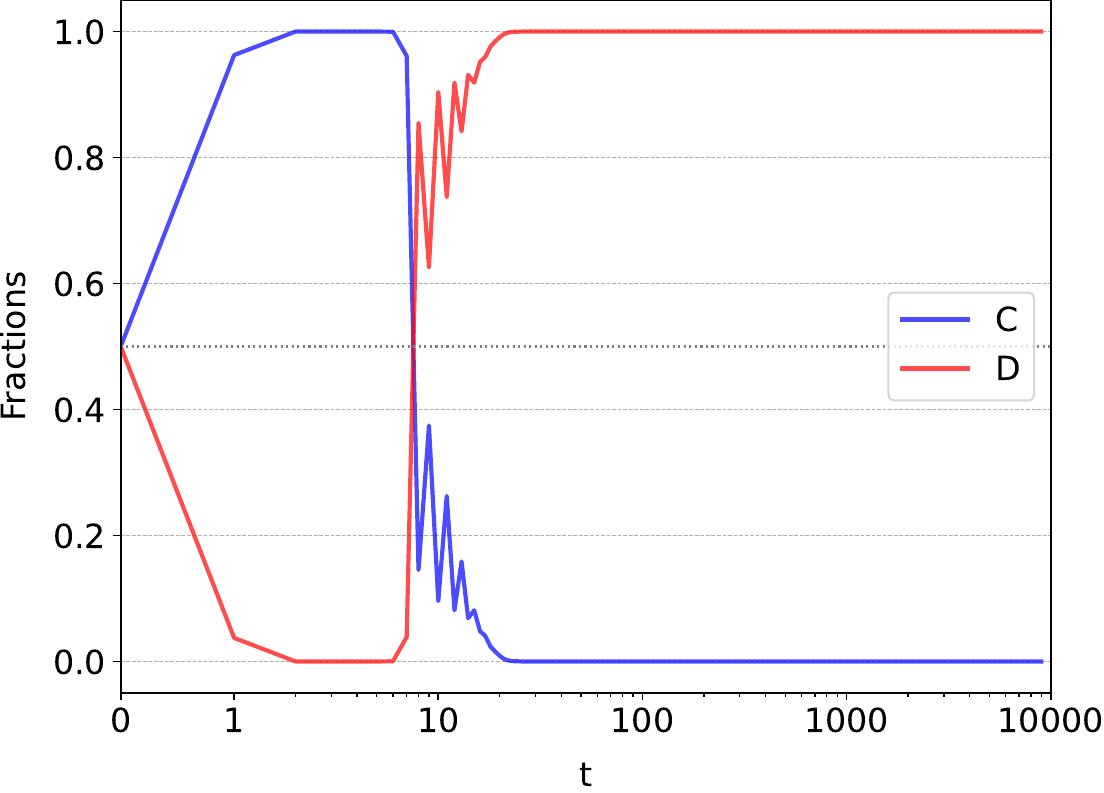}\\
			\end{minipage}
			\vspace{1mm}
			\\
			\begin{minipage}{0.188\linewidth}
				\centering
				\fbox{\includegraphics[width=\linewidth]{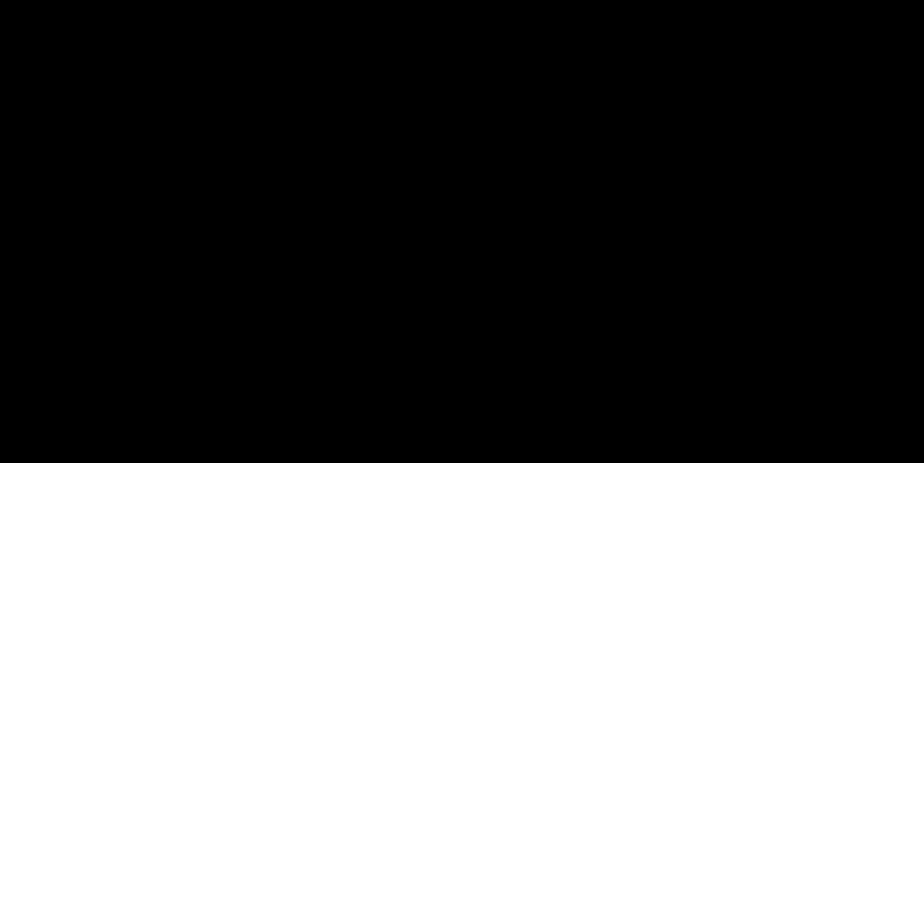}}\\
				\vspace{-2mm}
				{\footnotesize t=0}
			\end{minipage}
			\begin{minipage}{0.188\linewidth}
				\centering
				\fbox{\includegraphics[width=\linewidth]{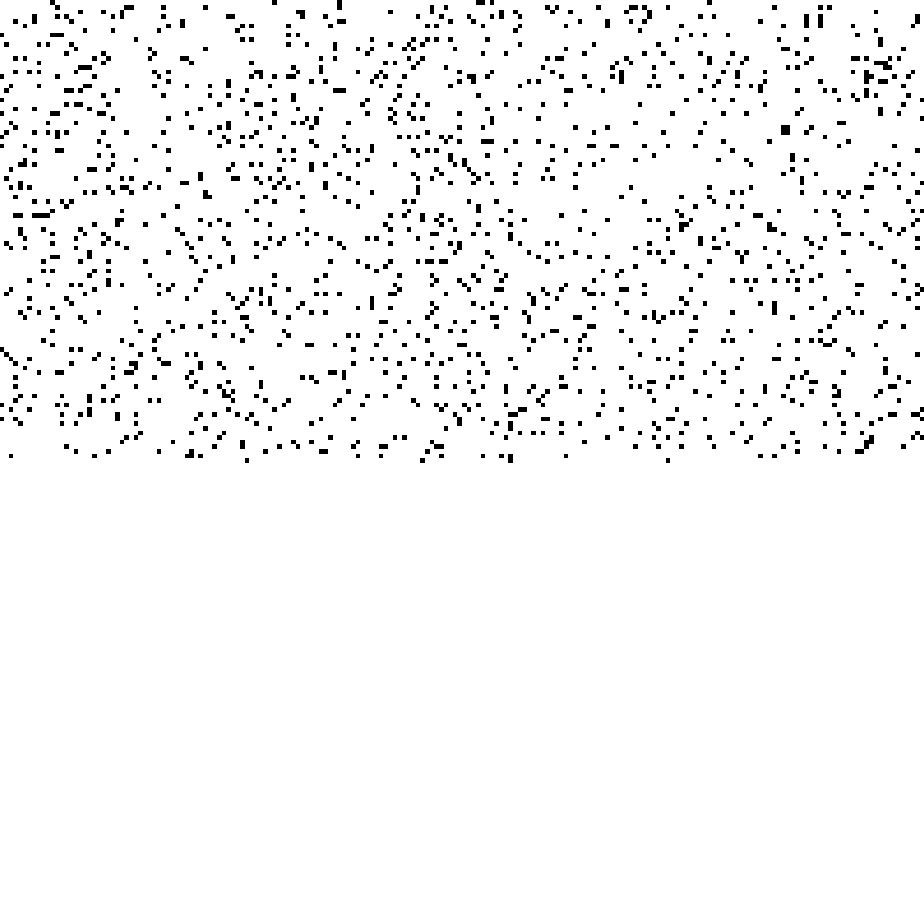}}\\
				\vspace{-2mm}
				{\footnotesize t=1}
			\end{minipage}
			\begin{minipage}{0.188\linewidth}
				\centering
				\fbox{\includegraphics[width=\linewidth]{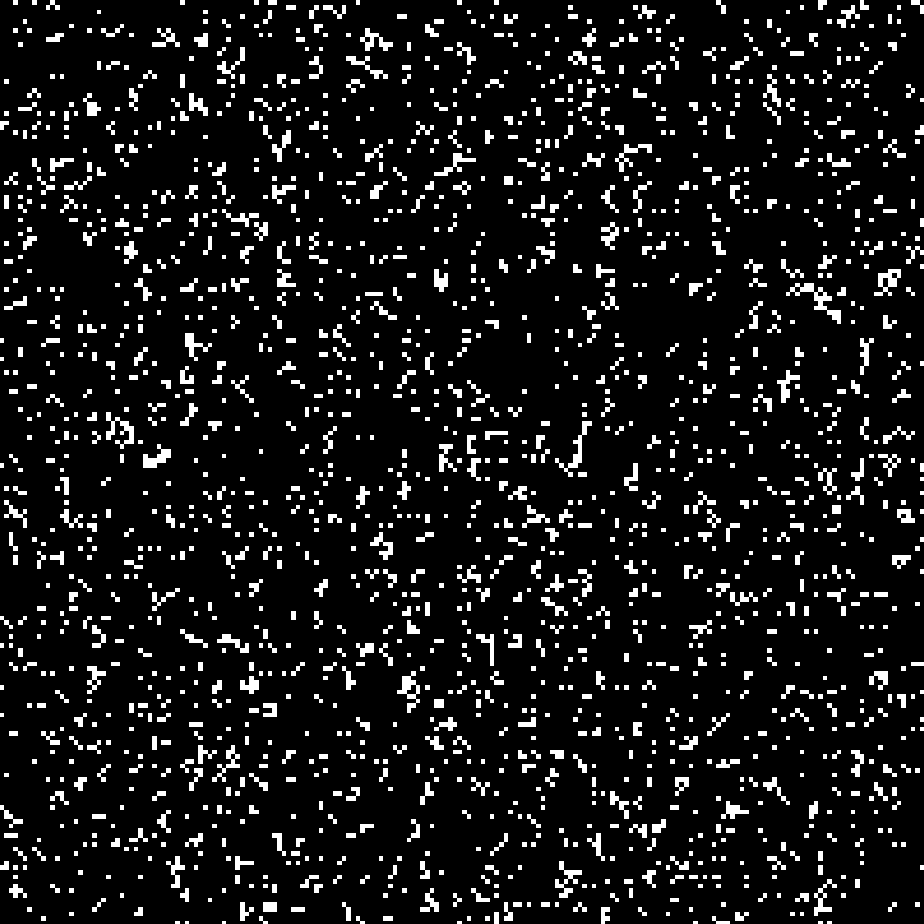}}\\
				\vspace{-2mm}
				{\footnotesize t=10}
			\end{minipage}
			\begin{minipage}{0.188\linewidth}
				\centering
				\fbox{\includegraphics[width=\linewidth]{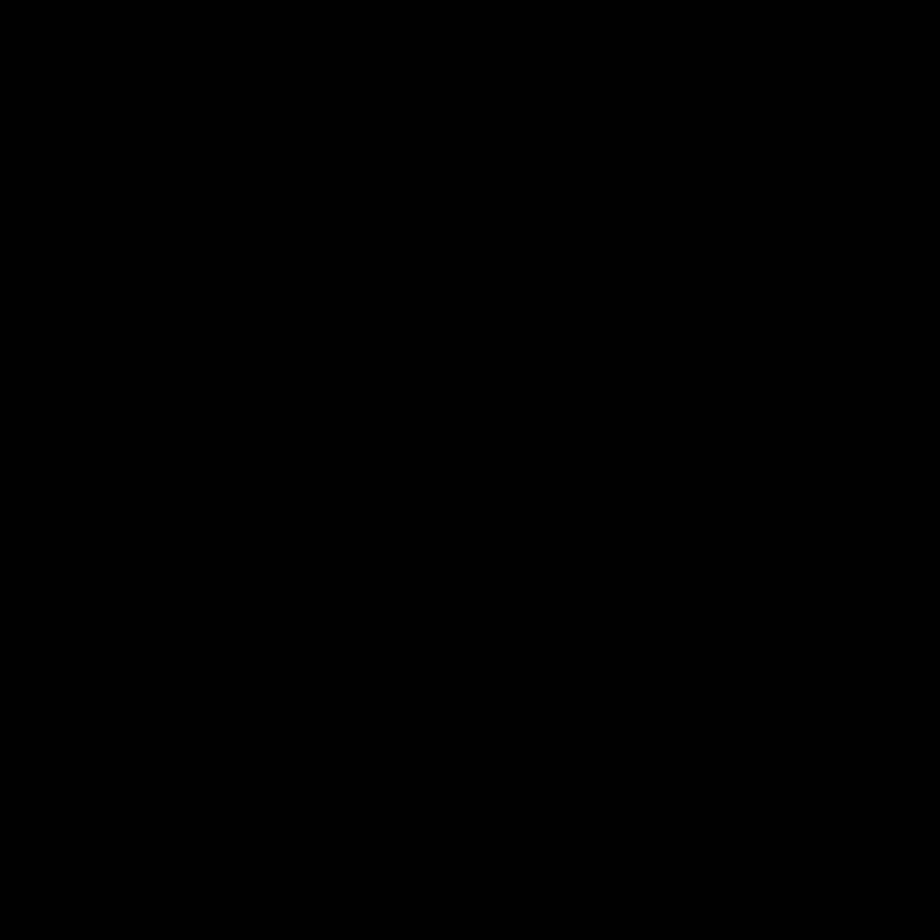}}\\
				\vspace{-2mm}
				{\footnotesize t=100}
			\end{minipage}
			\begin{minipage}{0.188\linewidth}
				\centering
				\fbox{\includegraphics[width=\linewidth]{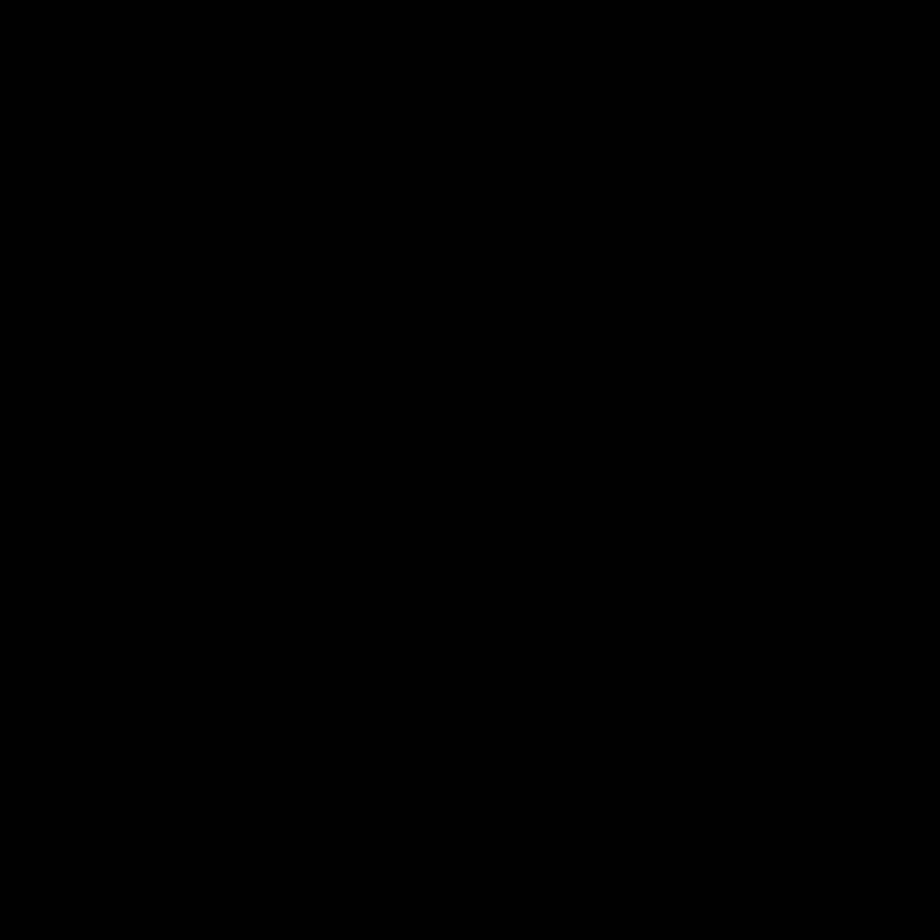}}\\
				\vspace{-2mm}
				{\footnotesize t=1000}
			\end{minipage}
			\vspace{-2mm}
			\caption*{\footnotesize (c) r=3.0 (Phase 2)}
		\end{minipage}
		\hfill
		\begin{minipage}{0.45\linewidth}
			\begin{minipage}{\linewidth}
				\centering
				\includegraphics[width=\linewidth]{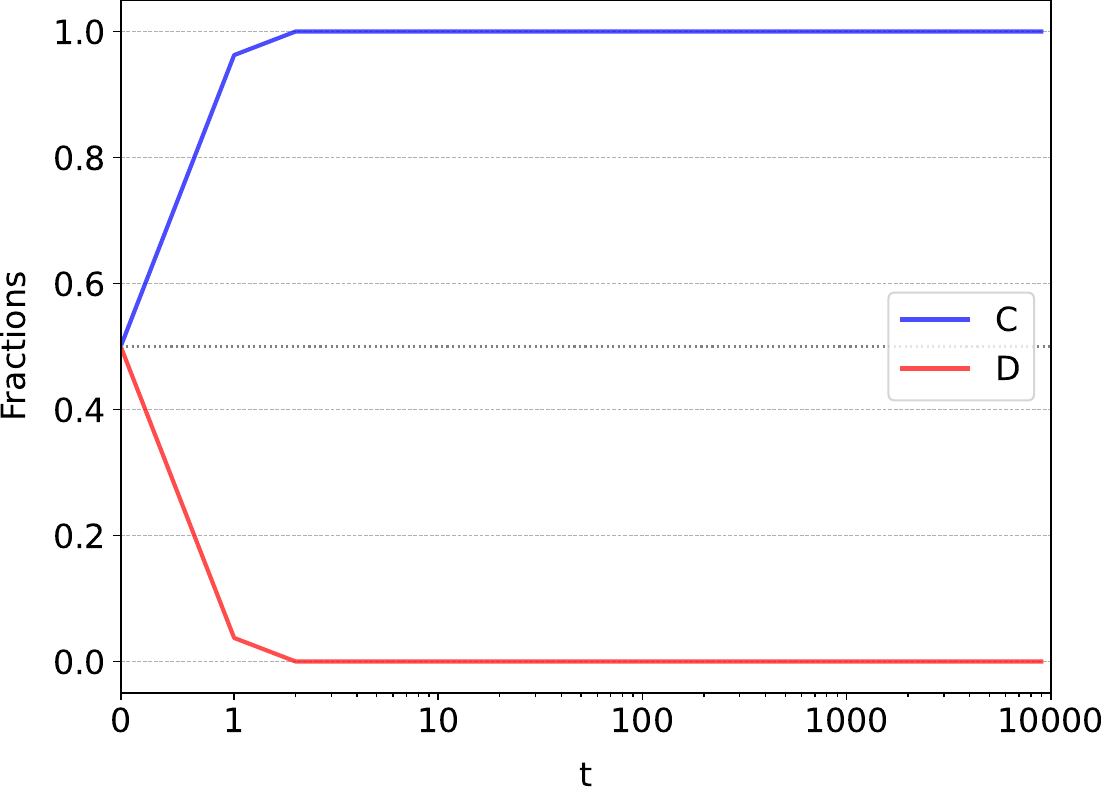}\\
			\end{minipage}
			\vspace{1mm}
			\\
			\begin{minipage}{0.188\linewidth}
				\centering
				\fbox{\includegraphics[width=\linewidth]{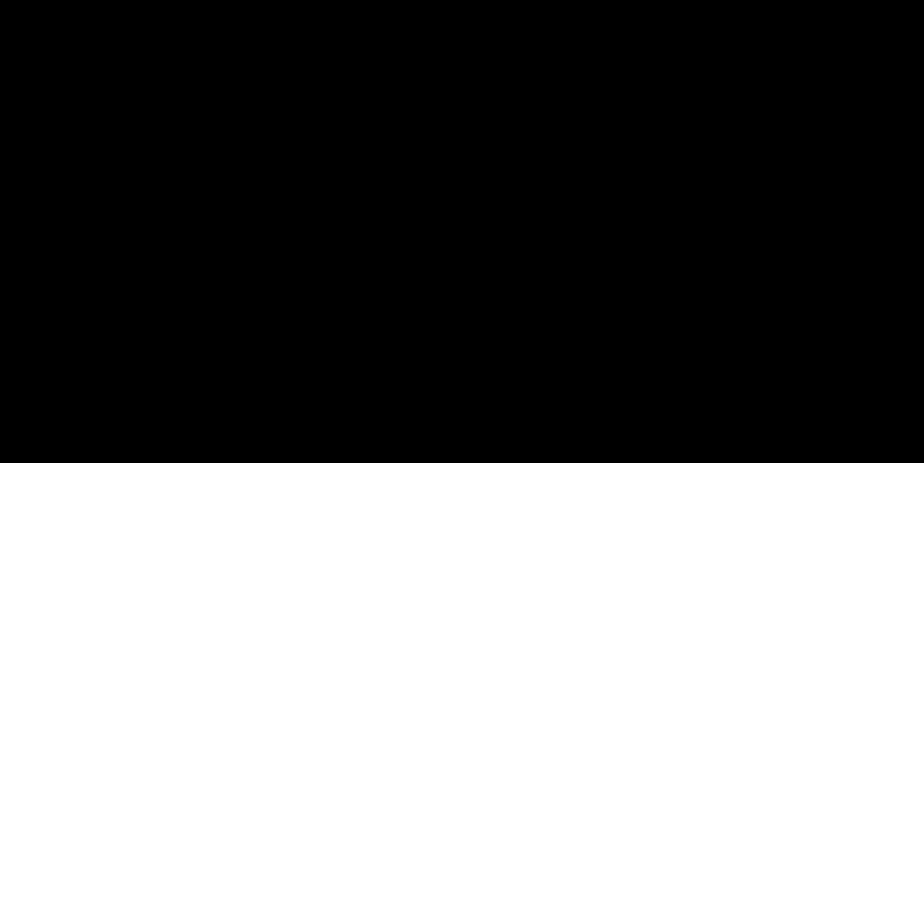}}\\
				\vspace{-2mm}
				{\footnotesize t=0}
			\end{minipage}
			\begin{minipage}{0.188\linewidth}
				\centering
				\fbox{\includegraphics[width=\linewidth]{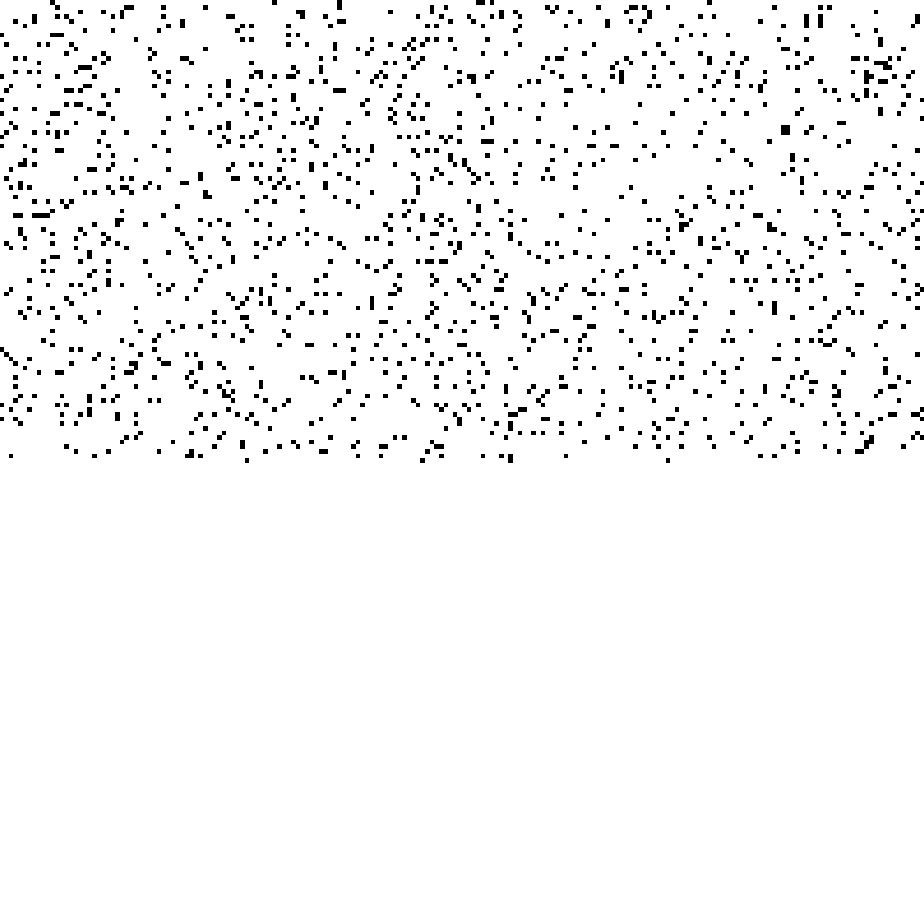}}\\
				\vspace{-2mm}
				{\footnotesize t=1}
			\end{minipage}
			\begin{minipage}{0.188\linewidth}
				\centering
				\fbox{\includegraphics[width=\linewidth]{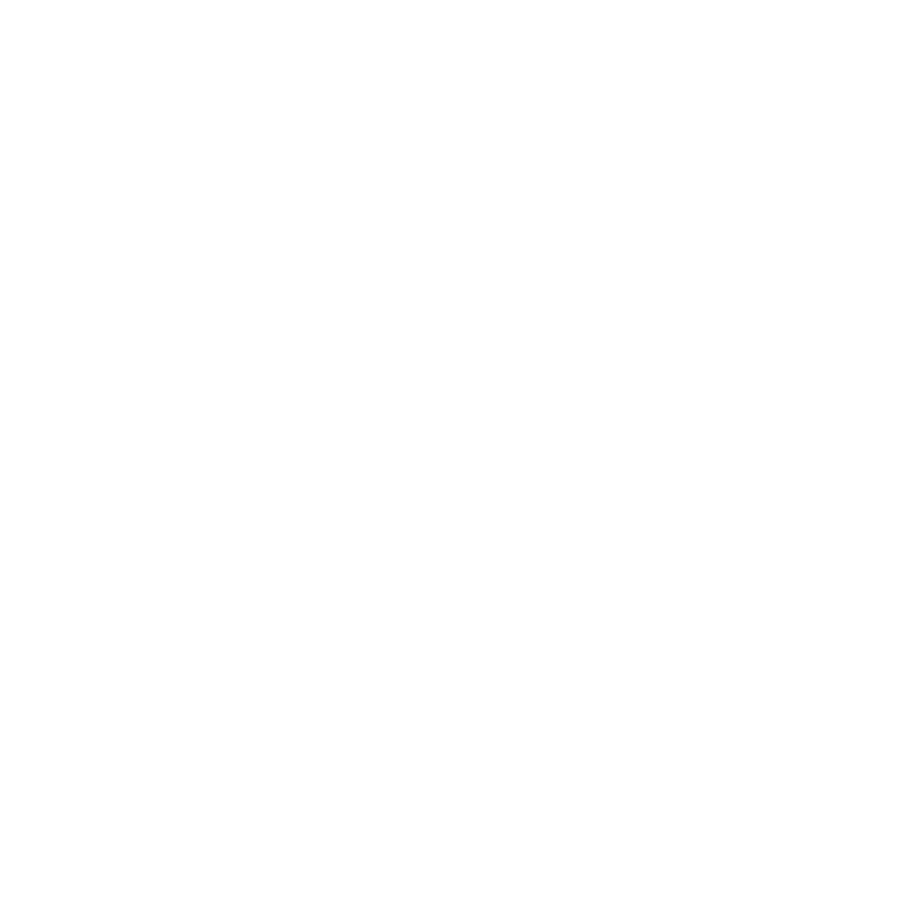}}\\
				\vspace{-2mm}
				{\footnotesize t=10}
			\end{minipage}
			\begin{minipage}{0.188\linewidth}
				\centering
				\fbox{\includegraphics[width=\linewidth]{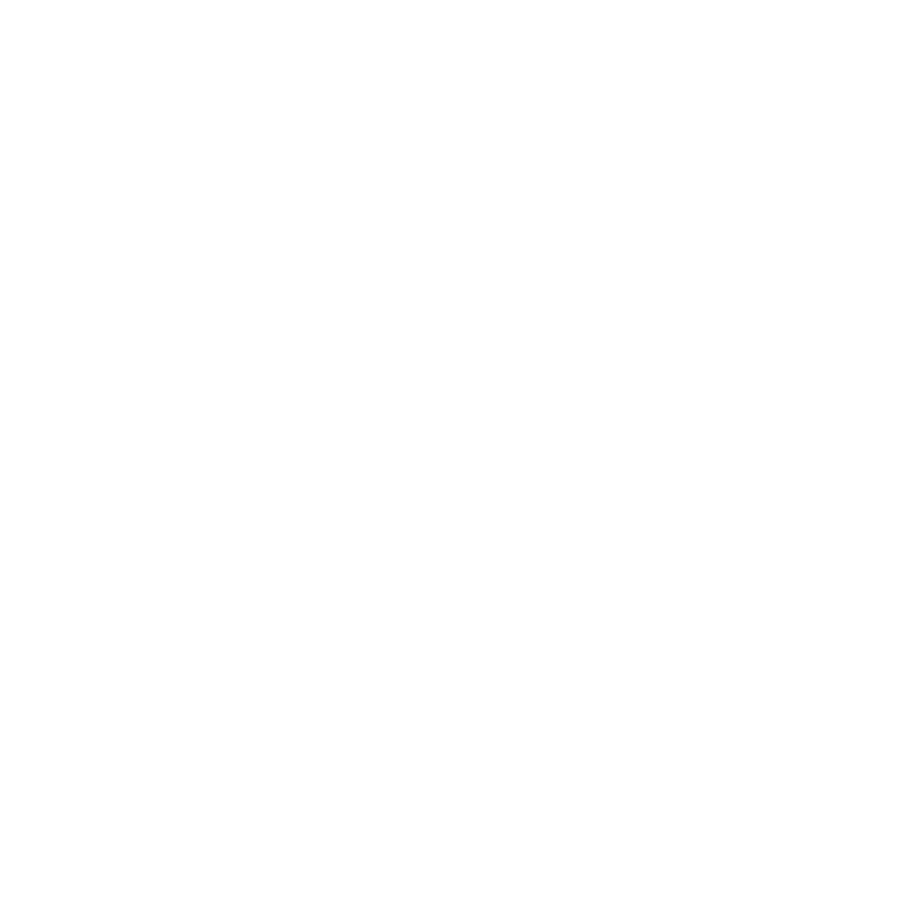}}\\
				\vspace{-2mm}
				{\footnotesize t=100}
			\end{minipage}
			\begin{minipage}{0.188\linewidth}
				\centering
				\fbox{\includegraphics[width=\linewidth]{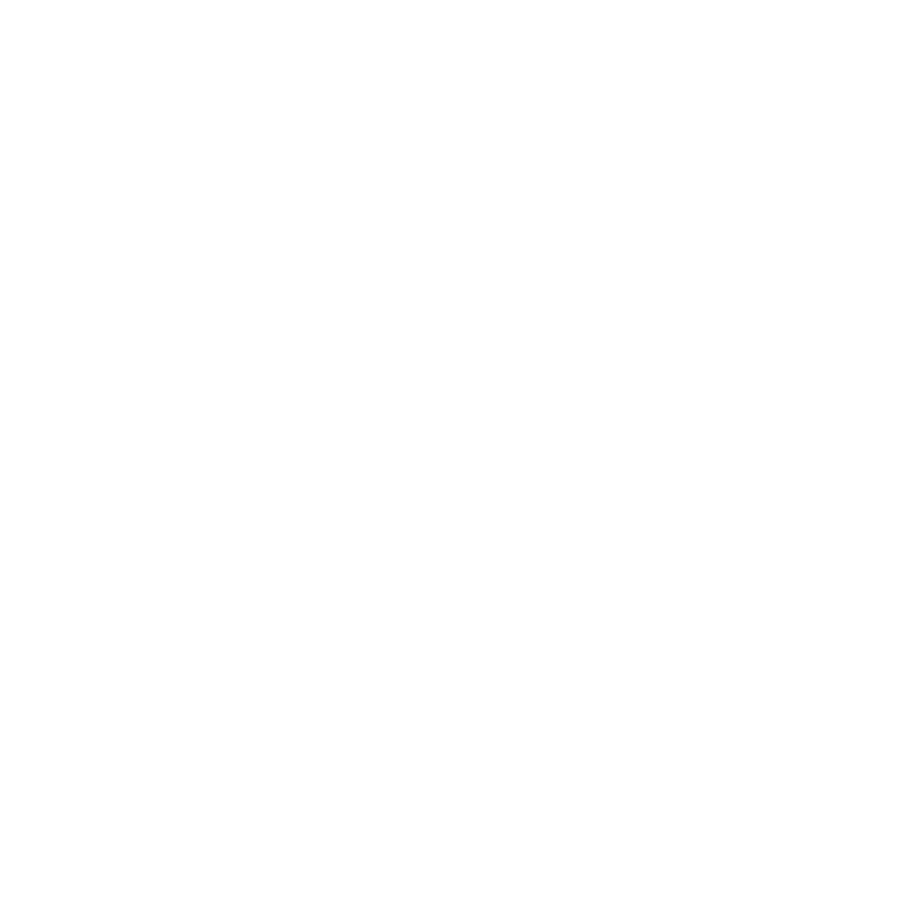}}\\
				\vspace{-2mm}
				{\footnotesize t=1000}
			\end{minipage}
			\vspace{-2mm}
			\caption*{\footnotesize (d) r=4.0 (Phase 2)}
		\end{minipage}
		\caption{Evolution of cooperative behavior in PPO-ACT under spatial heterogeneity. (a,c) Temporal evolution under enhancement factors $r=3.0$ (a) and $r=4.0$ (c) showing the complete Phase 1 + Phase 2 training process, with strategy space snapshots (white: cooperator, black: defector) at iterations $\{0, 10, 100, 1000, 10000\}$. (b,d) Corresponding Phase 2-only results for $r=3.0$ (b) and $r=4.0$ (d) with snapshots at $\{0, 1, 10, 100, 1000\}$ iterations. Initial conditions: defectors (red) occupy the upper grid half, cooperators (blue) lower half. All panels share identical color mapping and spatial scale bars.
		The initial training phase with r=5.0 established a cooperative foundation, enabling agents to partially overcome payoff disadvantages in the second phase. Evolutionary trajectories in panels (c) and (d) demonstrate that at r=3.0, the critically low cooperation payoff leads to complete dominance by defectors, while at r=4.0, the population rapidly converges to universal cooperation.}
		\label{fig:PPO-ACT_uDbC_matrix}
	\end{figure*}

	Fig.~\ref{fig:PPO-ACT_uDbC_matrix} shows the dynamic evolutionary characteristics of the PPO-ACT model under spatially heterogeneous initial conditions. The experiment adopts a specific initial configuration to study cooperative dynamics. The initialization strategy places defectors in the upper half and cooperators in the lower half. This spatial separation allows for clear observation of strategy interactions.	The study systematically examines system behavior evolution across different enhancement factors ($r=3.0$ and $r=4.0$). These parameter variations enable comprehensive analysis of cooperation patterns under varying conditions. Each experimental result contains two components: temporal curves of strategy fractions and spatial distribution snapshots. The upper subfigure in each group shows temporal evolution curves (blue: cooperators, red: defectors) with iteration count $t$ on the horizontal axis and  fraction of collaborators and defectors on the vertical axis. The lower subfigure displays state snapshots (white: cooperators, black: defectors).
	
	As shown in Figs.~\ref{fig:PPO-ACT_uDbC_matrix} (a) and (b), Phase 1 (iterations 0-999) at $r=5.0$ involves limited policy maturation, where the Actor-network avoids unconditional cooperation due to insufficient training iterations. The Critic network acquires usable value estimation capabilities during Phase 1. Two concurrent changes occur when transitioning to Phase 2 at iteration 1000: the enhancement factor adjusts to target $r$ (3.0 or 4.0) and all agent states reset. This allows observing how Phase 1 pretraining affects subsequent adaptation. Fig.~\ref{fig:PPO-ACT_uDbC_matrix}(c) specifically demonstrates the $r=3.0$ case in Phase 2. The cooperation ratio reaches $96.3\%$ at iteration 1, maintains $100\%$ through iterations 2-5, and then exhibits oscillatory decay terminating in full defection by iteration 27. In contrast, Fig.~\ref{fig:PPO-ACT_uDbC_matrix}(d) shows the $r=4.0$ scenario where the cooperation ratio similarly peaks at $96.3\%$ in iteration 1 but stabilizes at $100\%$ from iteration 2 onward. Spatial analysis confirms these temporal patterns. For the $r=4.0$ case in Fig.~\ref{fig:PPO-ACT_uDbC_matrix}(d), global strategy synchronization completes within two iterations. The $r=3.0$ system in Fig.~\ref{fig:PPO-ACT_uDbC_matrix}(c) sustains spatial pattern fluctuations until iteration 15, mirroring the delayed cooperation collapse observed in the temporal domain.
	
	\subsection{PPO-ACT with bernoulli random initialization}
	\label{exp_b}
\begin{figure*}[h]
		\begin{minipage}{0.45\linewidth}
			\begin{minipage}{\linewidth}
				\centering
				\includegraphics[width=\linewidth]{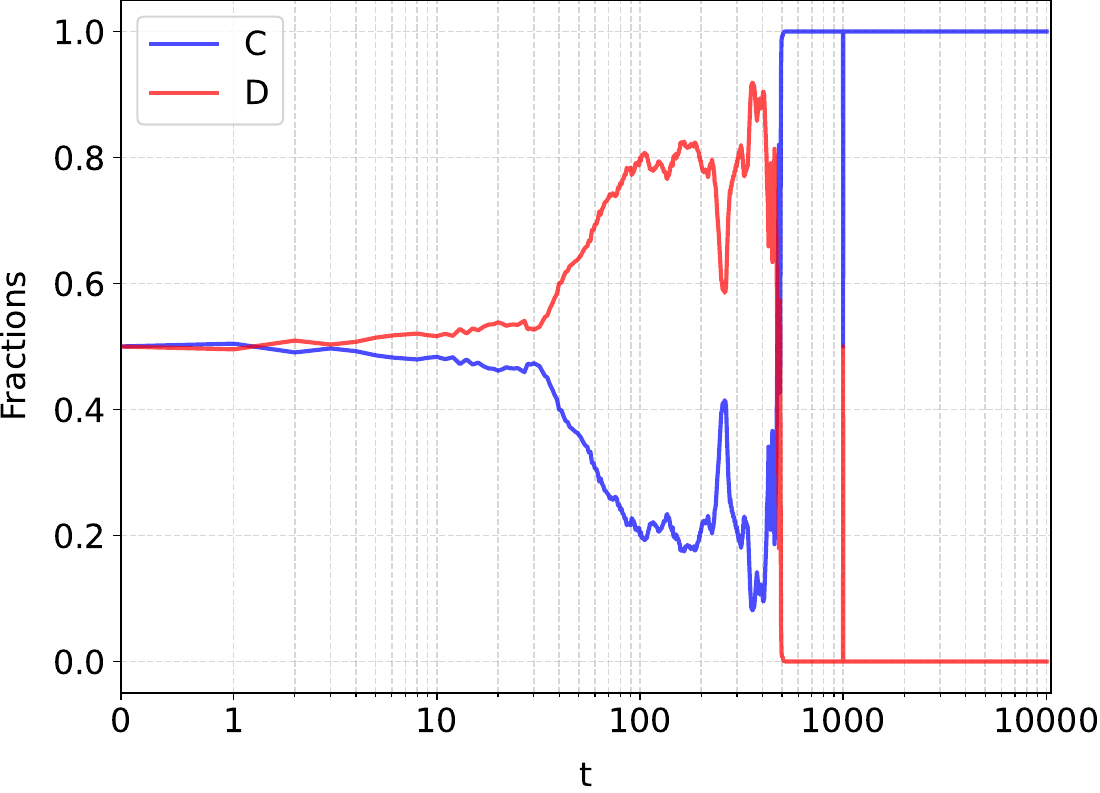}\\
			\end{minipage}
			\vspace{4mm}
			\\
			\begin{minipage}{0.188\linewidth}
				\centering
				\fbox{\includegraphics[width=\linewidth]{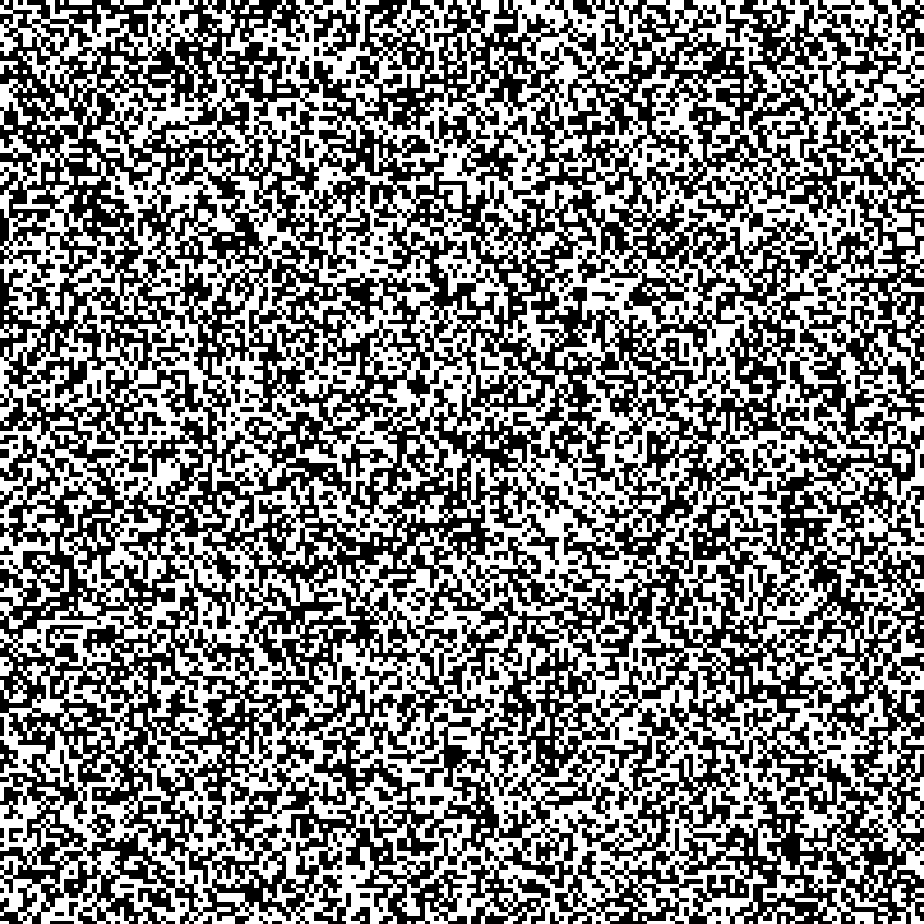}}\\
				\vspace{-2mm}
				{\footnotesize t=0}
			\end{minipage}
			\begin{minipage}{0.188\linewidth}
				\centering
				\fbox{\includegraphics[width=\linewidth]{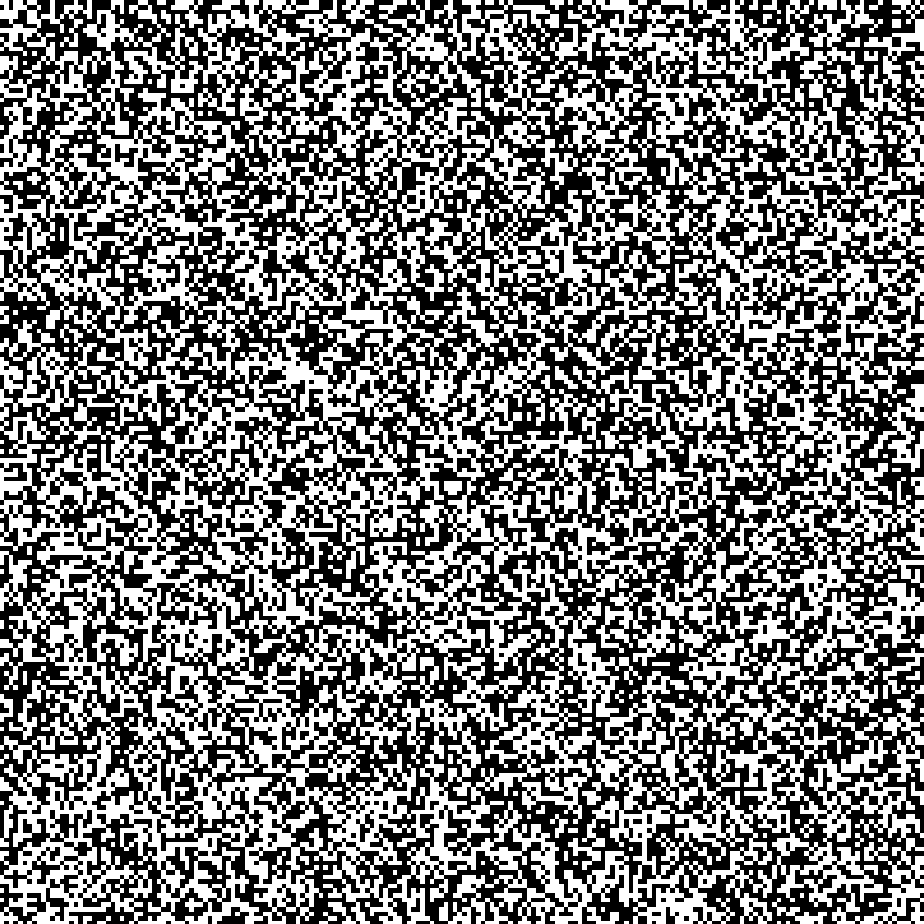}}\\
				\vspace{-2mm}
				{\footnotesize t=10}
			\end{minipage}
			\begin{minipage}{0.188\linewidth}
				\centering
				\fbox{\includegraphics[width=\linewidth]{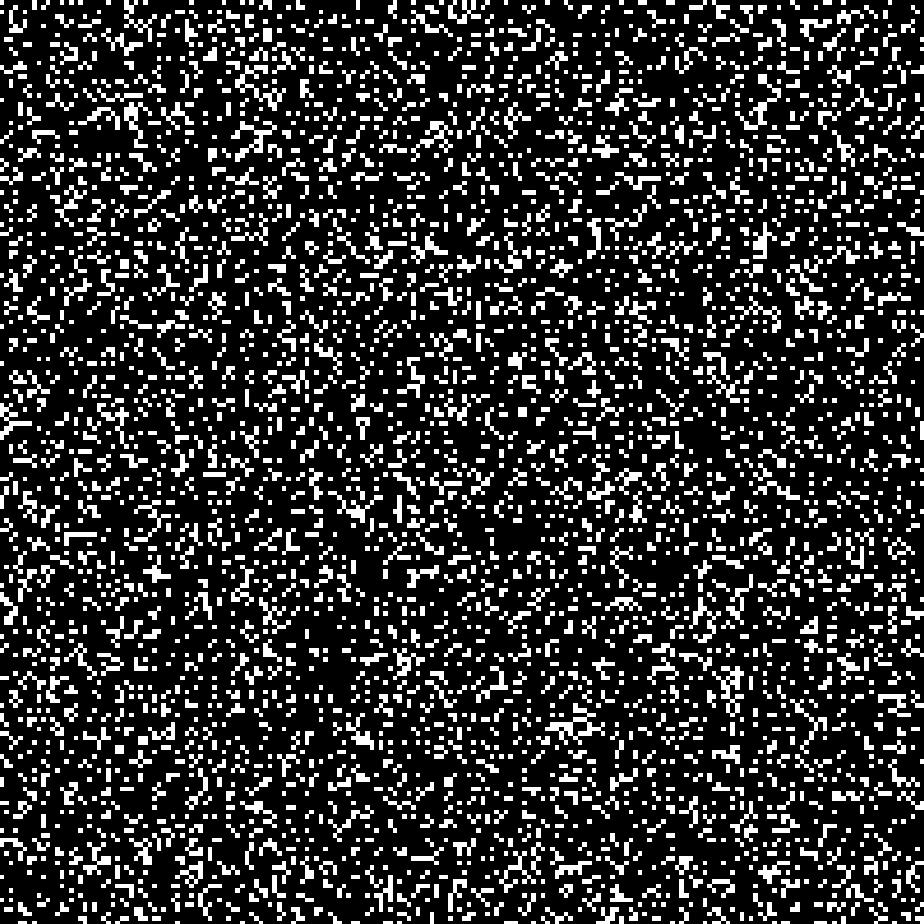}}\\
				\vspace{-2mm}
				{\footnotesize t=100}
			\end{minipage}
			\begin{minipage}{0.188\linewidth}
				\centering
				\fbox{\includegraphics[width=\linewidth]{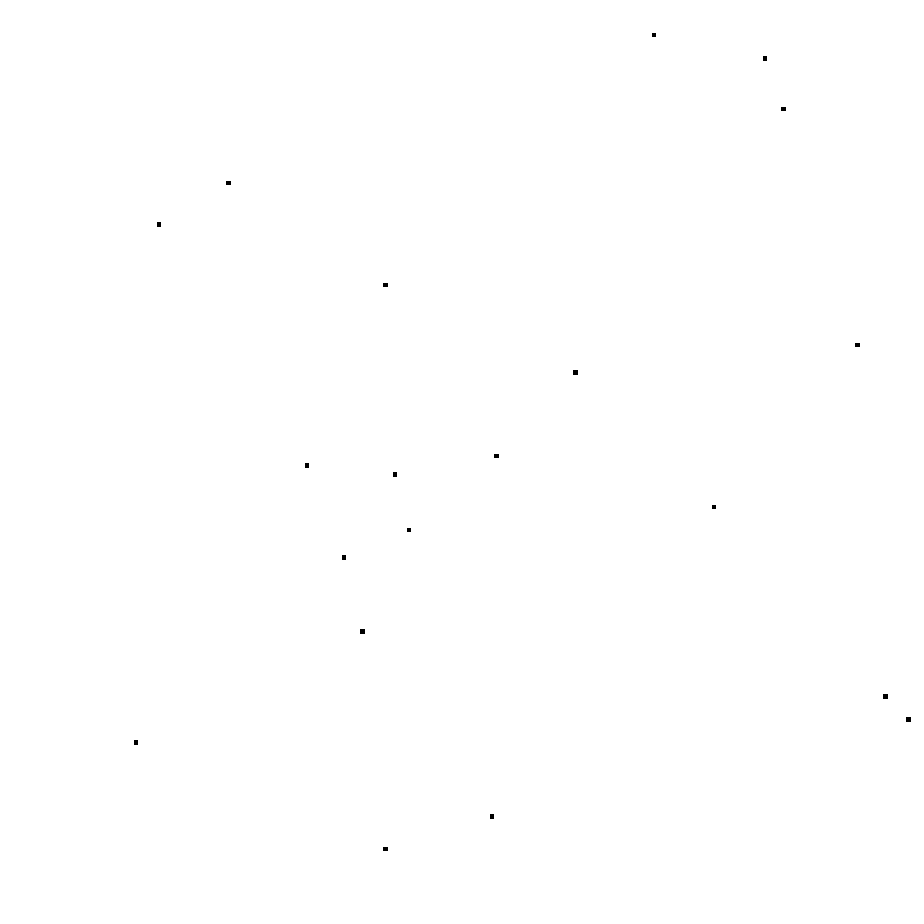}}\\
				\vspace{-2mm}
				{\footnotesize t=1000}
			\end{minipage}
			\begin{minipage}{0.188\linewidth}
				\centering
				\fbox{\includegraphics[width=\linewidth]{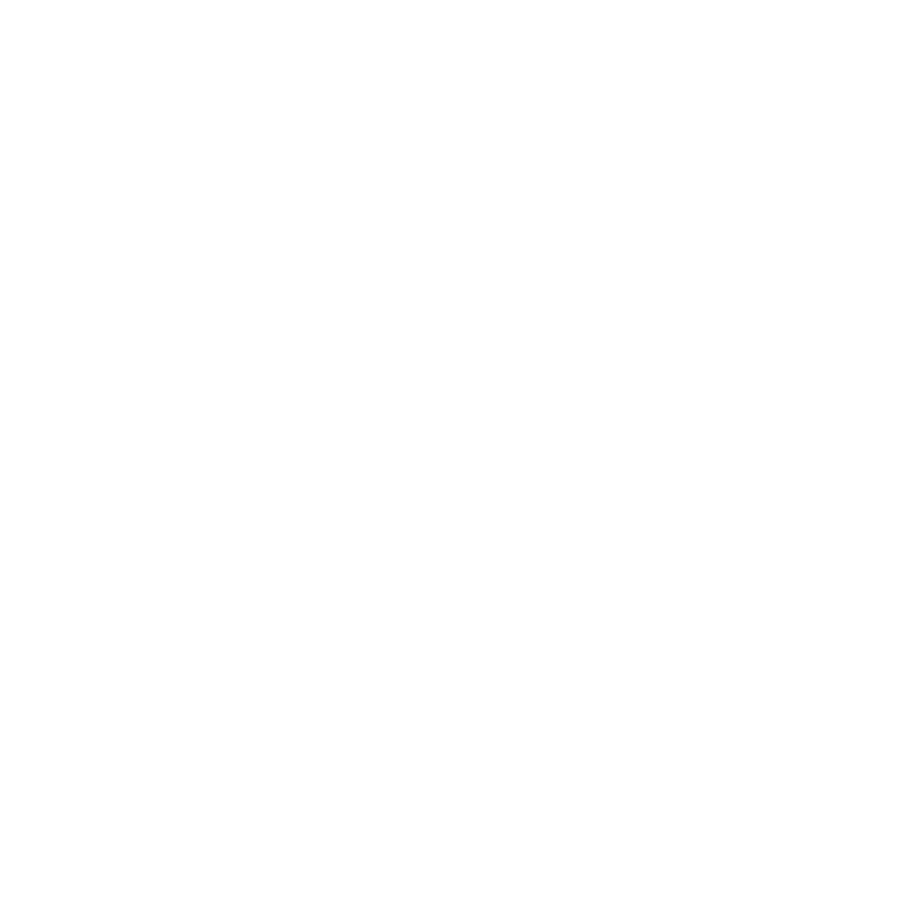}}\\
				\vspace{-2mm}
				{\footnotesize t=10000}
			\end{minipage}
			\vspace{-2mm}
			\caption*{\footnotesize (a) Phase 1+2}
		\end{minipage}
		\hfill
		\begin{minipage}{0.45\linewidth}
			\begin{minipage}{\linewidth}
				\centering
				\includegraphics[width=\linewidth]{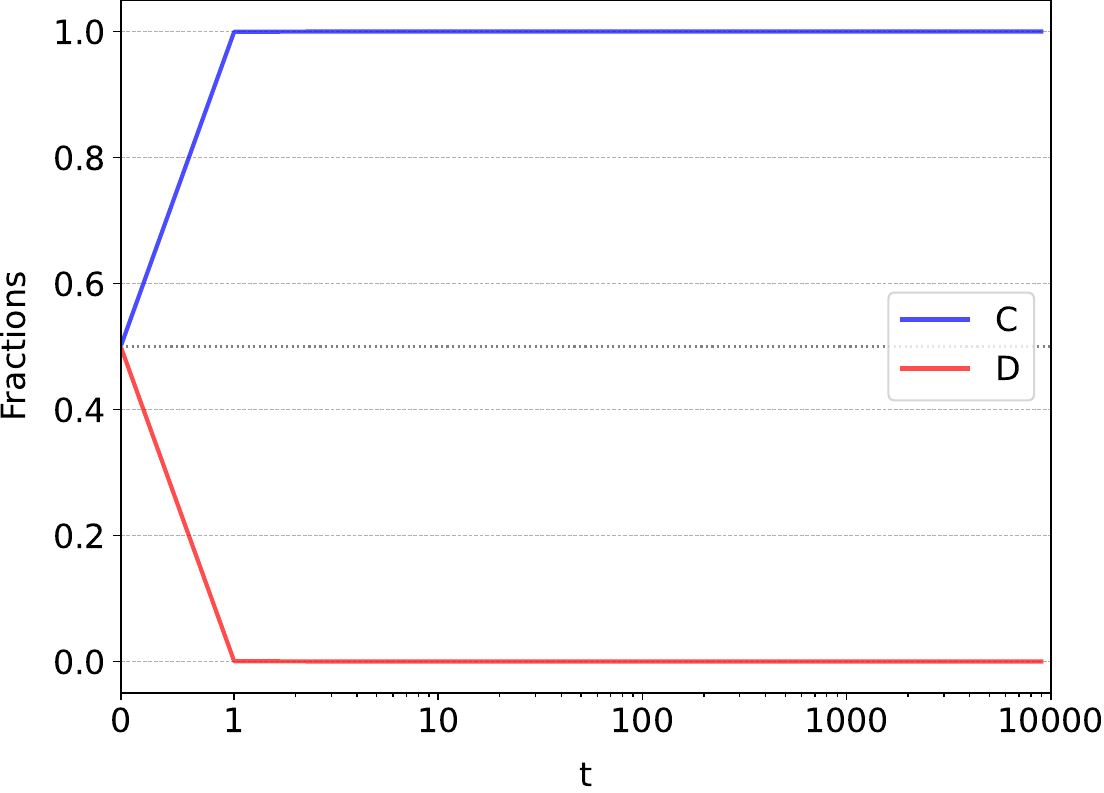}\\
			\end{minipage}
			\vspace{4mm}
			\\
			\begin{minipage}{0.188\linewidth}
				\centering
				\fbox{\includegraphics[width=\linewidth]{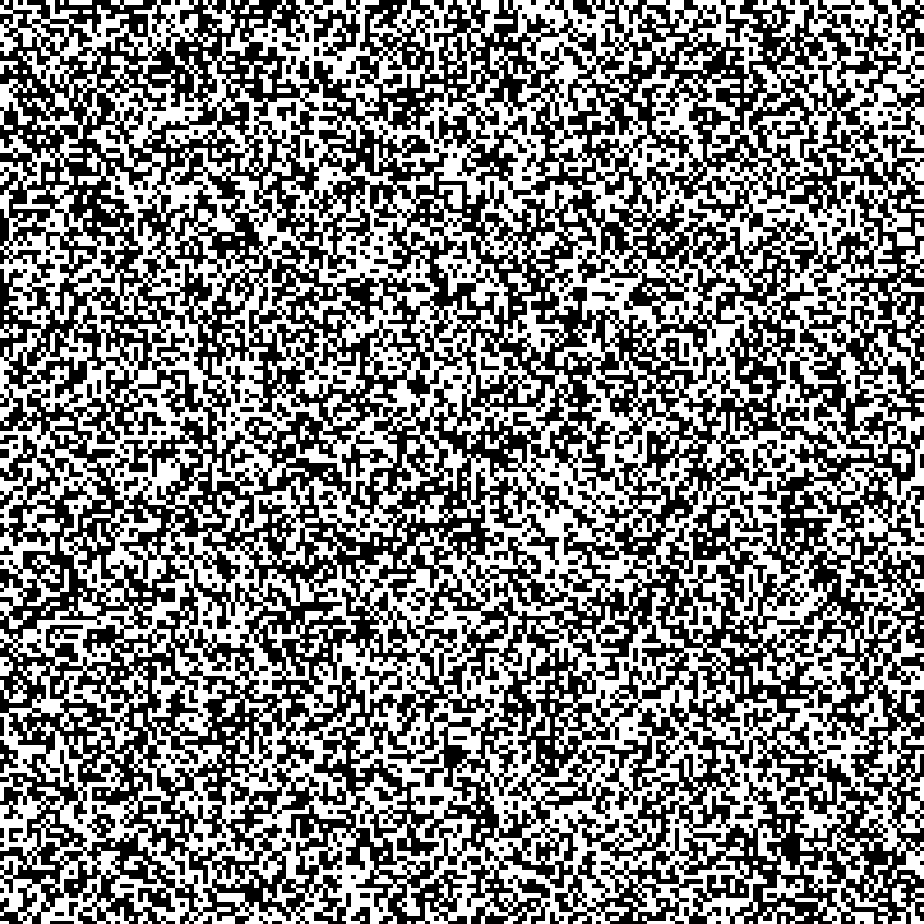}}\\
				\vspace{-2mm}
				{\footnotesize t=0}
			\end{minipage}
			\begin{minipage}{0.188\linewidth}
				\centering
				\fbox{\includegraphics[width=\linewidth]{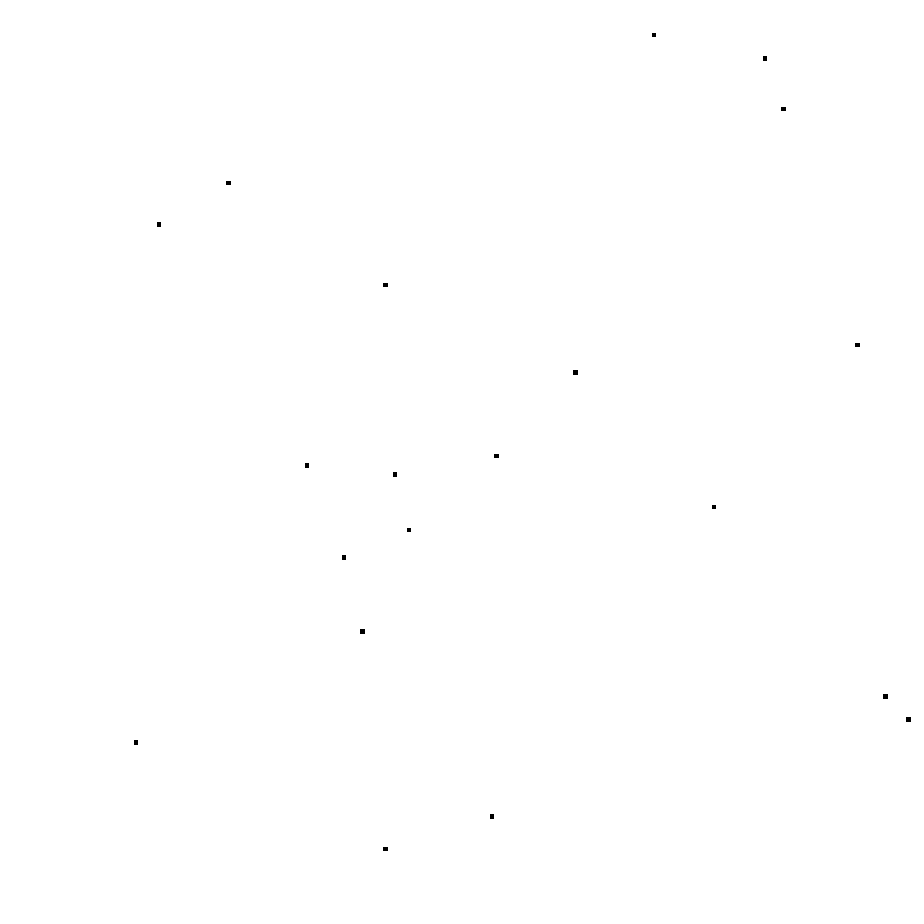}}\\
				\vspace{-2mm}
				{\footnotesize t=1}
			\end{minipage}
			\begin{minipage}{0.188\linewidth}
				\centering
				\fbox{\includegraphics[width=\linewidth]{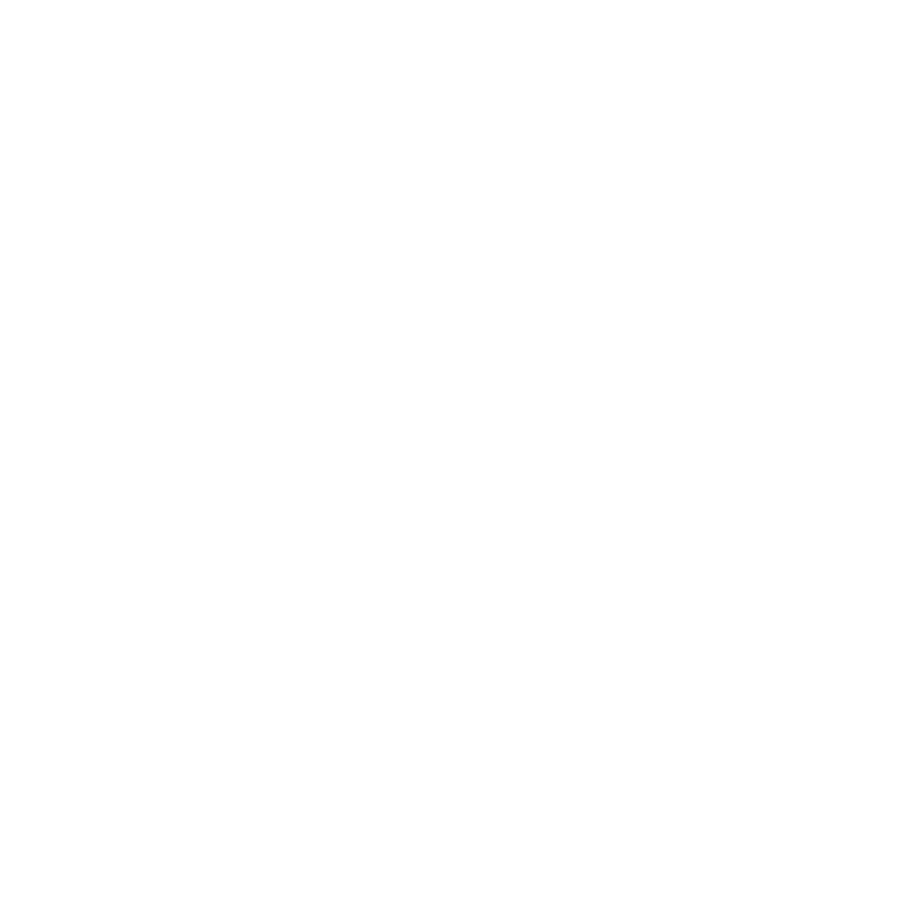}}\\
				\vspace{-2mm}
				{\footnotesize t=10}
			\end{minipage}
			\begin{minipage}{0.188\linewidth}
				\centering
				\fbox{\includegraphics[width=\linewidth]{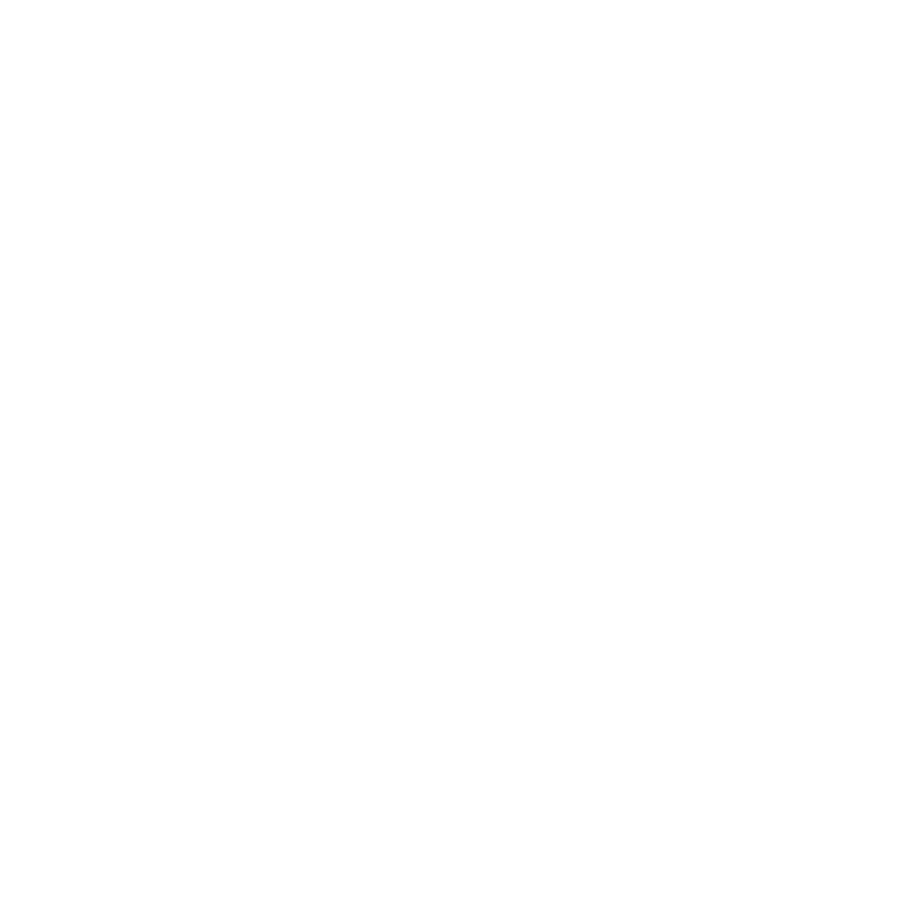}}\\
				\vspace{-2mm}
				{\footnotesize t=100}
			\end{minipage}
			\begin{minipage}{0.188\linewidth}
				\centering
				\fbox{\includegraphics[width=\linewidth]{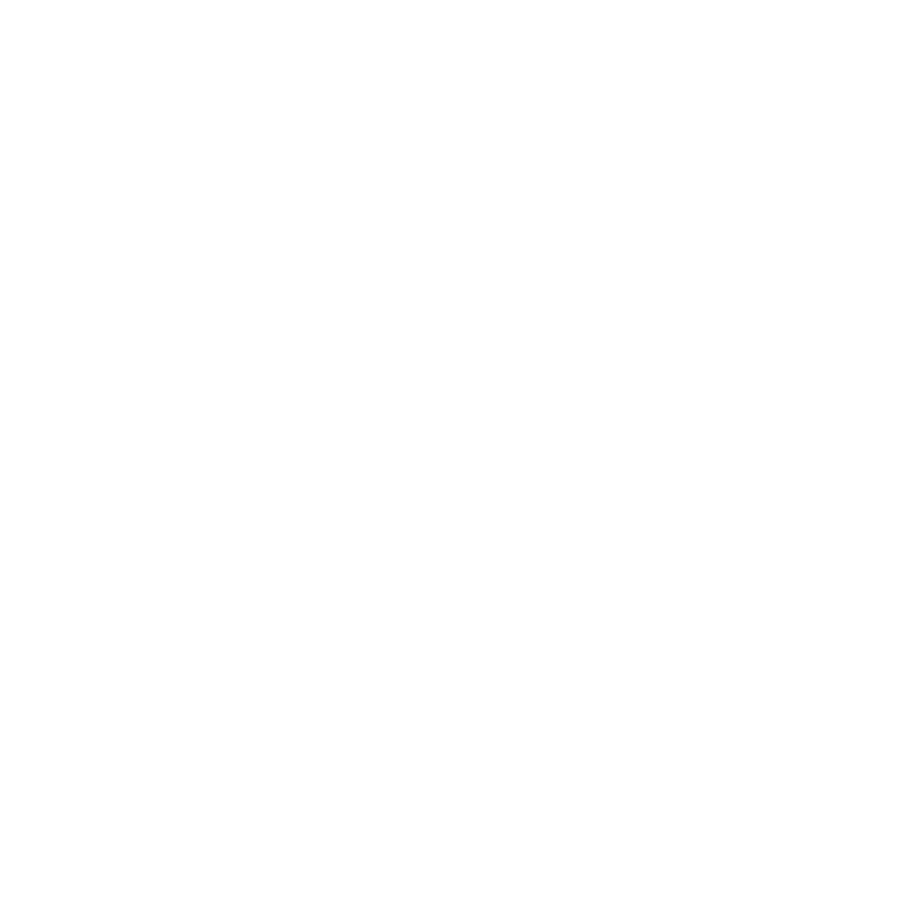}}\\
				\vspace{-2mm}
				{\footnotesize t=1000}
			\end{minipage}
			\vspace{-2mm}
			\caption*{\footnotesize (b) Phase 2}
		\end{minipage}
		\caption{Evolution of cooperative behavior under Bernoulli initialization ($p=0.5$). (a) Complete training process with Phase 1 ($r=5.0$, 0-999 iterations) and Phase 2 ($r=4.0$, from iteration 1000). Snapshots captured at $t=[0, 10, 100, 1000, 10000]$ (white: cooperator, black: defector). (b) Phase 2-only dynamics after strategy reinitialization. Snapshots at $t=[0, 1, 10, 100, 1000]$ demonstrate immediate cooperation fixation. Bernoulli initialization ($p=0.5$) enables faster cooperative convergence than spatially segregated setups, with Phase 2 achieving instant ($t=1$) global cooperation fixation.}
		\label{fig:PPO-ACT_bernoulli_matrix}
	\end{figure*}

	This experiment initializes the strategy space using Bernoulli distribution with equal $50\%$ probabilities for both cooperation and defection strategies. As shown in Fig.~\ref{fig:PPO-ACT_bernoulli_matrix}, the PPO-ACT framework demonstrates significant evolutionary differences under various enhancement factors $r$. The subplots are divided into upper and lower sections. The upper portion presents evolution curves of cooperator (blue) and defector (red) fractions, with the horizontal axis indicating iteration count $t$ and the vertical axis showing the fraction of collaborators and defectors respectively. The lower portion provides strategy space snapshots at selected time points, using white pixels for cooperators and black pixels for defectors.
	
	As shown in Fig.~\ref{fig:PPO-ACT_bernoulli_matrix}(a), the complete training process exhibits distinct strategic convergence characteristics. During Phase 1 (r=5.0, iterations 0-999), the cooperation rate exhibits oscillatory decay followed by a rapid transition to complete cooperation (all agents becoming cooperators) around iteration 500. Spatial snapshots demonstrate rapid stabilization to complete cooperation coverage, with full spatial uniformity achieved by approximately iteration 500. When transitioning to Phase 2 (iteration $1000$, $r=4.0$), despite both reducing the enhancement factor and reinitializing the policy matrix, all agents re-synchronize to cooperative strategies within a single iteration. The spatial configuration temporarily reverts to random distribution upon reset ($t=1000$), but immediately recovers global cooperation in the subsequent iteration ($t=1001$). Notably, the independent Phase 2 experiment in Fig.~\ref{fig:PPO-ACT_bernoulli_matrix}(b) validates this phenomenon: Post-reset, all agents achieve complete cooperation by the first iteration ($t=1$) and maintain this state permanently. This confirms that the coordination mechanism established during Phase 1 training exhibits parameter robustness, sustaining cooperation despite a moderate reduction in enhancement factor.
	
	\subsection{PPO-ACT with all-defectors initialization}
	\label{exp_ad}
	
\begin{figure*}[htbp!]
		\begin{minipage}{0.45\linewidth}
			\begin{minipage}{\linewidth}
				\centering
				\includegraphics[width=\linewidth]{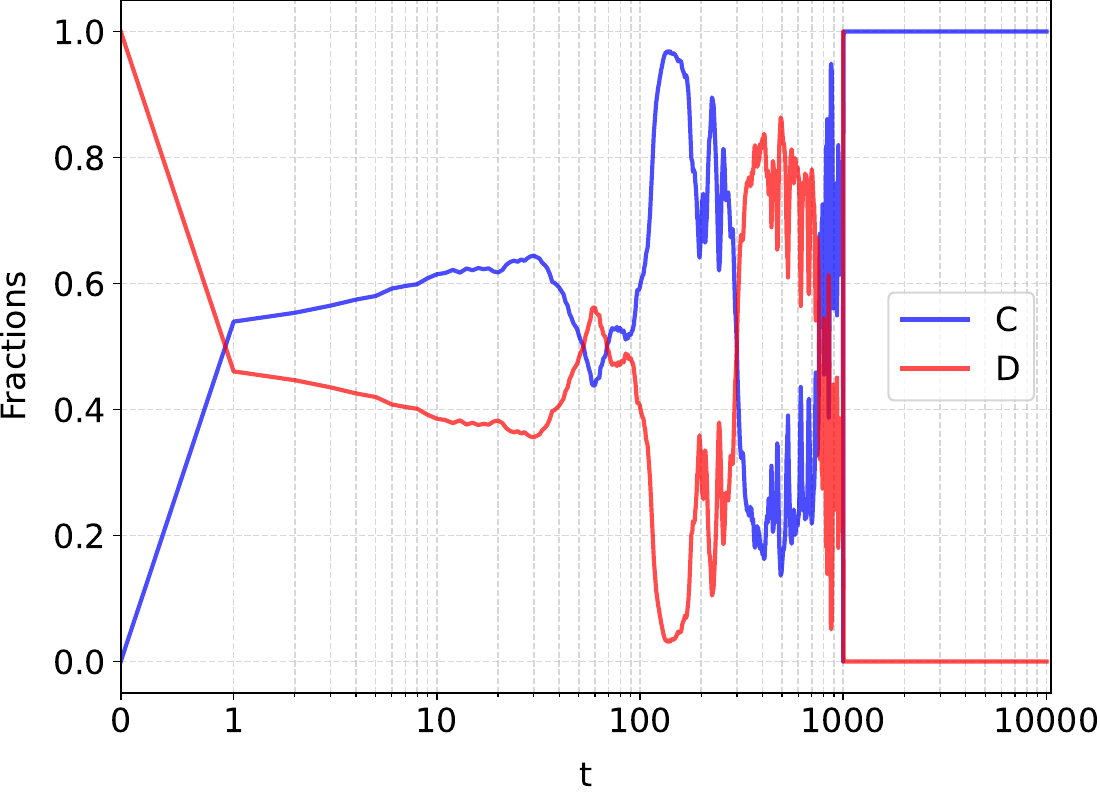}\\
			\end{minipage}
			\vspace{2mm}
			\\
			\begin{minipage}{0.188\linewidth}
				\centering
				\fbox{\includegraphics[width=\linewidth]{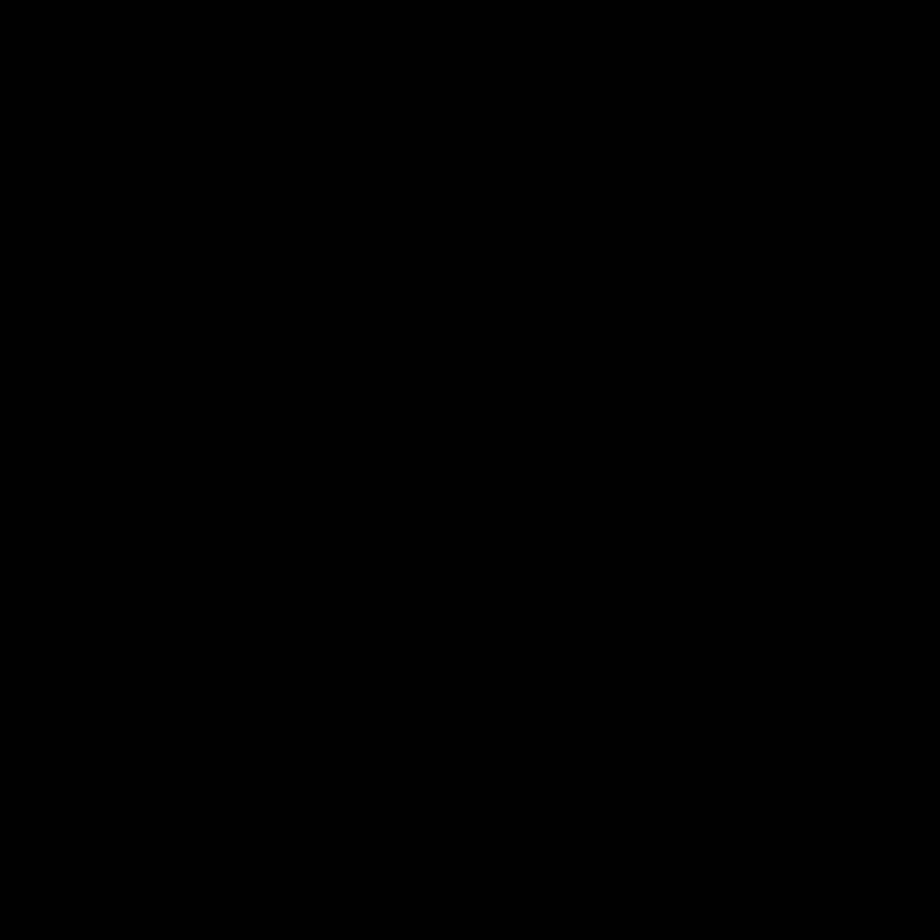}}\\
				\vspace{-2mm}
				{\footnotesize t=0}
			\end{minipage}
			\begin{minipage}{0.188\linewidth}
				\centering
				\fbox{\includegraphics[width=\linewidth]{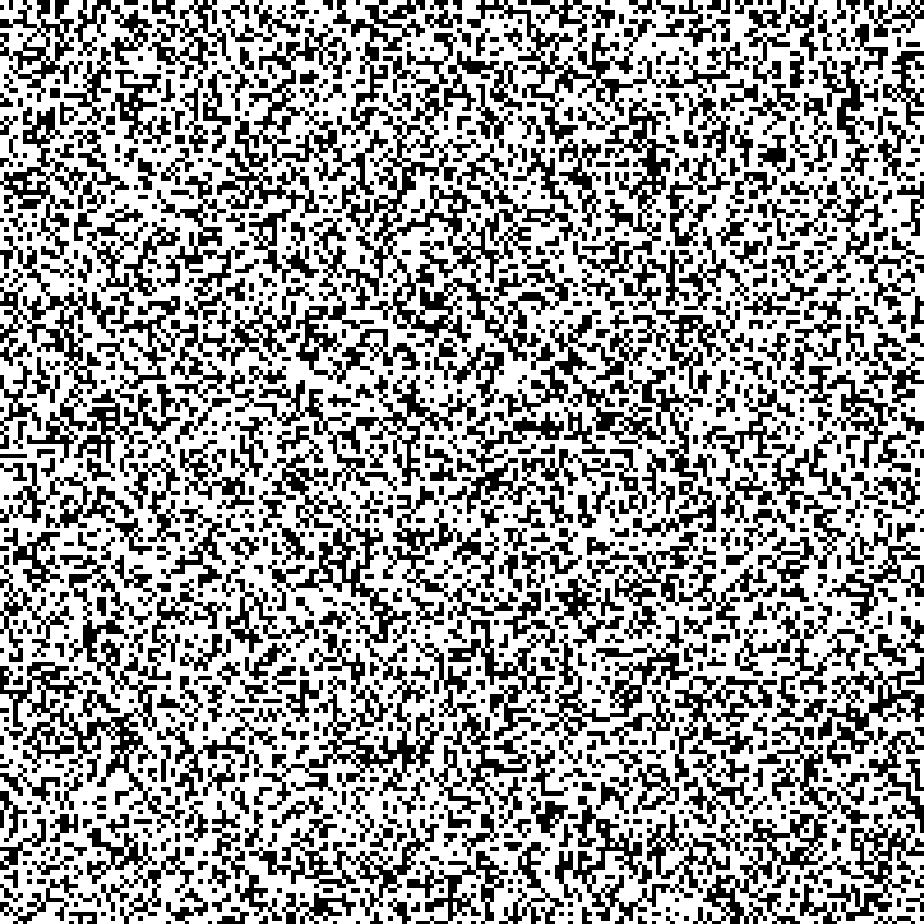}}\\
				\vspace{-2mm}
				{\footnotesize t=10}
			\end{minipage}
			\begin{minipage}{0.188\linewidth}
				\centering
				\fbox{\includegraphics[width=\linewidth]{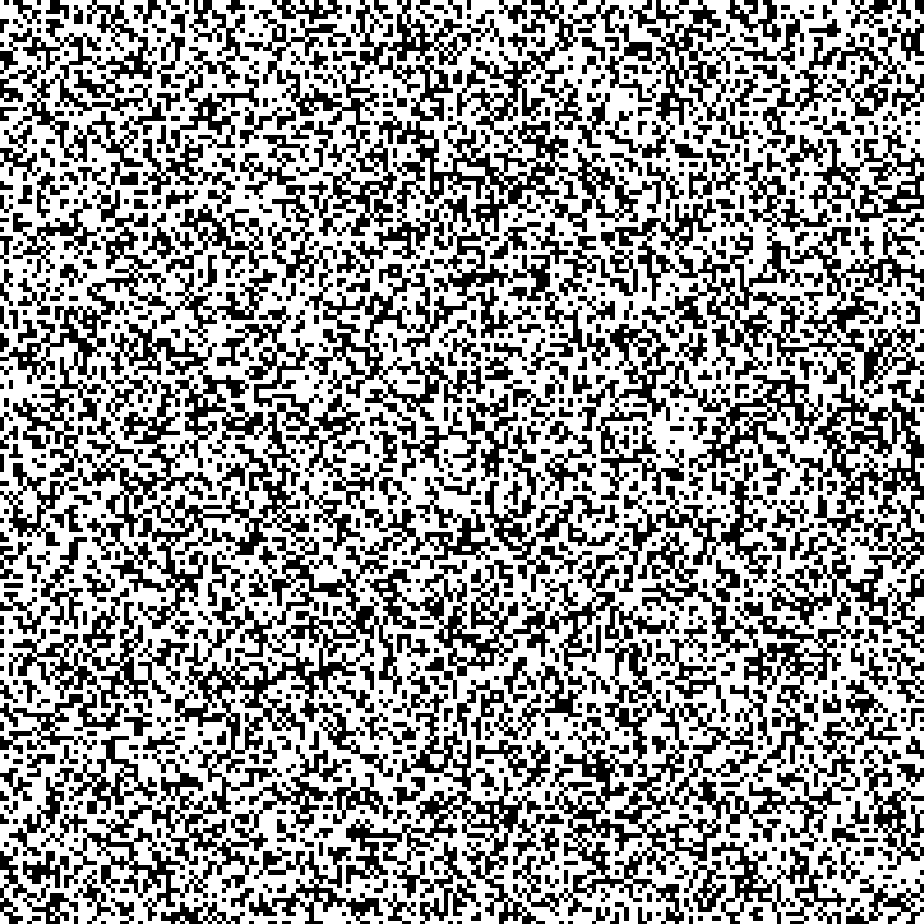}}\\
				\vspace{-2mm}
				{\footnotesize t=100}
			\end{minipage}
			\begin{minipage}{0.188\linewidth}
				\centering
				\fbox{\includegraphics[width=\linewidth]{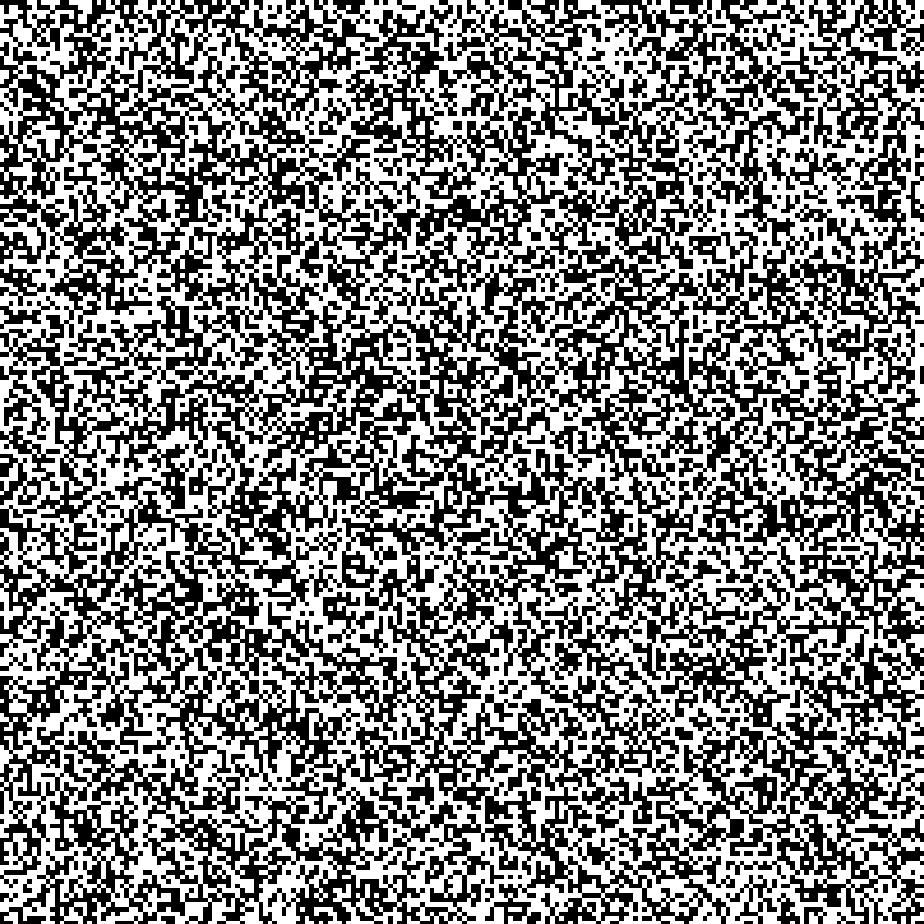}}\\
				\vspace{-2mm}
				{\footnotesize t=1000}
			\end{minipage}
			\begin{minipage}{0.188\linewidth}
				\centering
				\fbox{\includegraphics[width=\linewidth]{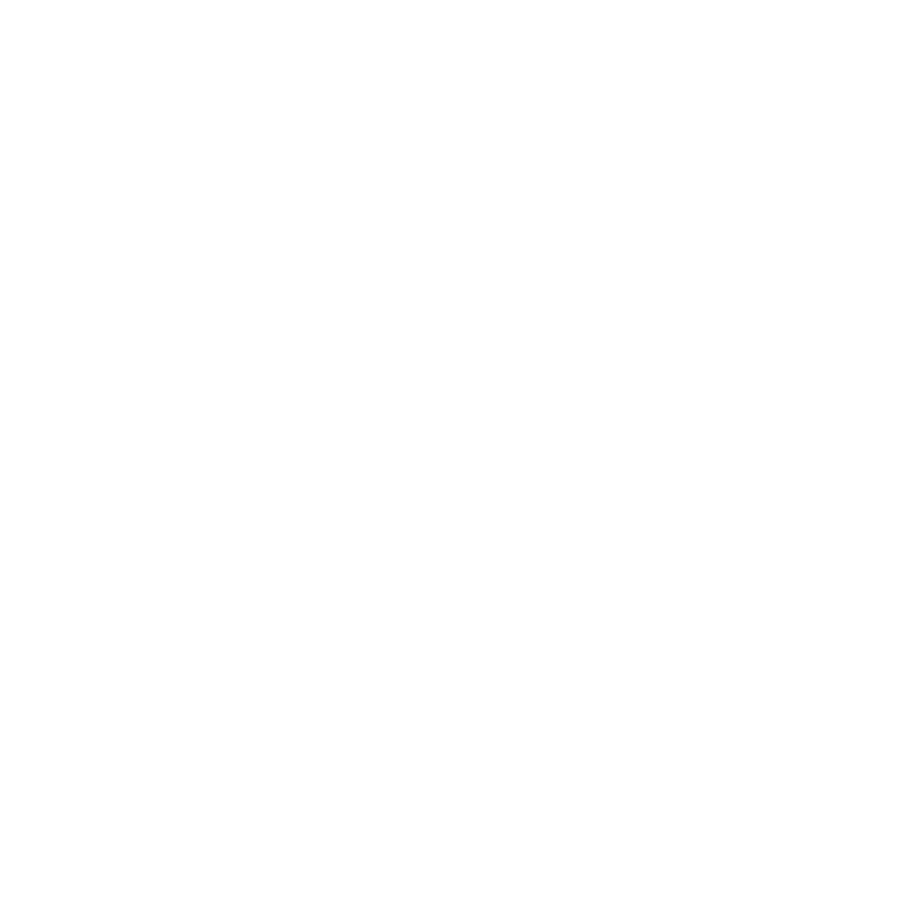}}\\
				\vspace{-2mm}
				{\footnotesize t=10000}
			\end{minipage}
			\vspace{-2mm}
			\caption*{\footnotesize (a) Phase 1+2}
		\end{minipage}
		\hfill
		\begin{minipage}{0.45\linewidth}
			\begin{minipage}{\linewidth}
				\centering
				\includegraphics[width=\linewidth]{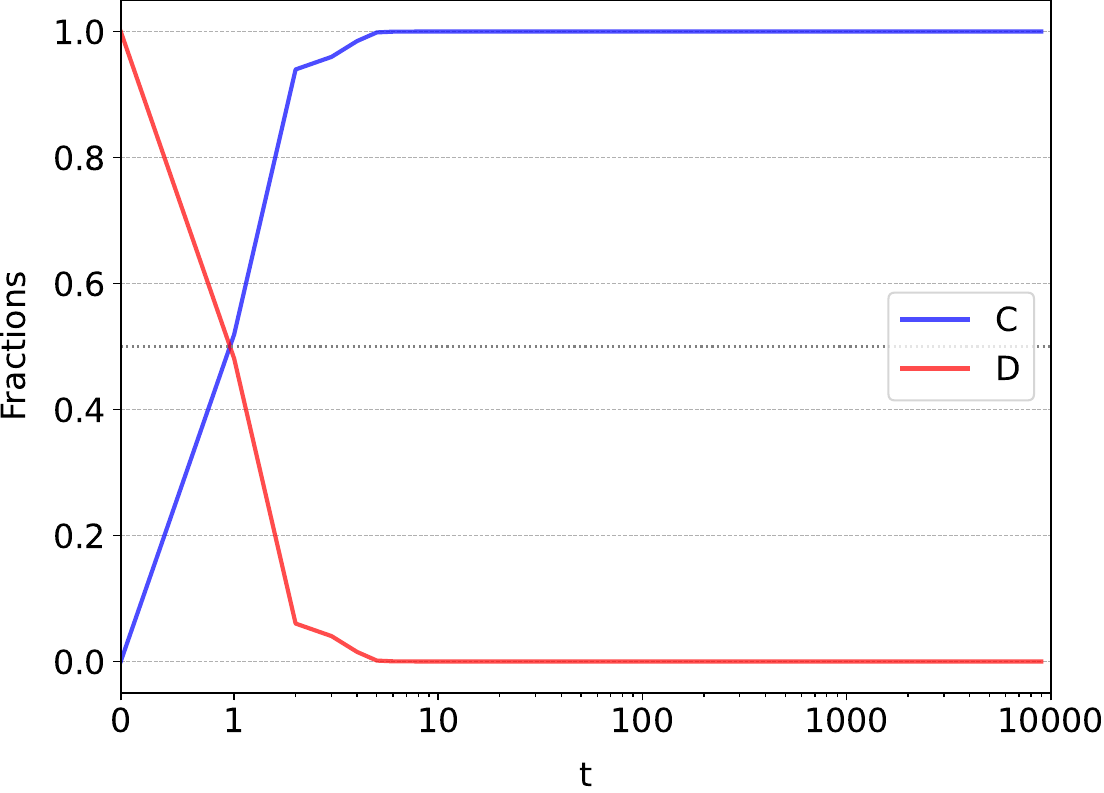}\\
			\end{minipage}
			\vspace{2mm}
			\\
			\begin{minipage}{0.188\linewidth}
				\centering
				\fbox{\includegraphics[width=\linewidth]{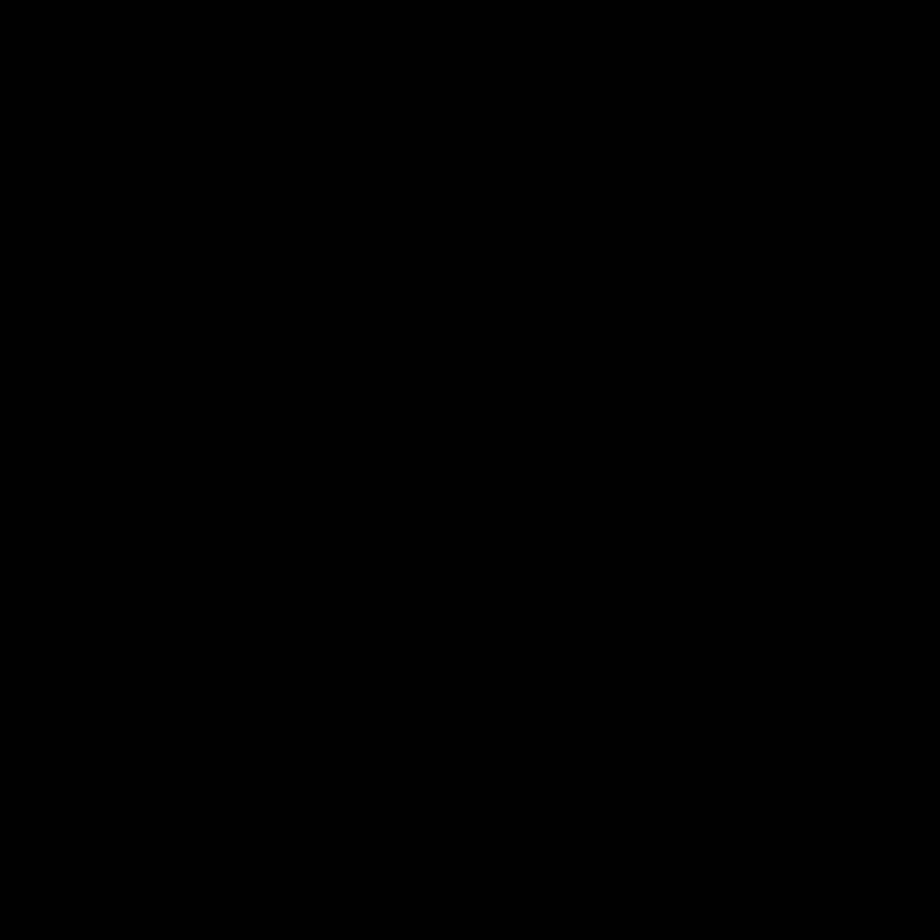}}\\
				\vspace{-2mm}
				{\footnotesize t=0}
			\end{minipage}
			\begin{minipage}{0.188\linewidth}
				\centering
				\fbox{\includegraphics[width=\linewidth]{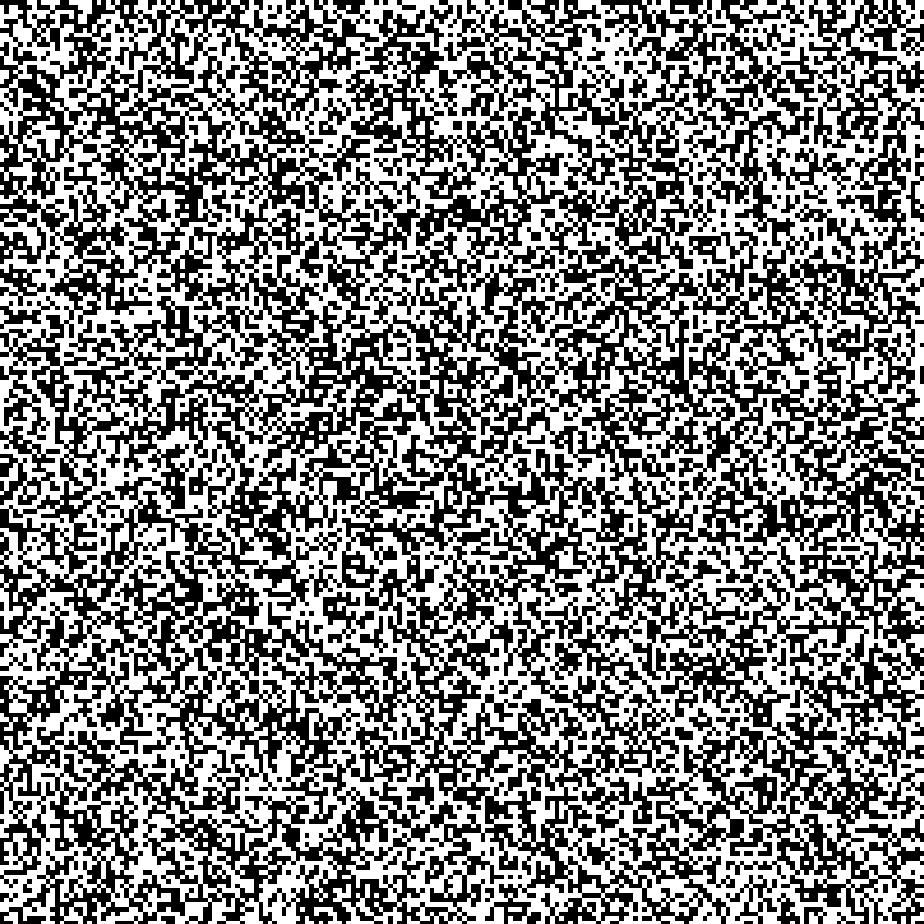}}\\
				\vspace{-2mm}
				{\footnotesize t=1}
			\end{minipage}
			\begin{minipage}{0.188\linewidth}
				\centering
				\fbox{\includegraphics[width=\linewidth]{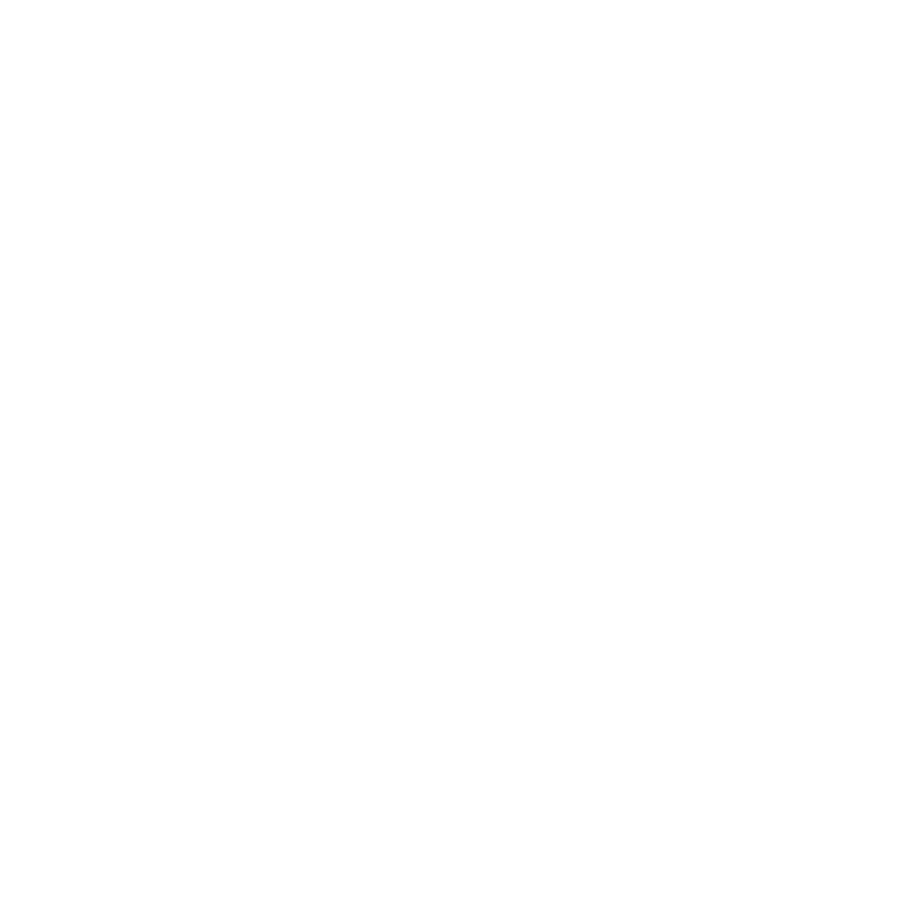}}\\
				\vspace{-2mm}
				{\footnotesize t=10}
			\end{minipage}
			\begin{minipage}{0.188\linewidth}
				\centering
				\fbox{\includegraphics[width=\linewidth]{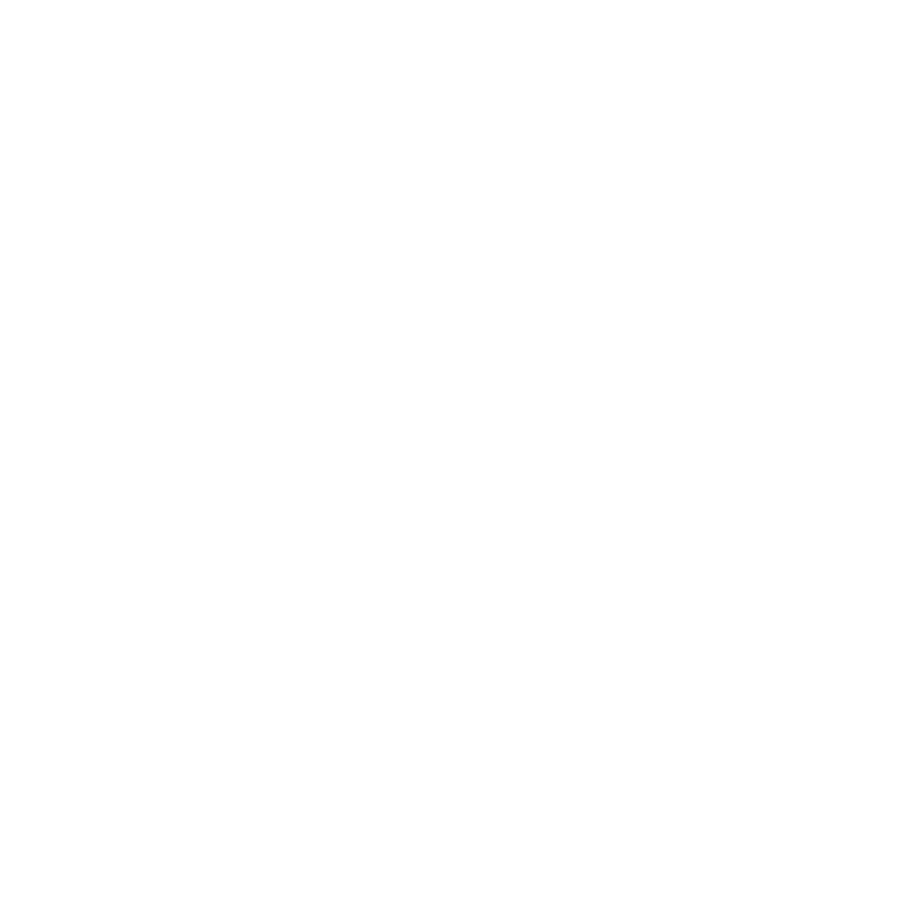}}\\
				\vspace{-2mm}
				{\footnotesize t=100}
			\end{minipage}
			\begin{minipage}{0.188\linewidth}
				\centering
				\fbox{\includegraphics[width=\linewidth]{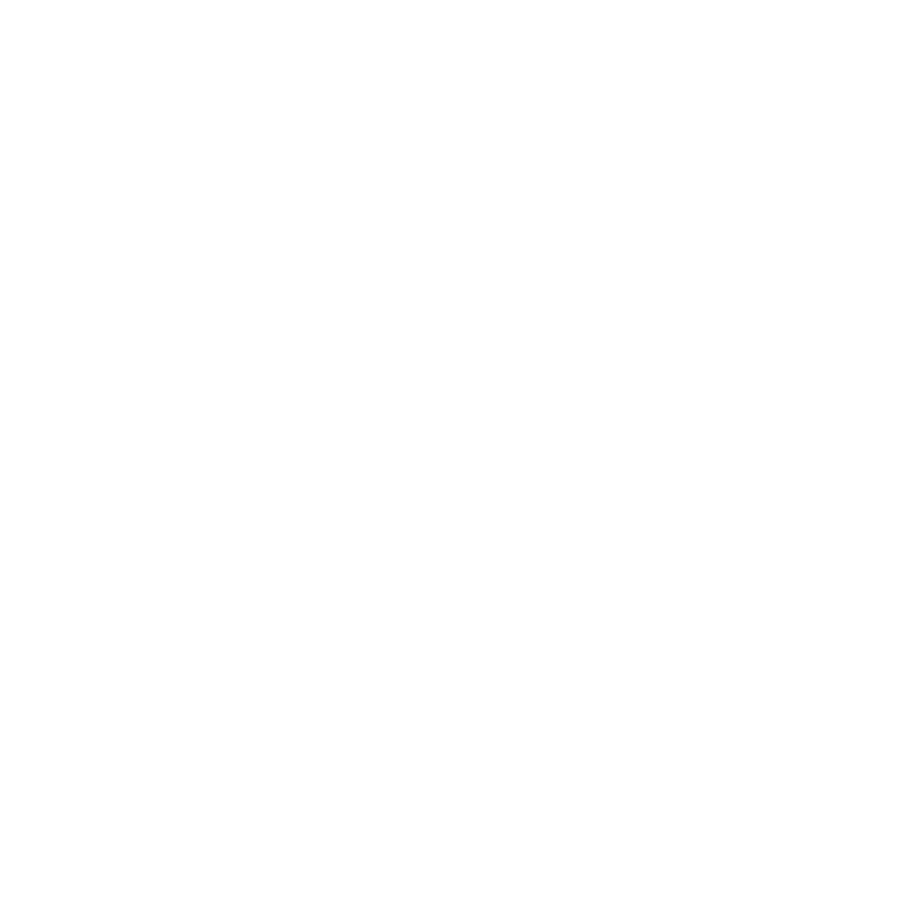}}\\
				\vspace{-2mm}
				{\footnotesize t=1000}
			\end{minipage}
			\vspace{-2mm}
			\caption*{\footnotesize (b) Phase 2}
		\end{minipage}
		\caption{Initial agent strategies were set with all agents as defectors. (a) Full process: Phase 1 ($r=5.0$, 0-999 iterations) and Phase 2 ($r=4.8$, from iteration 1000). Snapshots at $t=[0, 10, 100, 1000, 10000]$. (b) Phase 2-only process. Snapshots at $t=[0, 1, 10, 100, 1000]$. White indicates cooperators, and black denotes defectors. PPO-ACT can achieve global cooperation from all-defector initializations, though requiring higher enhancement factors $r$ compared to stochastic or spatially mixed initial conditions to reach stable cooperative equilibria.}
		\label{fig:PPO-ACT_unique_matrix}
	\end{figure*}

	This study systematically investigates the evolutionary dynamics of the PPO-ACT framework under all-defectors initial strategies. As shown in Fig.~\ref{fig:PPO-ACT_unique_matrix}, this experimental design effectively overcomes theoretical limitations of traditional imitation dynamics methods (e.g., Fermi update rule). Traditional methods face computational failures with all-defectors initial states due to neighbor strategy homogeneity. The PPO-ACT framework overcomes this challenge through its deep reinforcement learning architecture.
	
	The evolutionary dynamics under all-defector initialization reveal critical phase-dependent characteristics, as demonstrated in Fig.~\ref{fig:PPO-ACT_unique_matrix}. During Phase 1 training with $r=5.0$ (iterations 0-999), the PPO mechanism enables stochastic cooperation emergence through adaptive exploration-exploitation balance, cooperation rates show oscillatory growth patterns. Unlike Fermi rule-based updates that suffer from myopic decision-making, PPO's advantage estimation captures long-range spatial correlations, permitting intermittent cooperative cluster formation despite defective initialization. Fig.~\ref{fig:PPO-ACT_unique_matrix}(b) exclusively displays the Phase 2 evolutionary trajectory to better observe strategy adaptation under a reduced enhancement factor ($r=4.8$). During Phase 2 policy reinitialization with $r=4.8$, $50\%$ of agents adopt cooperation strategies at the first iteration through retained knowledge in the Actor-Critic network. All agents become cooperators within 10 iterations.

	\subsection{Comparative analysis of algorithms}
	\label{exp:compare}
	
\begin{figure*}[htbp!]
		\begin{minipage}{\linewidth}
			\begin{minipage}{0.24\linewidth}
				\centering
				\includegraphics[width=\linewidth]{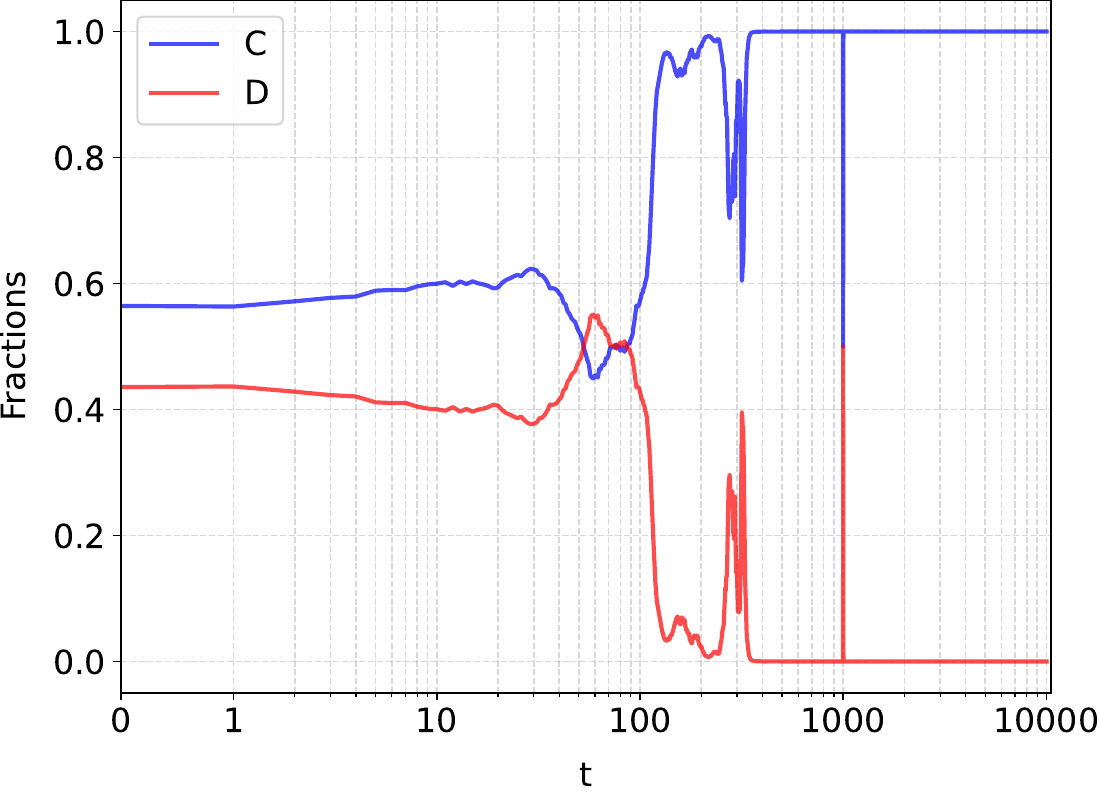}\\
			\end{minipage}
			\begin{minipage}{0.14\linewidth}
				\centering
				\fbox{\includegraphics[width=\linewidth]{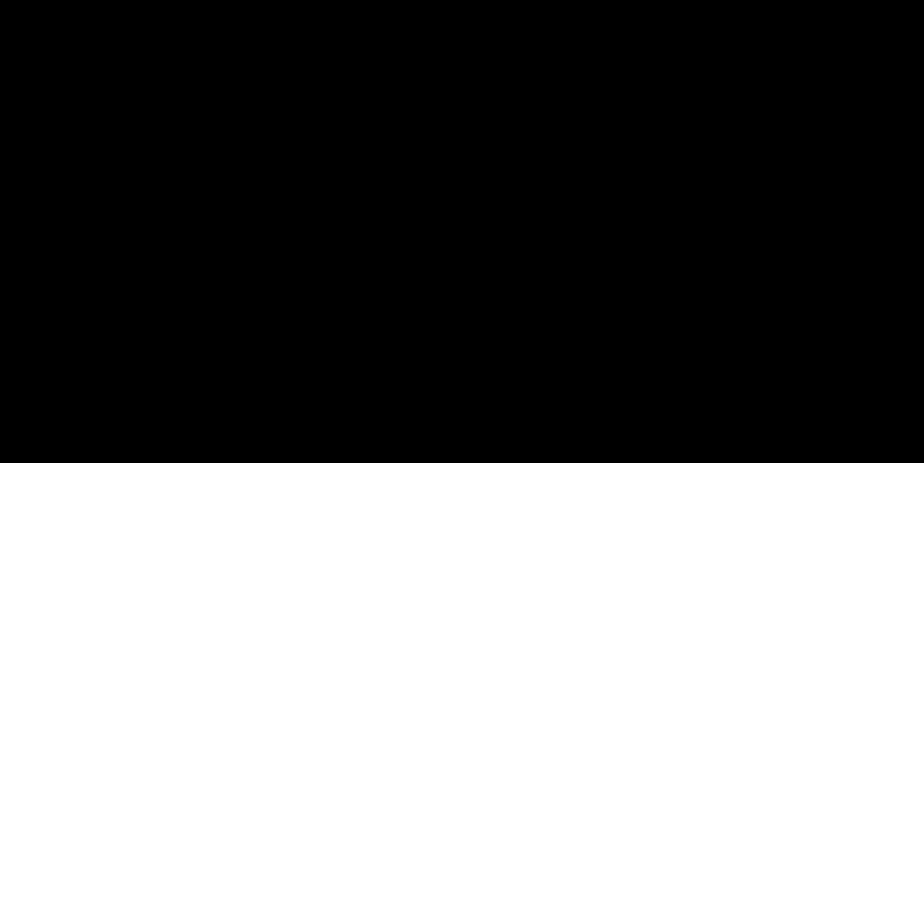}}\\
				\vspace{-2mm}
				{\footnotesize t=0}
			\end{minipage}
			\begin{minipage}{0.14\linewidth}
				\centering
				\fbox{\includegraphics[width=\linewidth]{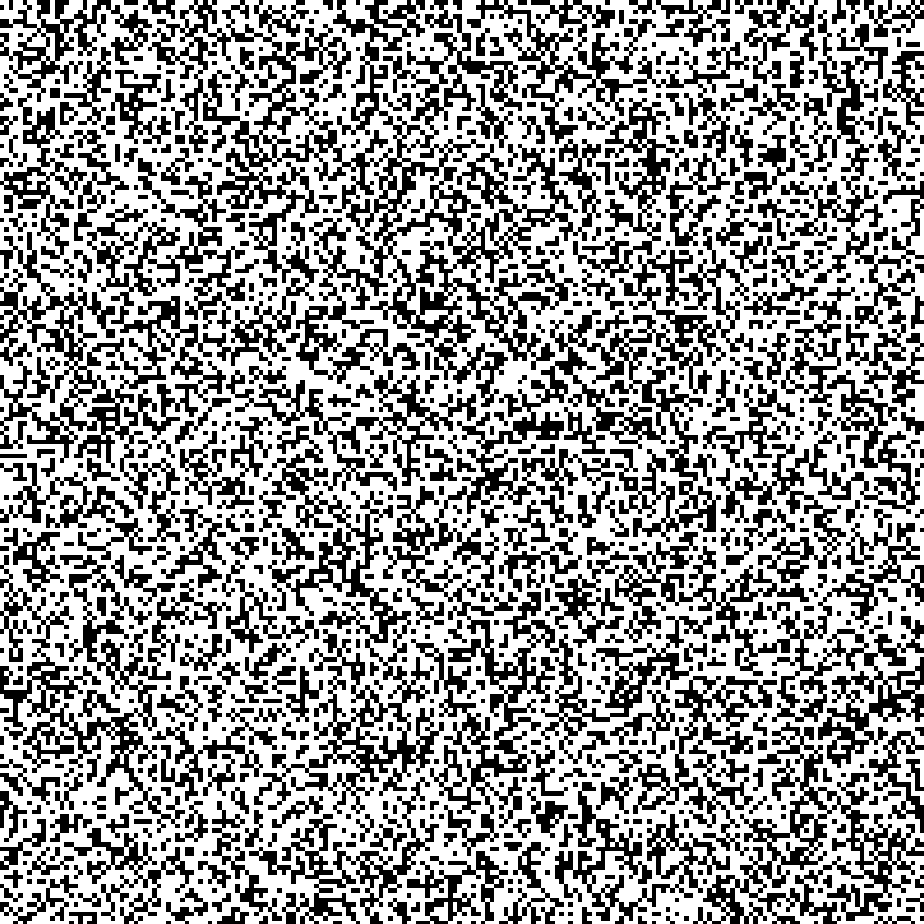}}\\
				\vspace{-2mm}
				{\footnotesize t=10}
			\end{minipage}
			\begin{minipage}{0.14\linewidth}
				\centering
				\fbox{\includegraphics[width=\linewidth]{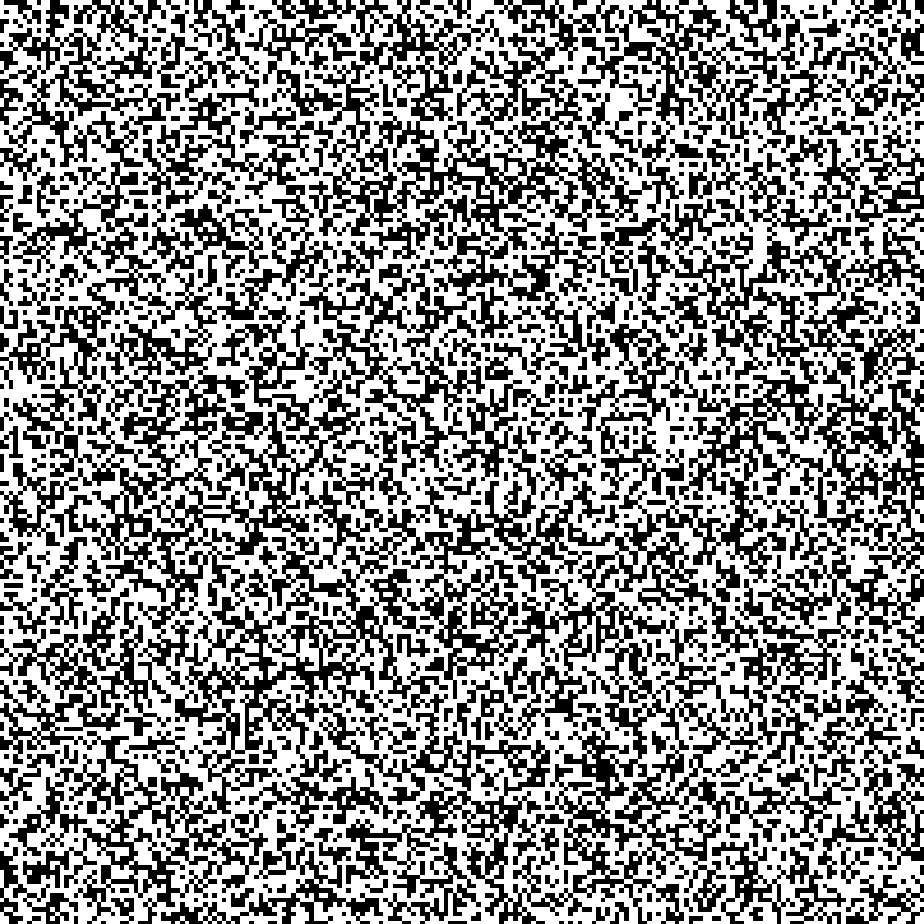}}\\
				\vspace{-2mm}
				{\footnotesize t=100}
			\end{minipage}
			\begin{minipage}{0.14\linewidth}
				\centering
				\fbox{\includegraphics[width=\linewidth]{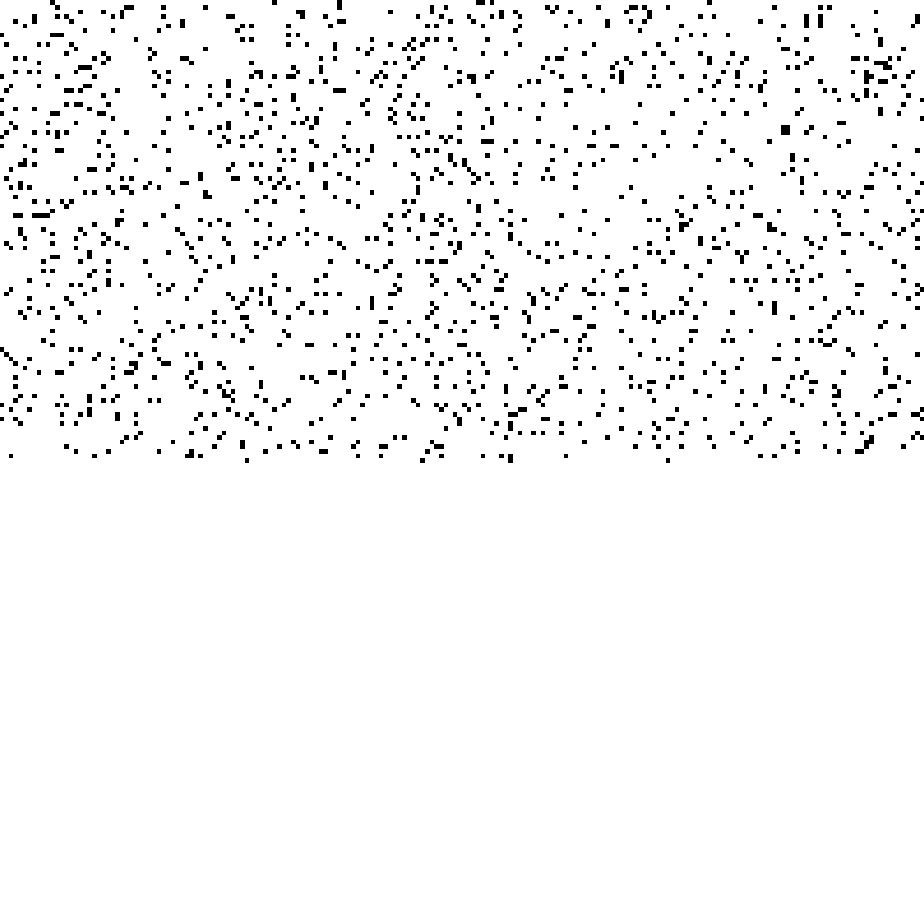}}\\
				\vspace{-2mm}
				{\footnotesize t=1000}
			\end{minipage}
			\begin{minipage}{0.14\linewidth}
				\centering
				\fbox{\includegraphics[width=\linewidth]{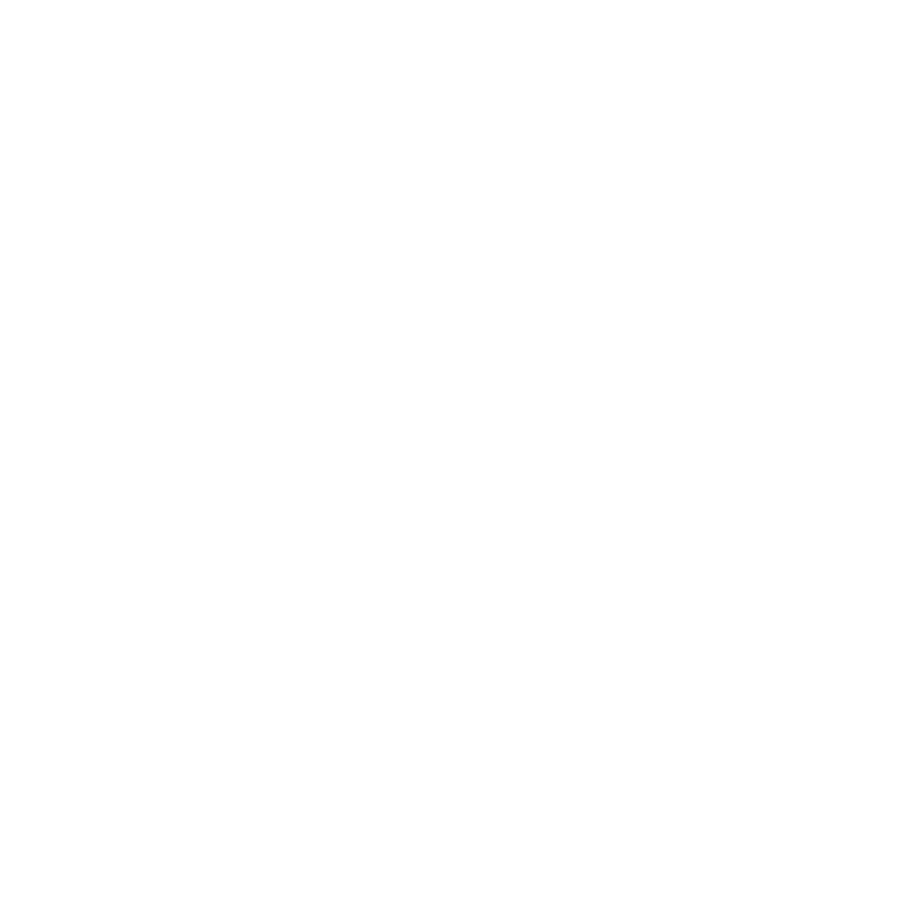}}\\
				\vspace{-2mm}
				{\footnotesize t=10000}
			\end{minipage}
			\vspace{-3mm}
			\caption*{\footnotesize (a)PPO-ACT}
		\end{minipage}
		\\[2mm]
		\begin{minipage}{\linewidth}
			\begin{minipage}{0.24\linewidth}
				\centering
				\includegraphics[width=\linewidth]{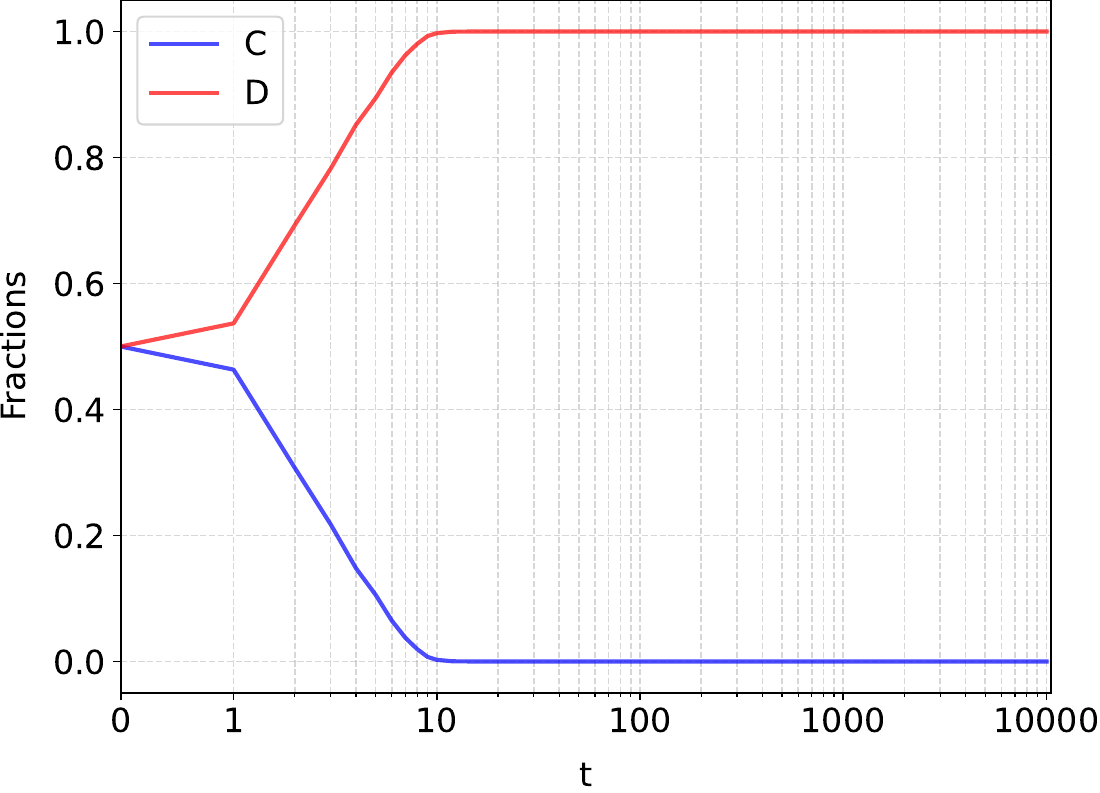}\\
			\end{minipage}
			\begin{minipage}{0.14\linewidth}
				\centering
				\fbox{\includegraphics[width=\linewidth]{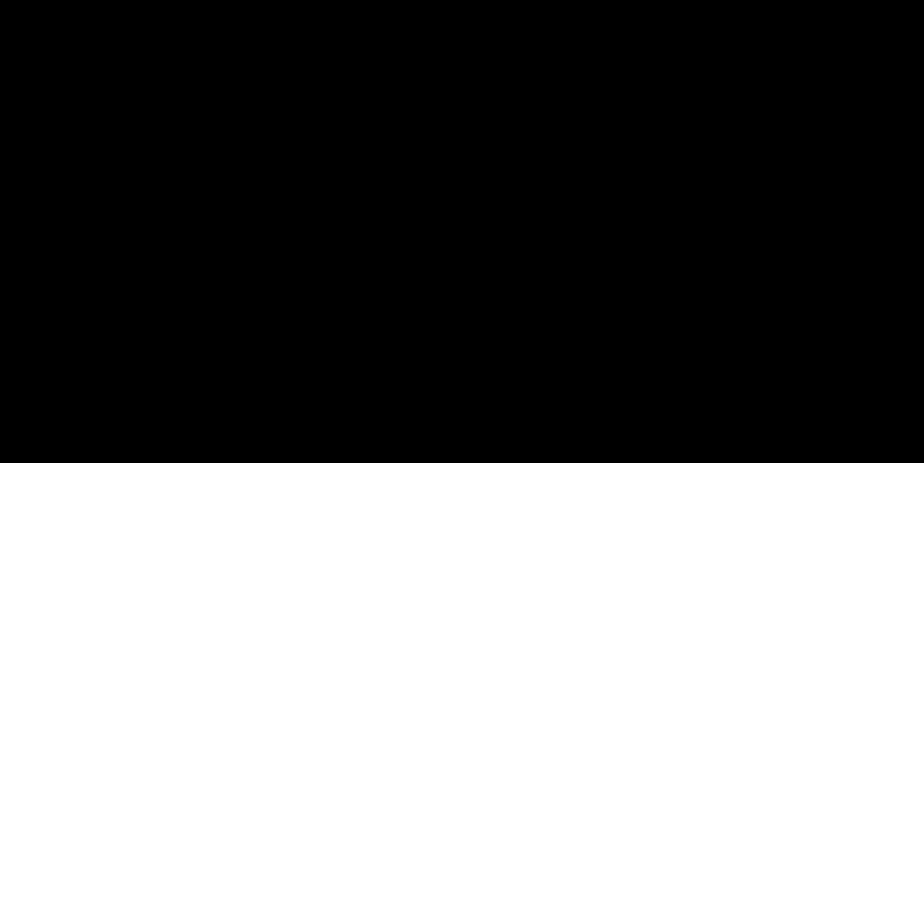}}\\
				\vspace{-2mm}
				{\footnotesize t=0}
			\end{minipage}
			\begin{minipage}{0.14\linewidth}
				\centering
				\fbox{\includegraphics[width=\linewidth]{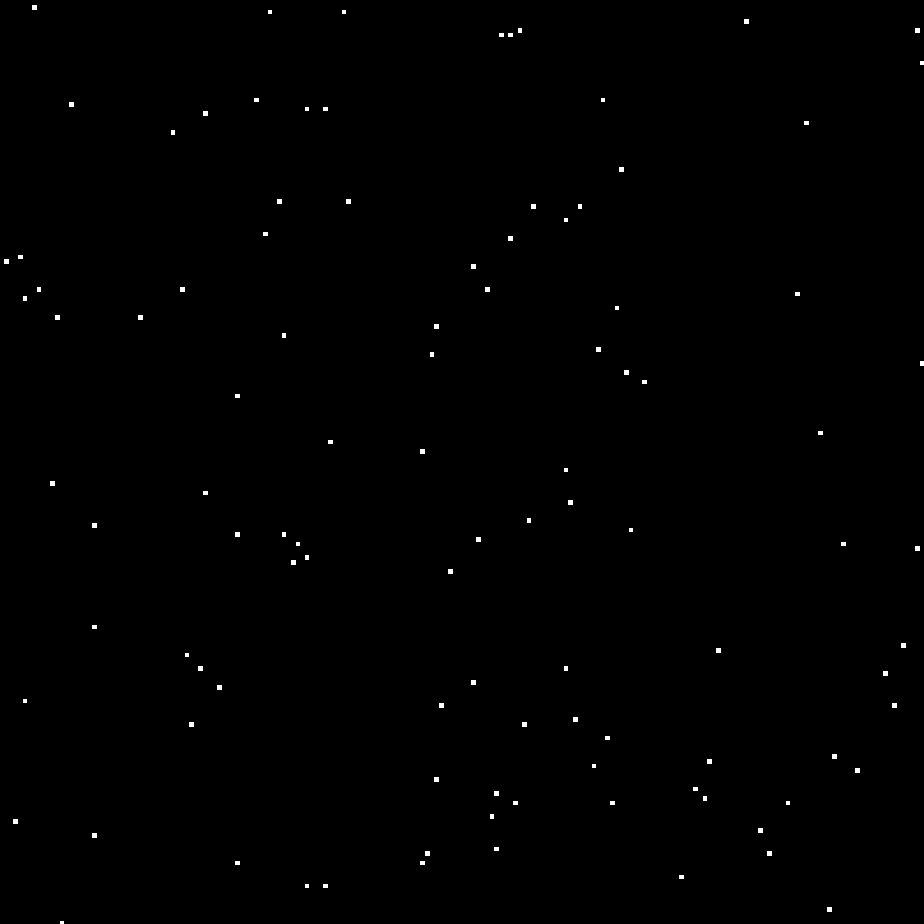}}\\
				\vspace{-2mm}
				{\footnotesize t=10}
			\end{minipage}
			\begin{minipage}{0.14\linewidth}
				\centering
				\fbox{\includegraphics[width=\linewidth]{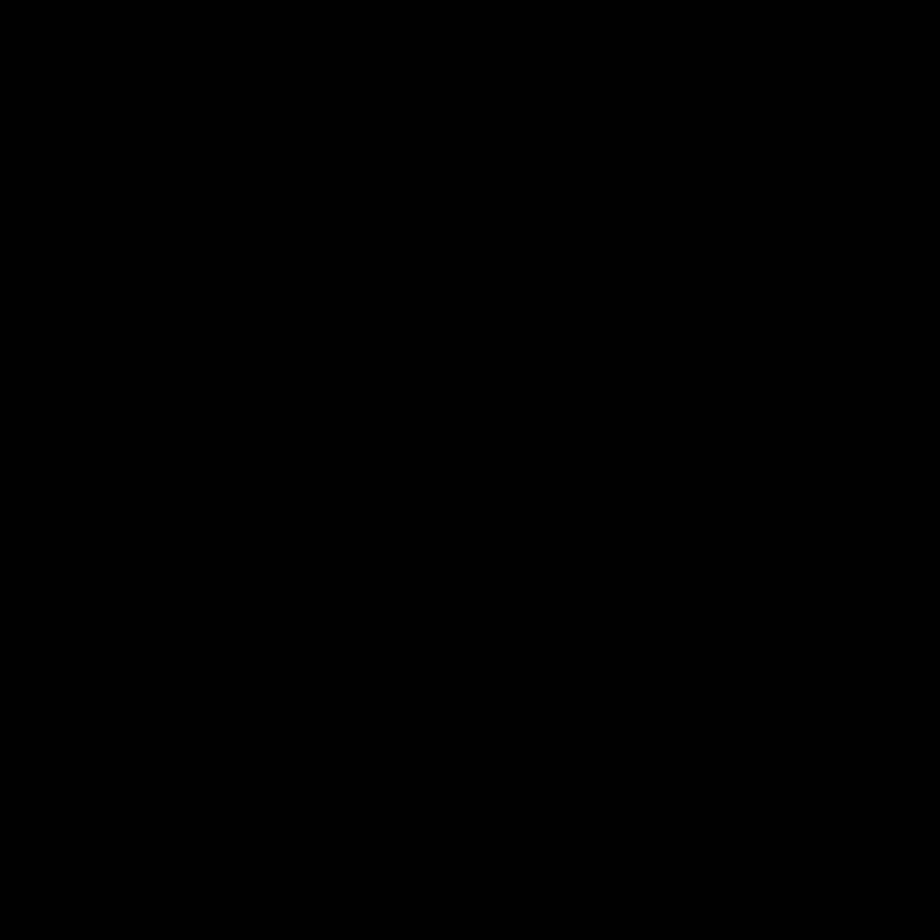}}\\
				\vspace{-2mm}
				{\footnotesize t=100}
			\end{minipage}
			\begin{minipage}{0.14\linewidth}
				\centering
				\fbox{\includegraphics[width=\linewidth]{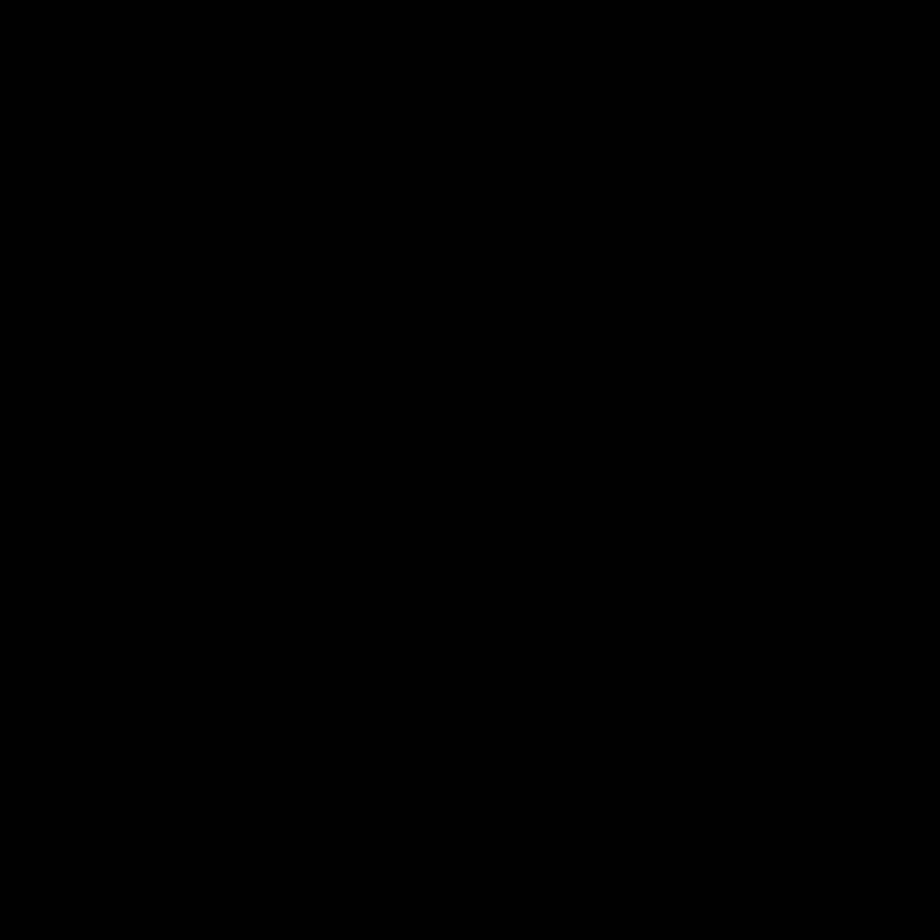}}\\
				\vspace{-2mm}
				{\footnotesize t=1000}
			\end{minipage}
			\begin{minipage}{0.14\linewidth}
				\centering
				\fbox{\includegraphics[width=\linewidth]{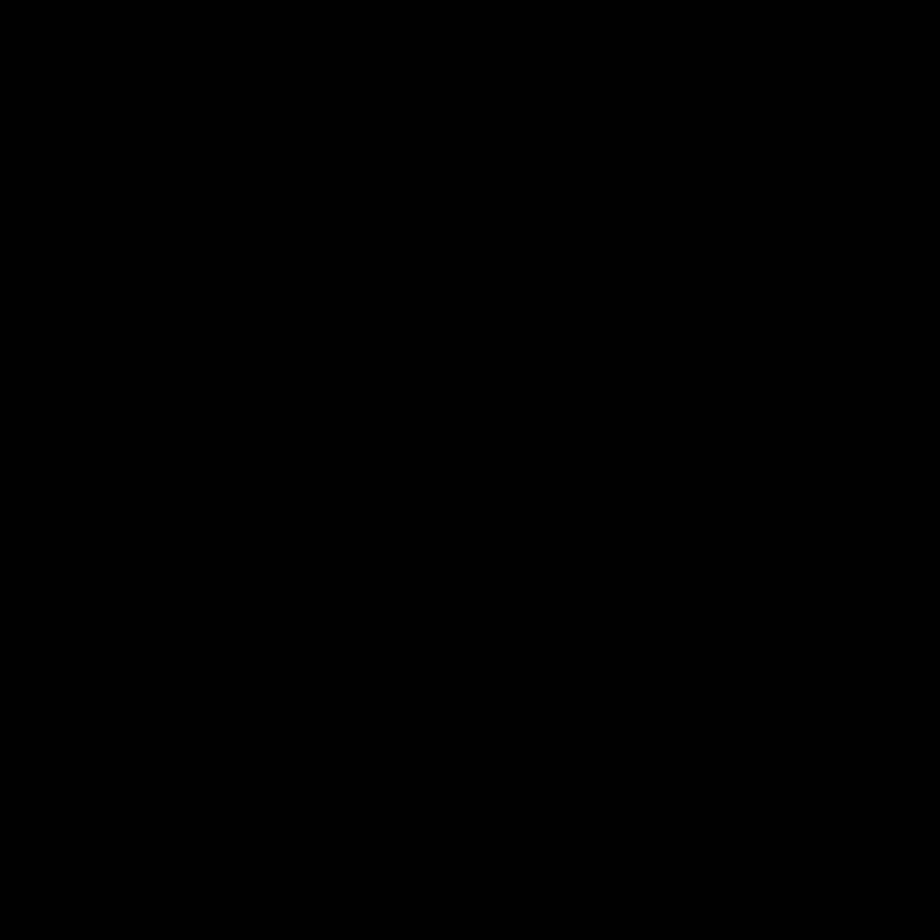}}\\
				\vspace{-2mm}
				{\footnotesize t=10000}
			\end{minipage}
			\vspace{-3mm}
			\caption*{\footnotesize (b)PPO}
		\end{minipage}
		\\[2mm]
		\begin{minipage}{\linewidth}
			\begin{minipage}{0.24\linewidth}
				\centering
				\includegraphics[width=\linewidth]{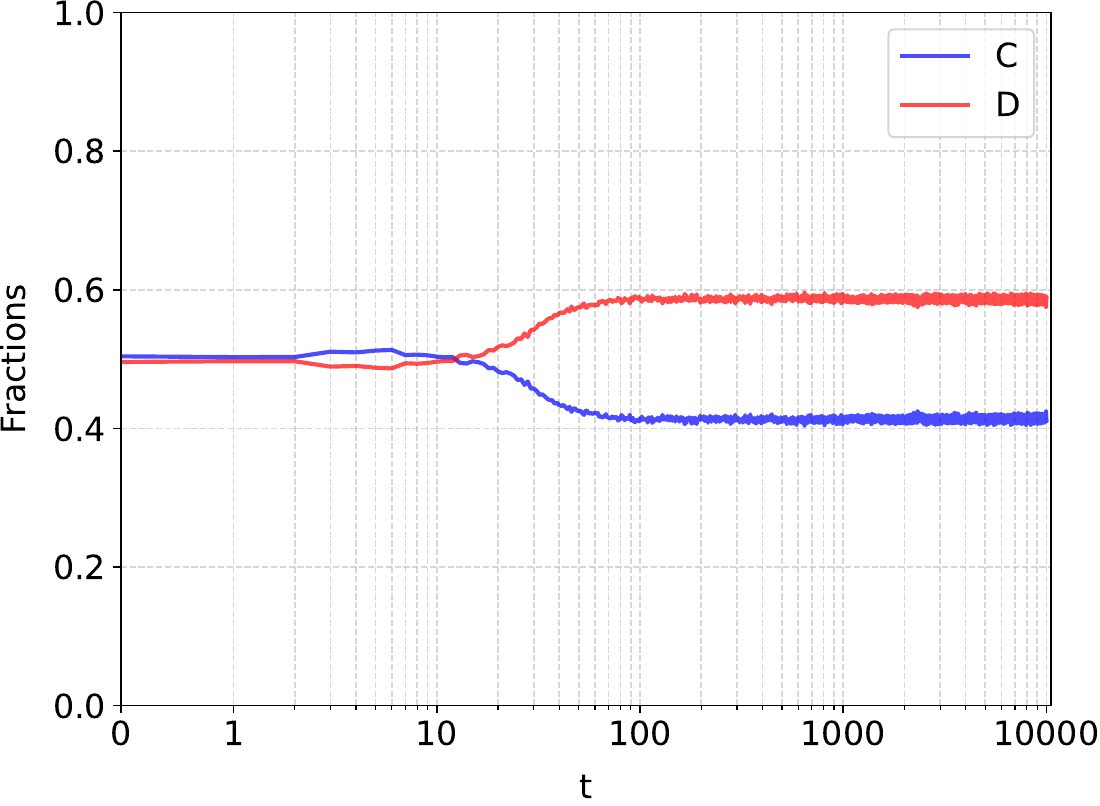}\\
			\end{minipage}
			\begin{minipage}{0.14\linewidth}
				\centering
				\fbox{\includegraphics[width=\linewidth]{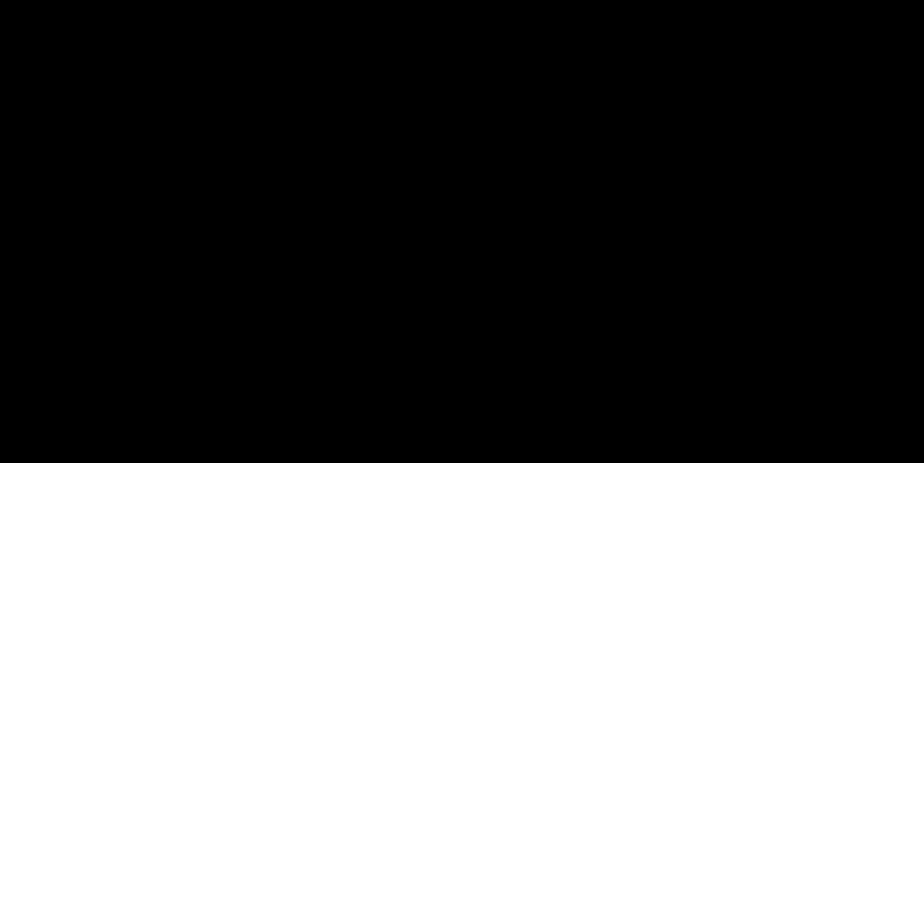}}\\
				\vspace{-2mm}
				{\footnotesize t=0}
			\end{minipage}
			\begin{minipage}{0.14\linewidth}
				\centering
				\fbox{\includegraphics[width=\linewidth]{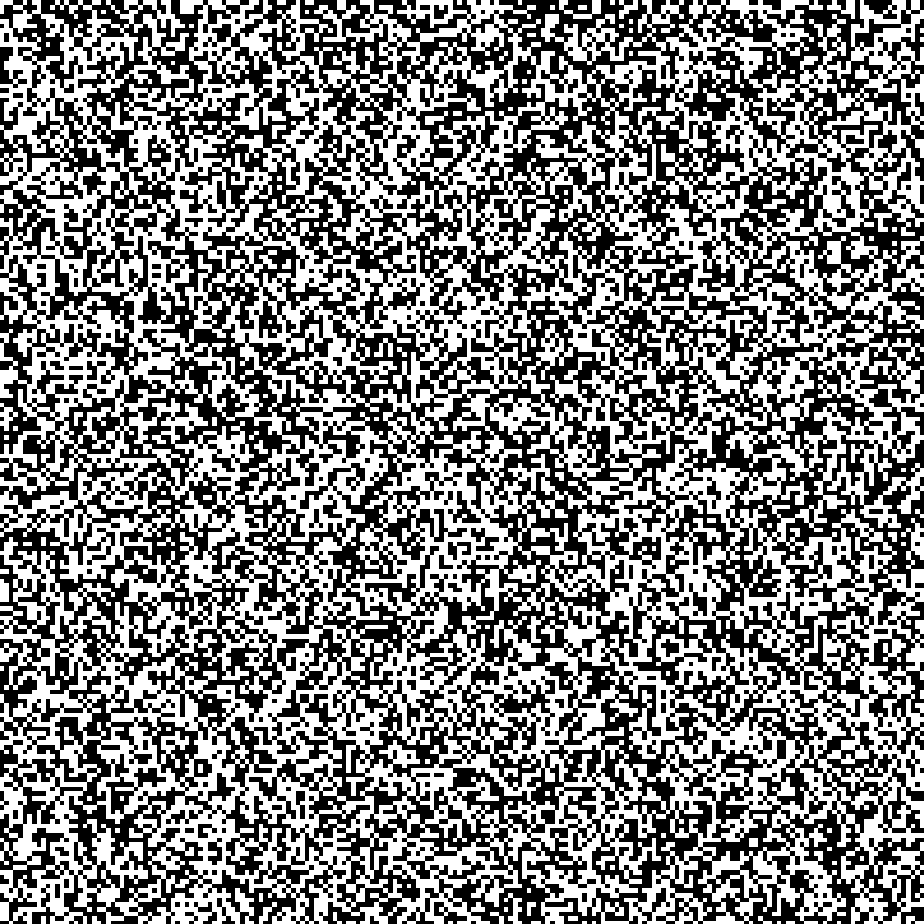}}\\
				\vspace{-2mm}
				{\footnotesize t=10}
			\end{minipage}
			\begin{minipage}{0.14\linewidth}
				\centering
				\fbox{\includegraphics[width=\linewidth]{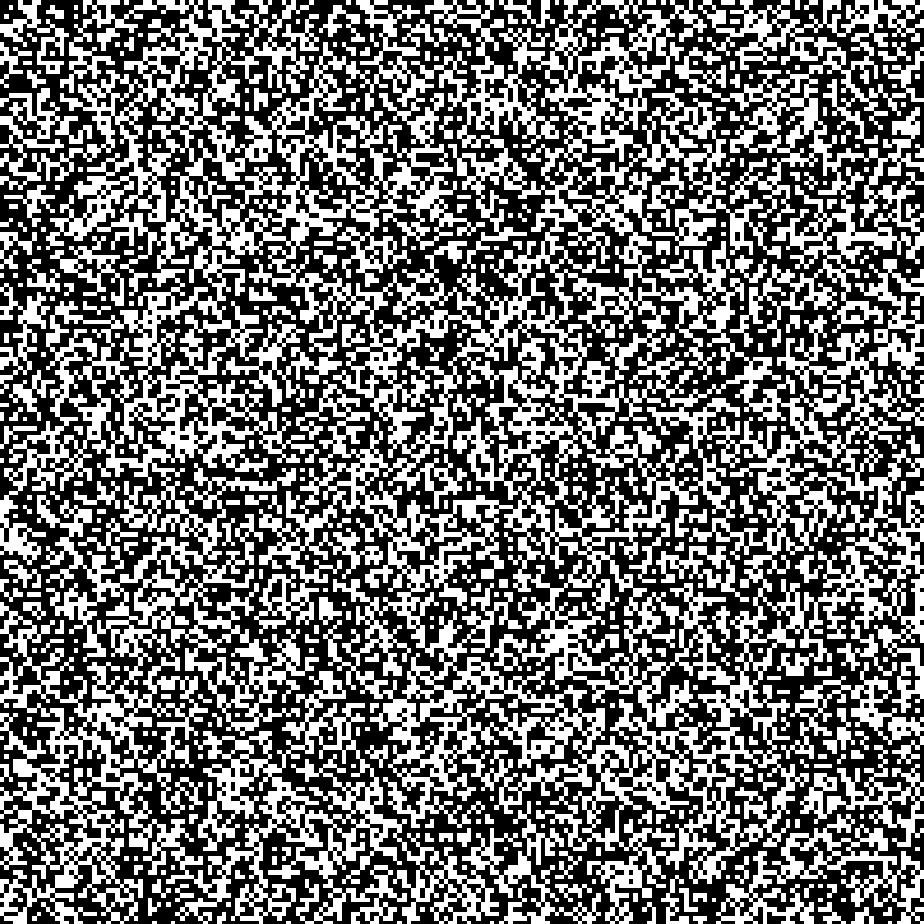}}\\
				\vspace{-2mm}
				{\footnotesize t=100}
			\end{minipage}
			\begin{minipage}{0.14\linewidth}
				\centering
				\fbox{\includegraphics[width=\linewidth]{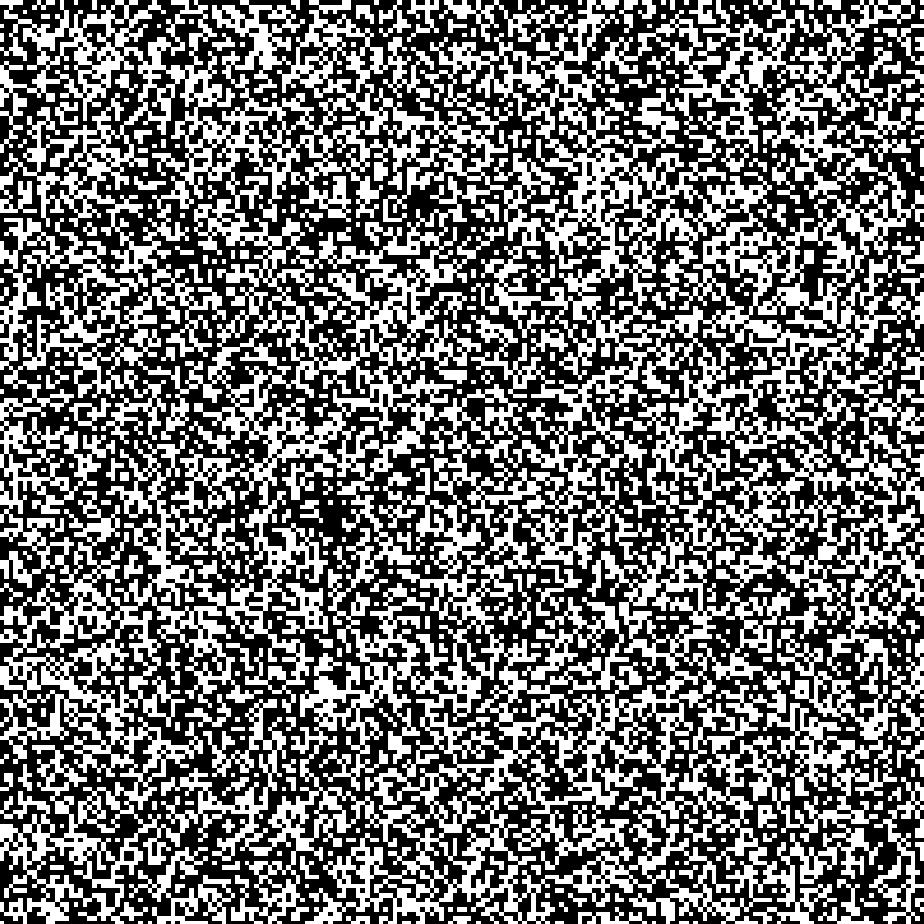}}\\
				\vspace{-2mm}
				{\footnotesize t=1000}
			\end{minipage}
			\begin{minipage}{0.14\linewidth}
				\centering
				\fbox{\includegraphics[width=\linewidth]{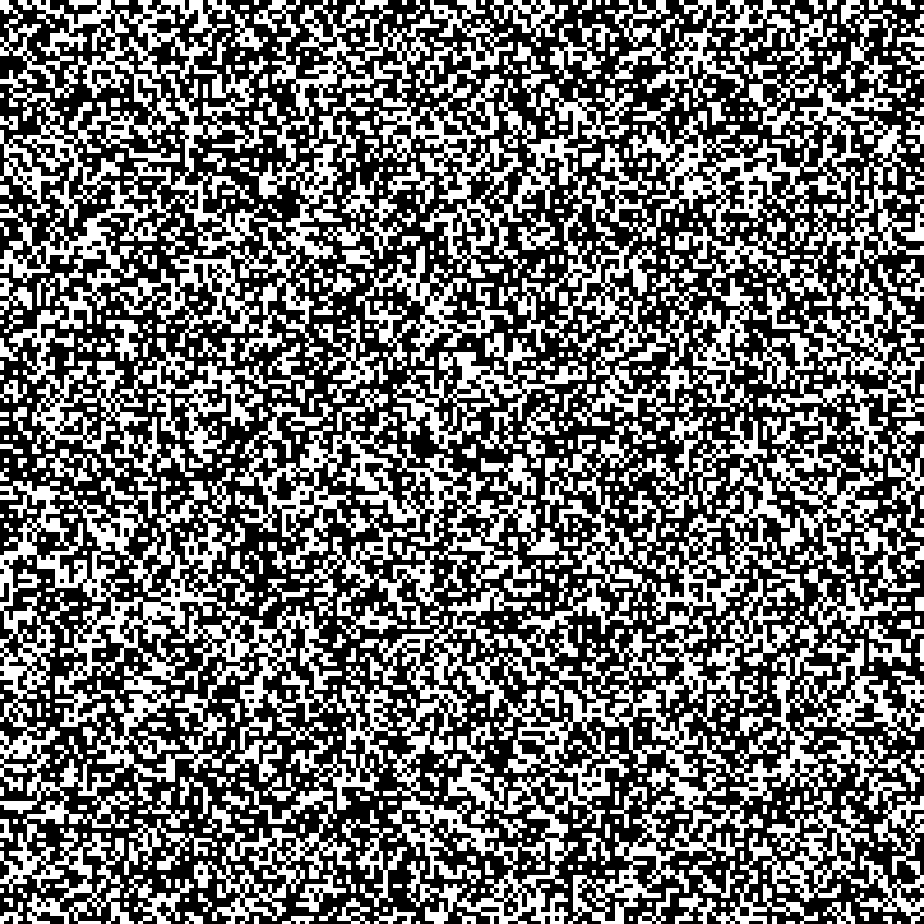}}\\
				\vspace{-2mm}
				{\footnotesize t=10000}
			\end{minipage}
			\vspace{-3mm}
			\caption*{\footnotesize (c)Q-learning}
		\end{minipage}	
		\\[2mm]
		\begin{minipage}{\linewidth}
			\begin{minipage}{0.24\linewidth}
				\centering
				\includegraphics[width=\linewidth]{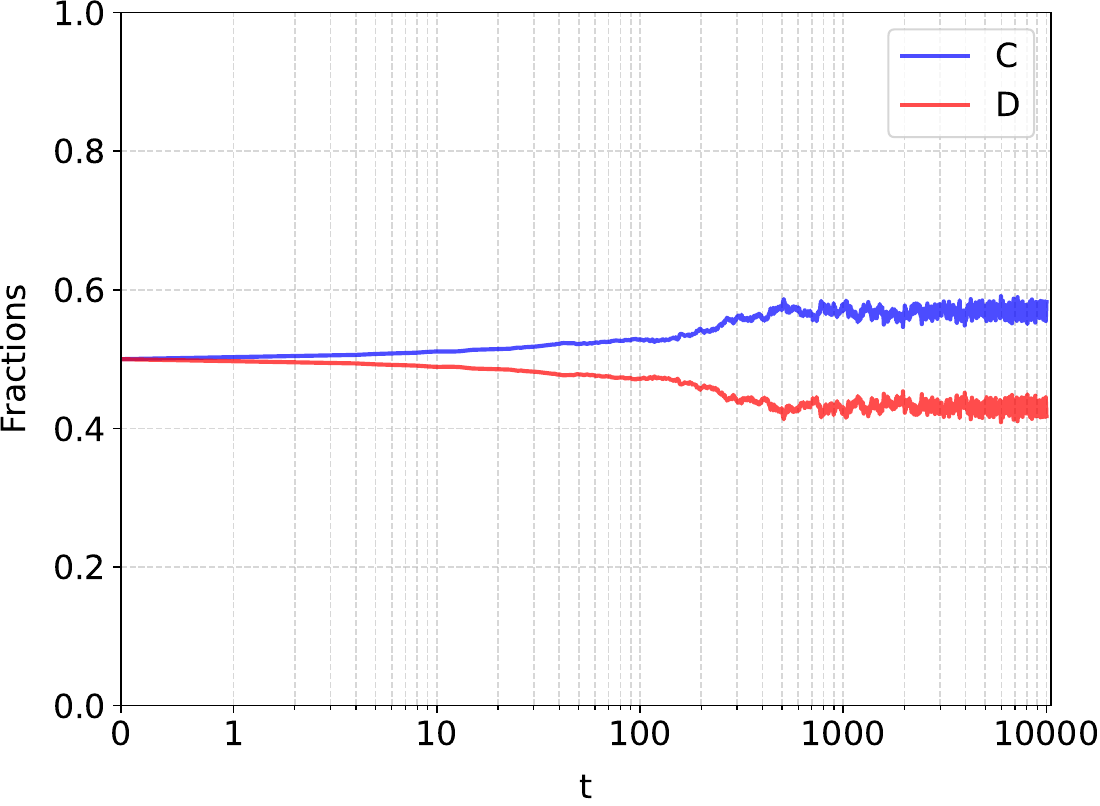}\\
			\end{minipage}
			\begin{minipage}{0.14\linewidth}
				\centering
				\fbox{\includegraphics[width=\linewidth]{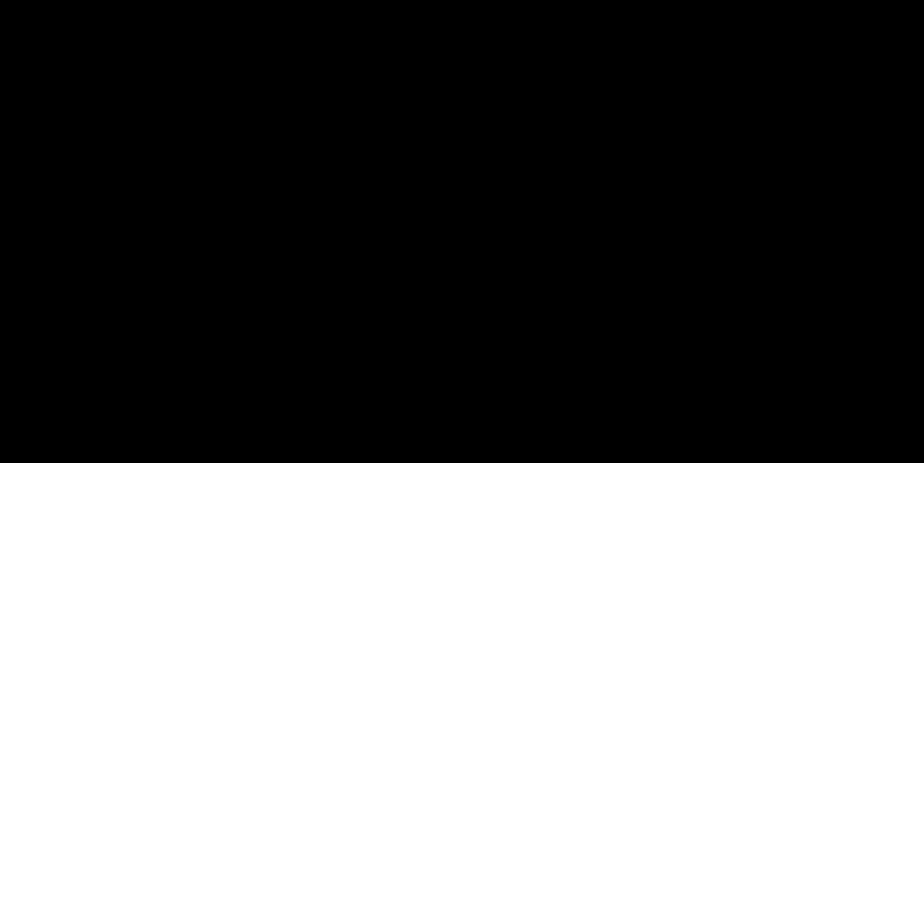}}\\
				\vspace{-2mm}
				{\footnotesize t=0}
			\end{minipage}
			\begin{minipage}{0.14\linewidth}
				\centering
				\fbox{\includegraphics[width=\linewidth]{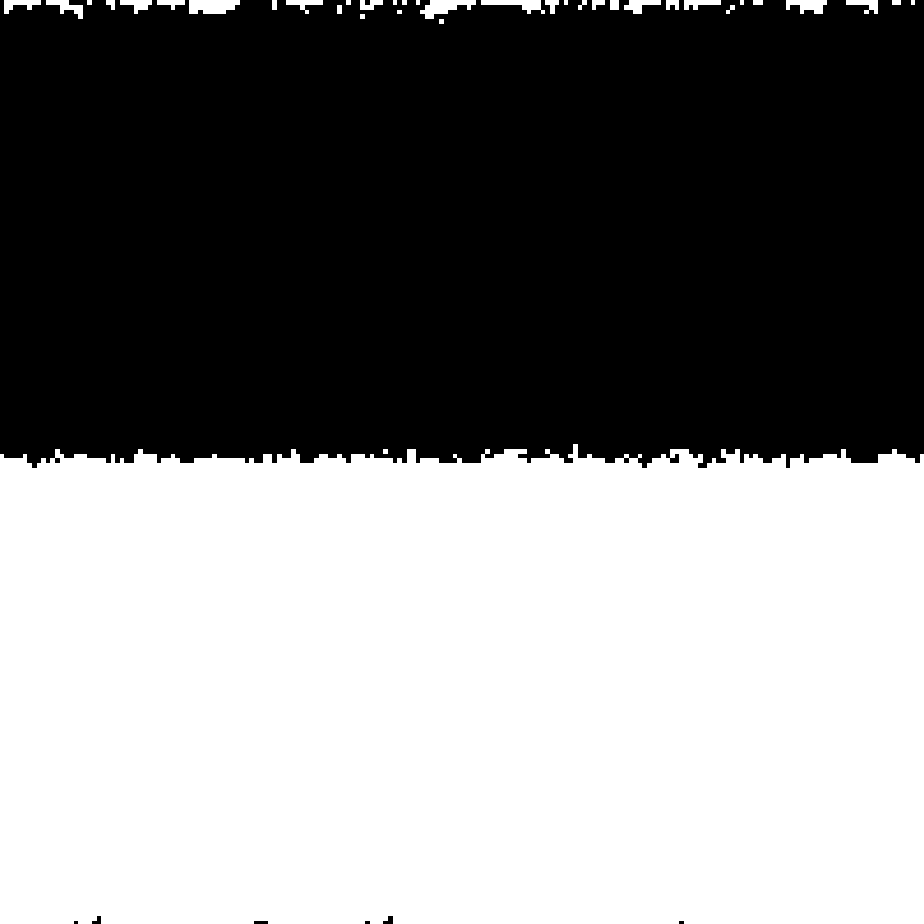}}\\
				\vspace{-2mm}
				{\footnotesize t=10}
			\end{minipage}
			\begin{minipage}{0.14\linewidth}
				\centering
				\fbox{\includegraphics[width=\linewidth]{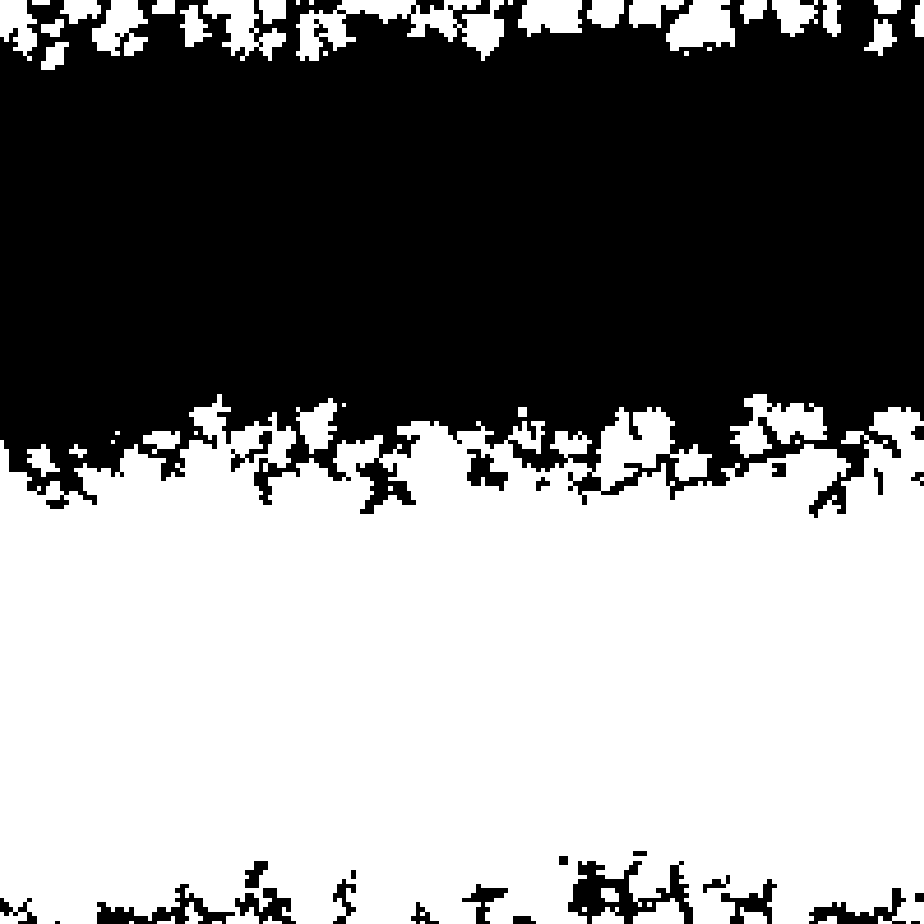}}\\
				\vspace{-2mm}
				{\footnotesize t=100}
			\end{minipage}
			\begin{minipage}{0.14\linewidth}
				\centering
				\fbox{\includegraphics[width=\linewidth]{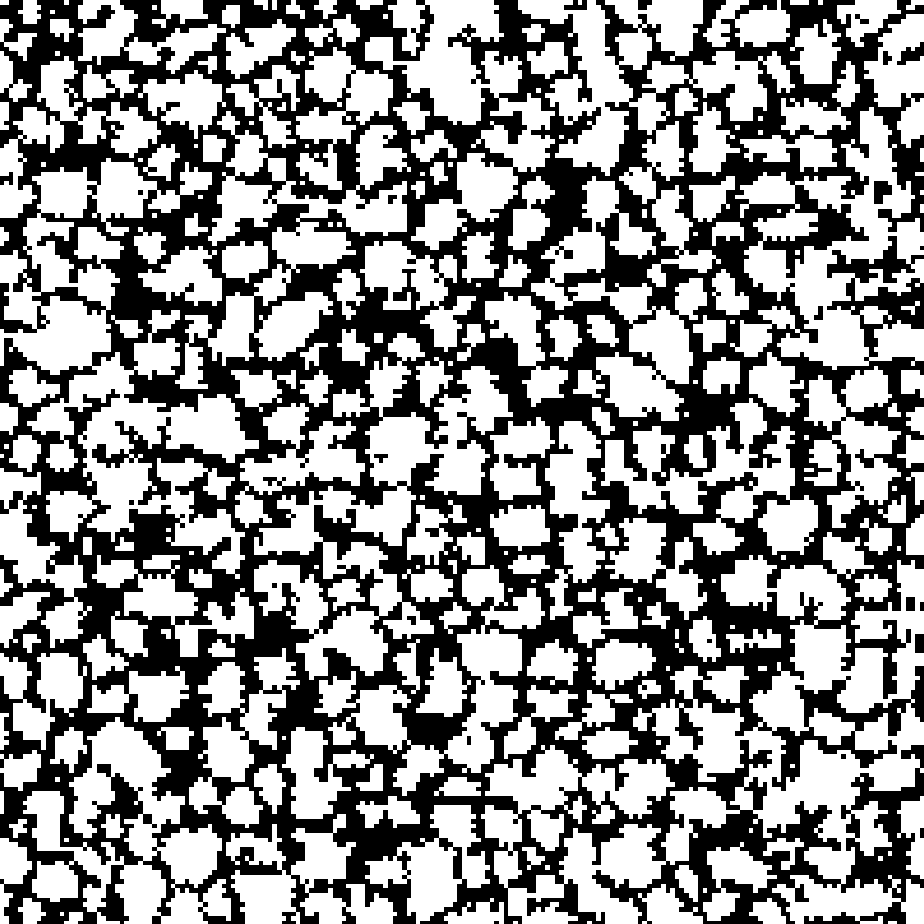}}\\
				\vspace{-2mm}
				{\footnotesize t=1000}
			\end{minipage}
			\begin{minipage}{0.14\linewidth}
				\centering
				\fbox{\includegraphics[width=\linewidth]{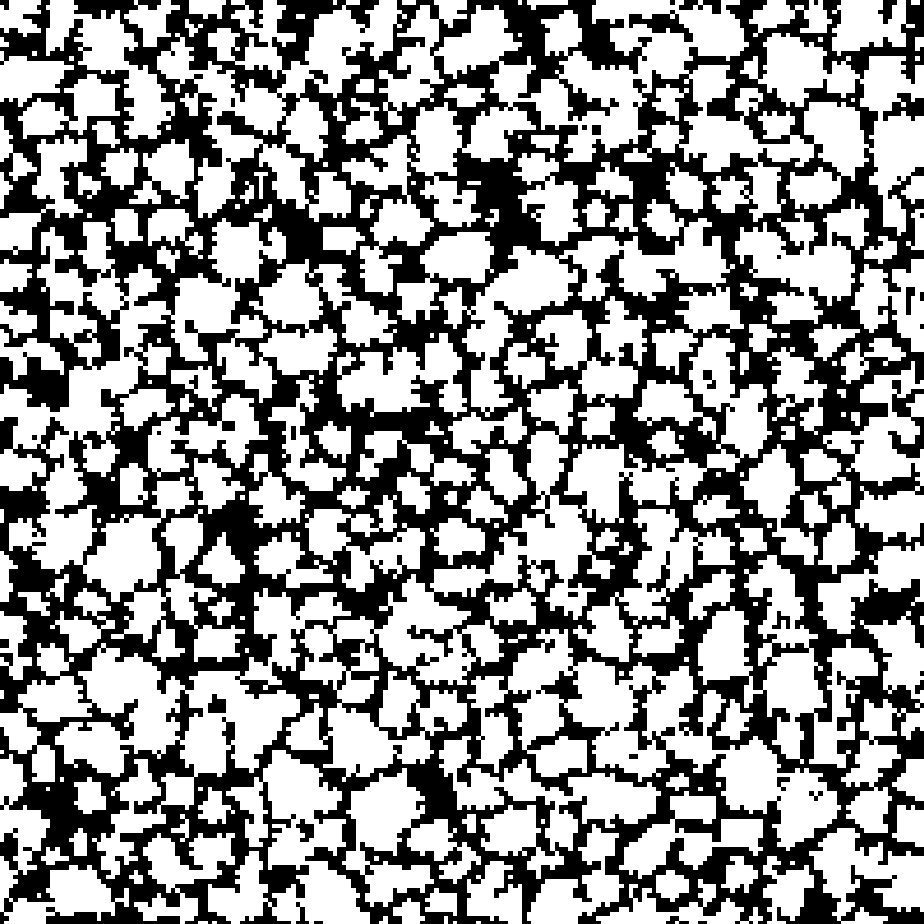}}\\
				\vspace{-2mm}
				{\footnotesize t=10000}
			\end{minipage}
			\vspace{-3mm}
			\caption*{\footnotesize (d)Fermi}
		\end{minipage}	
		\caption{Comparative analysis of PPO-ACT, PPO, Q-learning, and Fermi update rule: Temporal evolution curves of cooperators (blue) and defectors (red) at $r=4.0$, with corresponding snapshots of cooperators (white) and defectors (black). The initialization strategy places defectors in the upper half and cooperators in the lower half. Snapshots are shown chronologically from left to right (0, 10, 100, 1000, and 10000 iterations). Compared to other algorithms, our PPO-ACT can converge to a state where all agents are cooperators at a lower enhancement factor $r$.}
		\label{fig:PPO-ACT_uDbC_matrix_compare}
	\end{figure*}

	Figure \ref{fig:PPO-ACT_uDbC_matrix_compare} compares the evolutionary dynamics of four algorithms (PPO-ACT, PPO, Q-learning, and Fermi update rule) under enhancement factor $r=4.0$. The initialization strategy places defectors in the upper half and cooperators in the lower half. The leftmost subfigure shows temporal evolution curves with iteration count t on the horizontal axis and  fraction of collaborators (blue) and defectors (red) on the vertical axis. The remaining subfigures display state snapshots (white: cooperators, black: defectors).  Comparative analysis of temporal curves and spatial snapshots reveals clear performance differences among algorithms in critical parameter regions.
	
	Our PPO-ACT demonstrates optimal convergence characteristics, ultimately leading all agents to adopt cooperative strategies. The evolutionary process evolves through distinct temporal phases. During Phase 1 ($t < 1000$ iterations, $r=5.0$), the Actor-Critic network learns to output cooperation-preferring policies and achieves improved state estimation. With the existence of cooperation-preferring agents, all agents' strategies rapidly stabilized at cooperative policies during Phase 2 under $r=4.0$. In contrast, the standard PPO algorithm rapidly converges to complete defection during early iterations ($t < 10$). Snapshots reveal early global defection strategies diffusion. Q-learning exhibits persistent oscillations, with the fraction of cooperators fluctuating around $42\%$ throughout training without achieving stable strategy formation. Random spatial mixtures indicate ineffective spatial information utilization, reflecting limitations of value-function methods in high-dimensional continuous policy spaces. This primarily originates from inherent discrete state representation constraints: inability to capture spatial correlations causes delayed responses to local neighborhood changes and lacks global policy distribution modeling. Comparatively, neural-network-based PPO-ACT better models spatial correlations through parameter sharing, enabling stable convergence. As a classical imitation dynamics method, the Fermi update rule displays unique spatial clustering. In later iterations ($t>1000$), the system reaches a steady state with coexisting cooperator/defector clusters. The final fraction of cooperators stabilizes around $57\%$, with snapshots showing clear phase separation. This verifies spatial structure's role in promoting cooperation under local interaction rules while revealing global optimization deficiencies.
	
	\subsection{Statistical analysis of PPO-ACT}
	\label{exp:compare_stat}
	
	To verify the reliability of PPO-ACT and quantitatively evaluate the enhancement from the ACT module, we systematically tested enhancement factors $r \in [3.0,6.0]$ in 50 independent trials per value. Results are presented through three complementary statistical visualizations.
	
\begin{figure*}[htbp!]
		\begin{minipage}{0.48\linewidth}
			\centering
			\includegraphics[width=\linewidth]{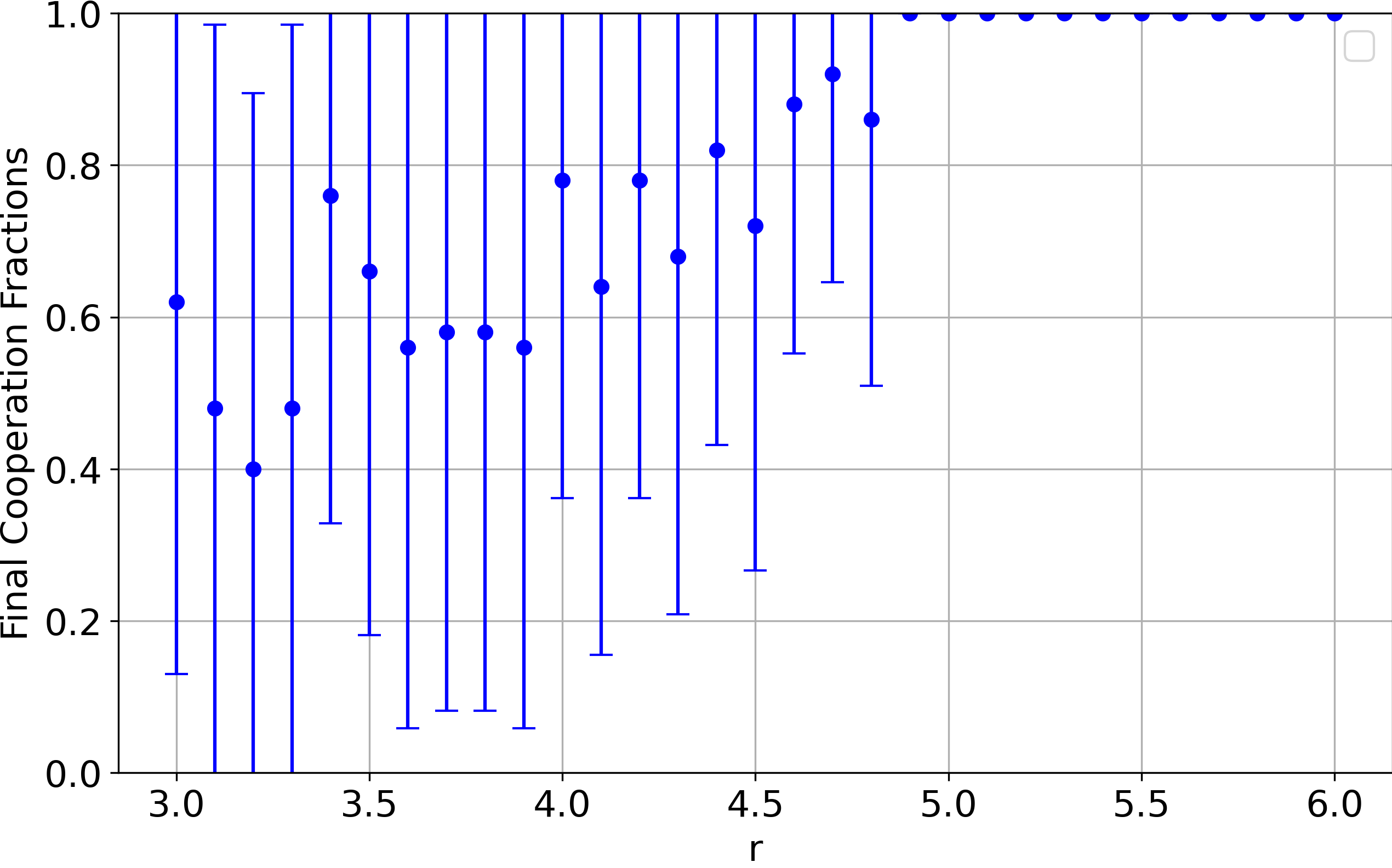}\\
			\vspace{-4mm}
			\caption*{\footnotesize (a) PPO-ACT}
		\end{minipage}
		\begin{minipage}{0.48\linewidth}
			\centering
			\includegraphics[width=\linewidth]{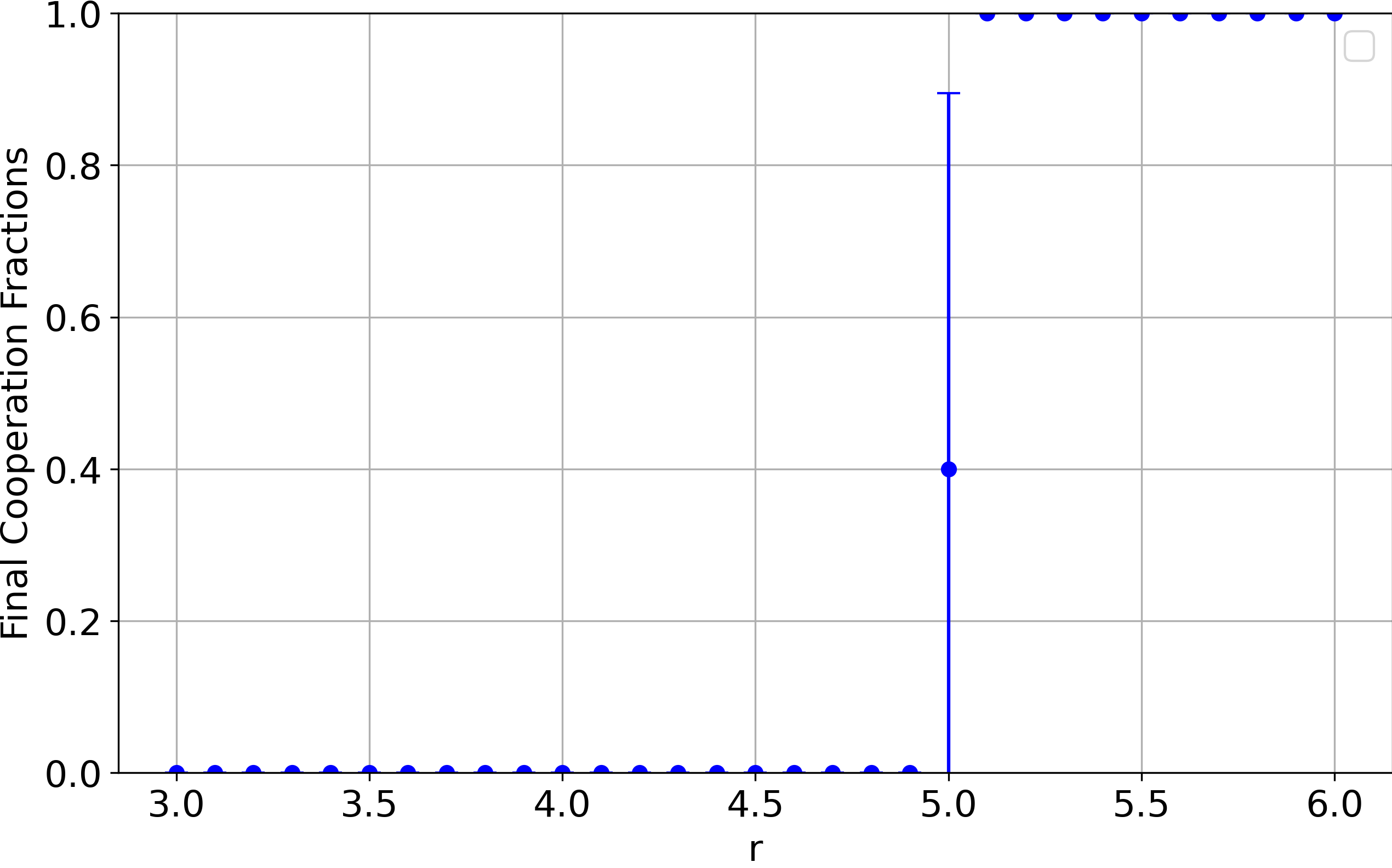}\\
			\vspace{-4mm}
			\caption*{\footnotesize (b) PPO}
		\end{minipage}
		\caption{Comparison of mean cooperation rates and standard deviation between PPO-ACT and PPO. Due to the effect of ACT, our PPO-ACT exhibits cooperative behavior at a smaller $r$ than PPO.}
		\label{fig:PPO-ACT_r_stat_err}
	\end{figure*}

	Error Bar Analysis (Fig.~\ref{fig:PPO-ACT_r_stat_err}) demonstrates the fundamental distinction between PPO-ACT and baseline PPO in terms of mean cooperation rates and their variability. Both algorithms exhibit binary outcomes (either 0\% or 100\% cooperation) in individual runs, resulting in substantial standard deviations across all $r$ values. Specifically, PPO-ACT achieves cooperation at lower $r$ values compared to PPO. It confirms the ACT's effectiveness in lowering the cooperation barrier.
	﻿
	
\begin{figure*}[htbp!]
		\centering
		\begin{minipage}{0.8\linewidth}
			\centering
			\includegraphics[width=\linewidth]{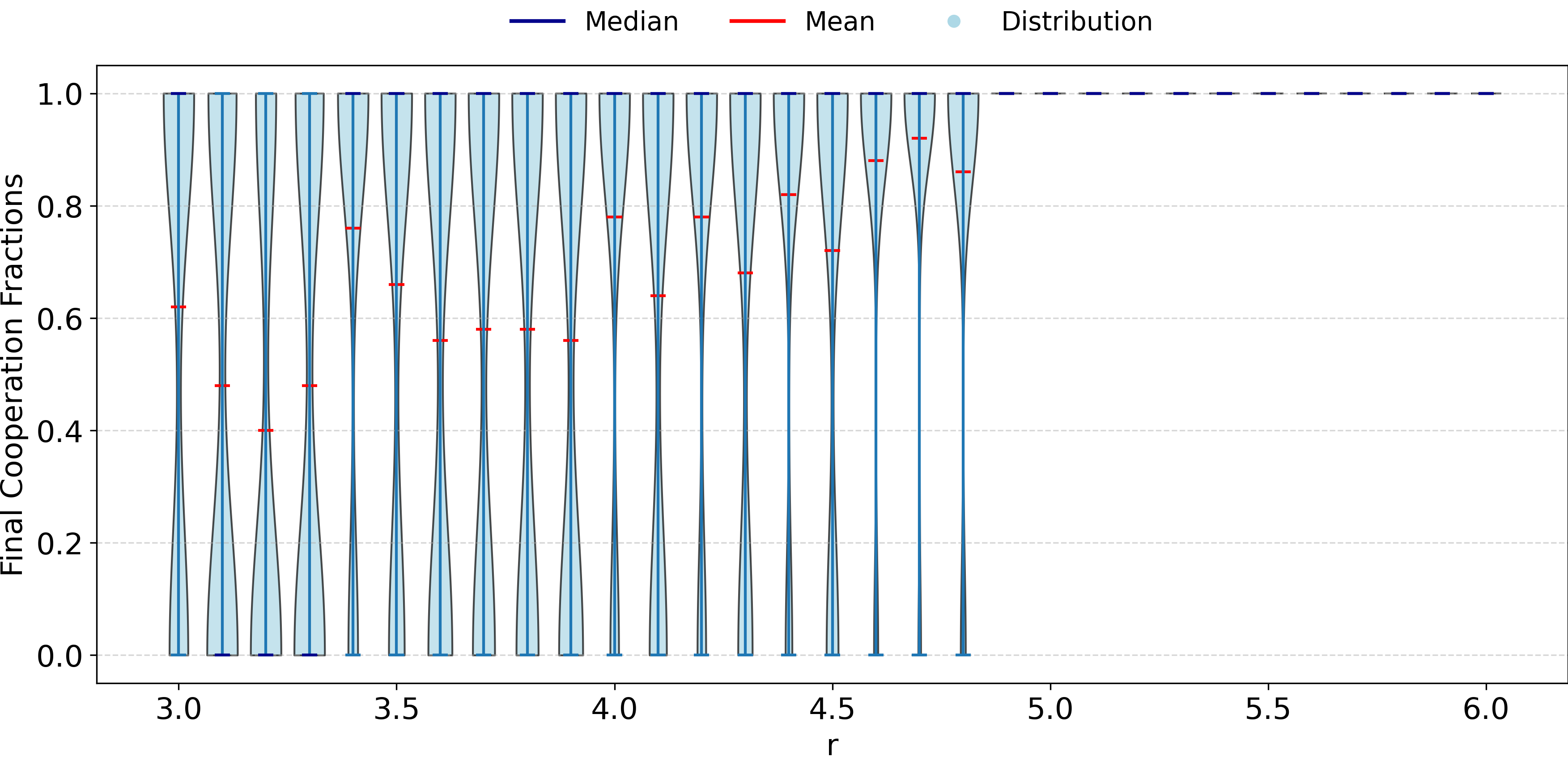}\\
			\vspace{-4mm}
			\caption*{\footnotesize (a) PPO-ACT}
		\end{minipage}
		\\[1mm]
		\begin{minipage}{0.8\linewidth}
			\centering
			\includegraphics[width=\linewidth]{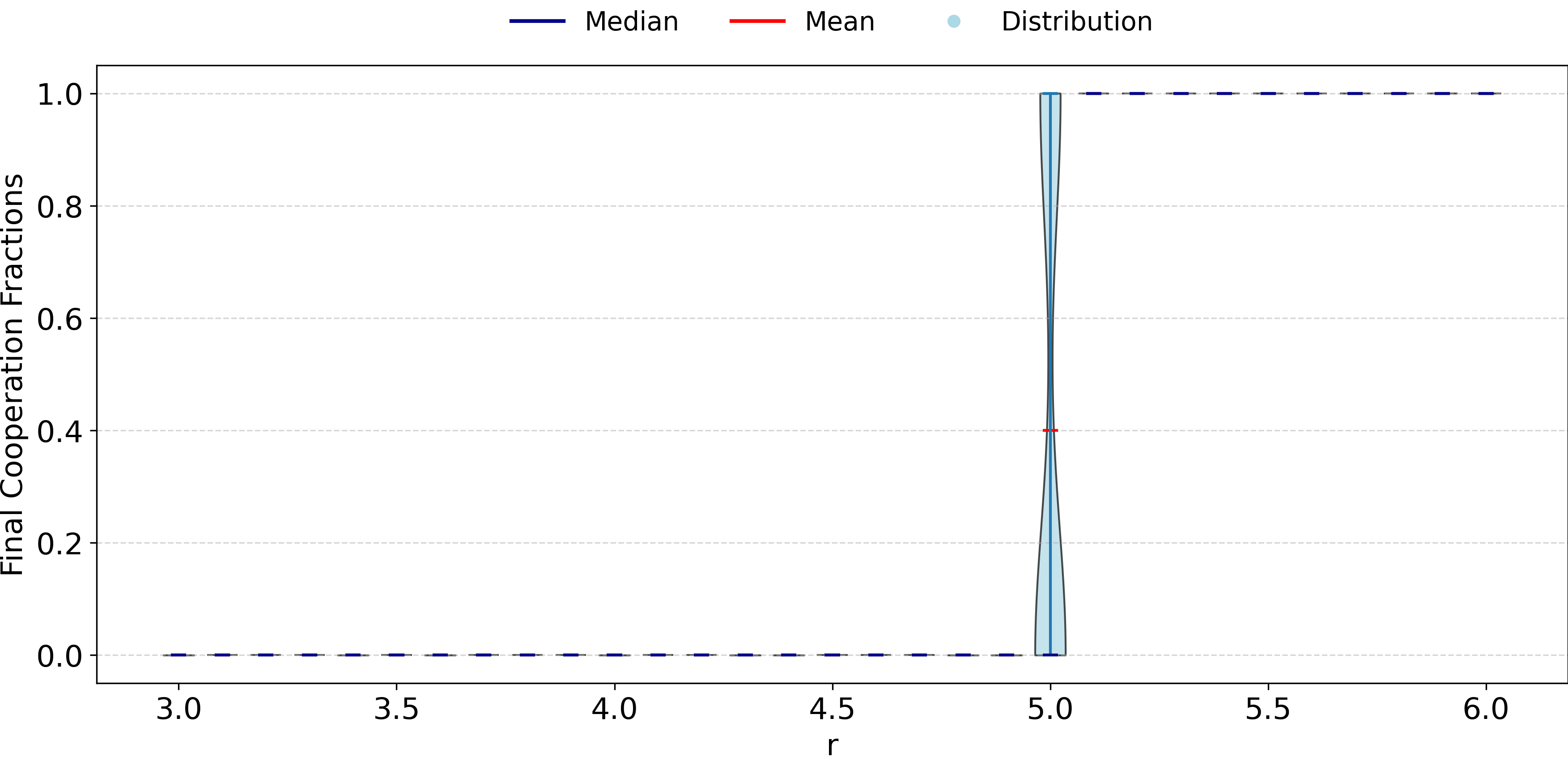}\\
			\vspace{-4mm}
			\caption*{\footnotesize (b) PPO}
		\end{minipage}	
		\caption{Probability density of cooperation fractions for PPO-ACT (left) and PPO (right) across enhancement factors $r \in [3.0,6.0]$. The violin plot indicates that PPO-ACT has a higher probability of converging to a state where all agents are collaborators even when r is small.}
		\label{fig:PPO-ACT_r_stat_vio}
	\end{figure*}

	Violin plot analysis (Fig.~\ref{fig:PPO-ACT_r_stat_vio}) reveals key findings through median, mean, and distribution examination.  Compared with PPO at $r = 5.0$, when the enhancement factor $r \in [3.0, 5.0]$, PPO-ACT demonstrates significantly superior performance in both median cooperation rates and mean values. Moreover, Fig.~\ref{fig:PPO-ACT_r_stat_vio}) visualizes the probability density distribution of cooperation rates across different r values. It clearly shows that PPO-ACT achieves earlier emergence and more stable high-density cooperative regions in its distribution.

	As shown in Table~\ref{tab:CI_comparison}, PPO-ACT and PPO exhibit distinct behavioral patterns in the spatial public goods game. PPO-ACT demonstrates progressive evolutionary characteristics, with its cooperation rate confidence intervals evolving smoothly from 0.48--0.76 at $r=3.0$ to 0.84--1.00 at $r=4.7$, reflecting the probabilistic decision-making nature of this strategy. In contrast, PPO displays a clear threshold transition behavior: maintaining complete defection when $r<5.0$, transitioning to partial cooperation at $r=5.0$, and achieving full cooperation when $r>5.0$.
		The confidence intervals marked as `nan' in the table indicate conditions where the strategy outputs were completely deterministic (e.g., PPO-ACT's full cooperation at $r\geq4.9$), making valid confidence interval computation impossible. The 0.00--0.00 intervals represent cases where the observed cooperation rates were extremely low but not absolutely zero. These results suggest that PPO-ACT is better suited for complex social dilemma scenarios, while PPO's deterministic strategy shows significant limitations in flexibility.
	
	\begin{table}[h]
		\centering
		\footnotesize
		\caption{$95\%$ confidence intervals comparison for cooperation fractions}
		\label{tab:CI_comparison}
		\resizebox{\textwidth}{!}{ 
			\begin{tabular}{@{}c*{7}{S[table-format=1.2]@{\,--\,}S[table-format=1.2]}@{}}
				\toprule
				r & \multicolumn{2}{c}{3.0} & \multicolumn{2}{c}{3.1} & \multicolumn{2}{c}{3.2} & \multicolumn{2}{c}{3.3} & \multicolumn{2}{c}{3.4} & \multicolumn{2}{c}{3.5} & \multicolumn{2}{c}{3.6} \\
				\cmidrule(lr){2-3} \cmidrule(lr){4-5} \cmidrule(lr){6-7} \cmidrule(lr){8-9} \cmidrule(lr){10-11} \cmidrule(lr){12-13} \cmidrule(lr){14-15}
				PPO-ACT & 0.48 & 0.76 & 0.34 & 0.62 & 0.26 & 0.54 & 0.34 & 0.62 & 0.64 & 0.88 & 0.52 & 0.80 & 0.42 & 0.70 \\
				PPO & 0.00 & 0.00 & nan & nan & nan & nan  & nan & nan  & 0.00 & 0.00 & nan & nan  & 0.00 & 0.00 \\
				\midrule
				
				r & \multicolumn{2}{c}{3.7} & \multicolumn{2}{c}{3.8} & \multicolumn{2}{c}{3.9} & \multicolumn{2}{c}{4.0} & \multicolumn{2}{c}{4.1} & \multicolumn{2}{c}{4.2} & \multicolumn{2}{c}{4.3} \\
				\cmidrule(lr){2-3} \cmidrule(lr){4-5} \cmidrule(lr){6-7} \cmidrule(lr){8-9} \cmidrule(lr){10-11} \cmidrule(lr){12-13} \cmidrule(lr){14-15}
				PPO-ACT & 0.44 & 0.72 & 0.44 & 0.72 & 0.42 & 0.70 & 0.66 & 0.90 & 0.50 & 0.78 & 0.66 & 0.90 & 0.55 & 0.81 \\
				PPO &nan & nan  & nan & nan  & nan & nan & 0.00 & 0.00 & nan & nan  & nan & nan  & 0.00 & 0.00 \\
				\midrule
				
				r & \multicolumn{2}{c}{4.4} & \multicolumn{2}{c}{4.5} & \multicolumn{2}{c}{4.6} & \multicolumn{2}{c}{4.7} & \multicolumn{2}{c}{4.8} & \multicolumn{2}{c}{4.9} & \multicolumn{2}{c}{5.0} \\
				\cmidrule(lr){2-3} \cmidrule(lr){4-5} \cmidrule(lr){6-7} \cmidrule(lr){8-9} \cmidrule(lr){10-11} \cmidrule(lr){12-13} \cmidrule(lr){14-15}
				PPO-ACT & 0.71 & 0.93 & 0.59 & 0.85 & 0.79 & 0.97 & 0.84 & 1.00 & 0.76 & 0.96 & nan & nan & nan & nan \\
				PPO & 0.00 & 0.00 & nan & nan  & nan & nan  & nan & nan & 0.00 & 0.00 & 0.00 & 0.00 & 0.26 & 0.54 \\
				\bottomrule
			\end{tabular}
		}
	\end{table}

	\subsection{Comparative analysis of algorithms across different $r$}
	\label{exp:compare_r}
	
\begin{figure*}[h]
		\begin{minipage}{0.48\linewidth}
			\centering
			\includegraphics[width=\linewidth]{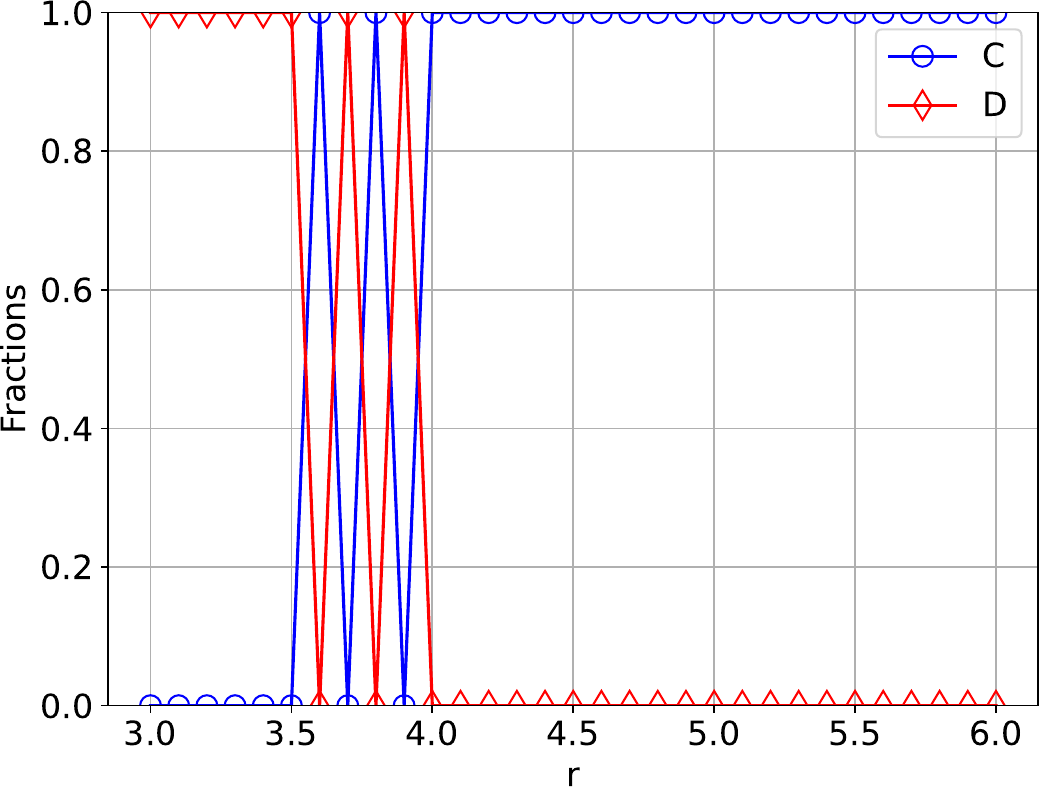}\\
			\vspace{-4mm}
			\caption*{\footnotesize (a) PPO-ACT}
		\end{minipage}
		\begin{minipage}{0.48\linewidth}
			\centering
			\includegraphics[width=\linewidth]{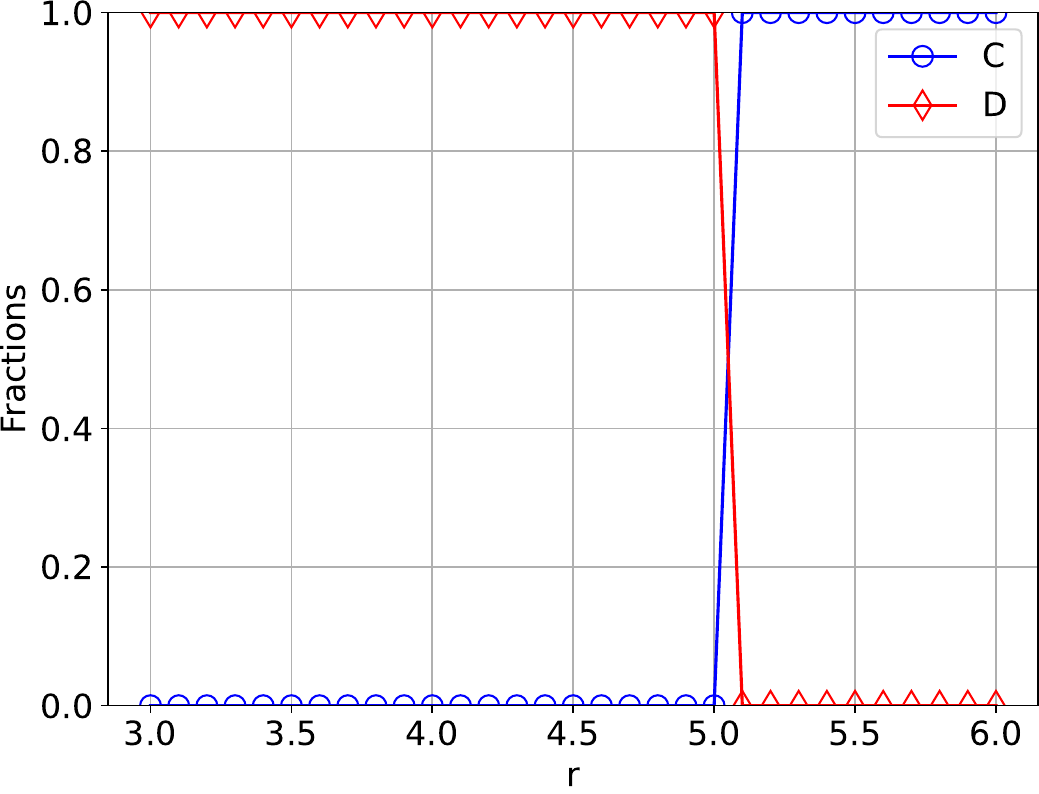}\\
			\vspace{-4mm}
			\caption*{\footnotesize (b) PPO}
		\end{minipage}
		\\
		[3mm]
		\begin{minipage}{0.48\linewidth}
			\centering
			\includegraphics[width=\linewidth]{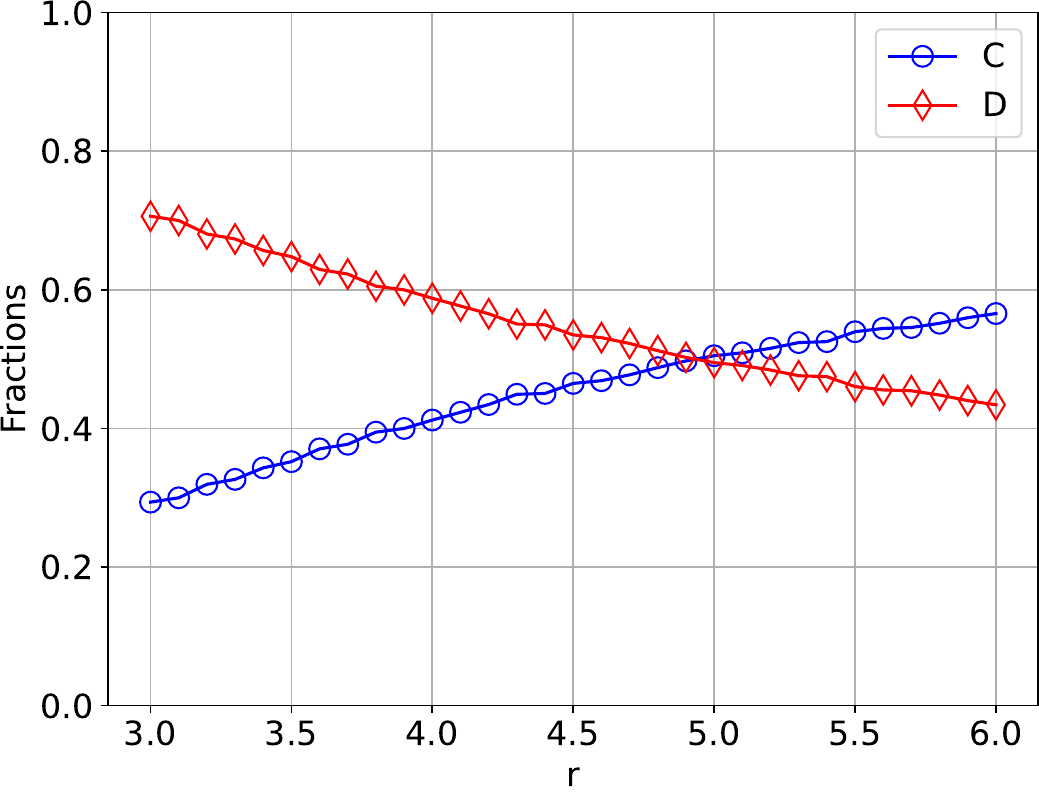}\\
			\vspace{-4mm}
			\caption*{\footnotesize (c) Q-learing}
		\end{minipage}	
		\begin{minipage}{0.48\linewidth}
			\centering
			\includegraphics[width=\linewidth]{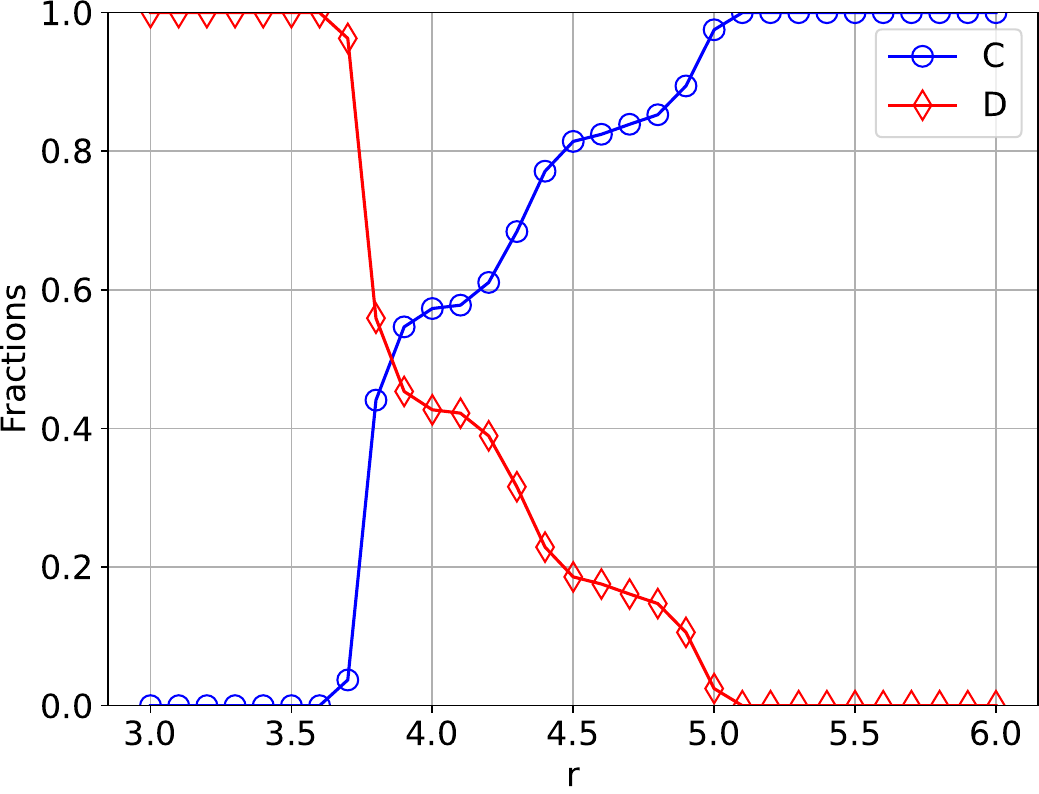}\\
			\vspace{-4mm}
			\caption*{\footnotesize (d) Fermi}
		\end{minipage}	
		\caption{Performance comparison of PPO-ACT, PPO, Q-learning, and Fermi update rule across different r-values: cooperators (blue) and defectors (red). Initial agent strategies were set with all defectors in the upper grid half and all cooperators in the lower half. Compared to other algorithms, starting from a lower enhancement factor $r$, our PPO-ACT can stably converge to a state where all agents' strategies are cooperative strategies.}
		\label{fig:PPO-ACT_r_uDbC_matrix_compare}
	\end{figure*}

	The experiment conducts a systematic performance comparison across four distinct algorithms: PPO-ACT, PPO, Q-learning, and the Fermi update rule. All methods are evaluated under identical spatially heterogeneous initial conditions, where defectors exclusively occupy the upper half of the domain while cooperators populate the lower half. The Fig.~\ref{fig:PPO-ACT_r_uDbC_matrix_compare} illustrates the relationship between enhancement factor r (horizontal axis) and the fraction of collaborators and defectors (vertical axis). This analysis reveals fundamental differences in how the four algorithms drive evolutionary dynamics.

	PPO-ACT maintains cooperative behavior under lower gain conditions ($r=4.0$). This stems from the Actor-Critic networks after Phase 1 training, which enhances spatial synergy effects and biases agents toward cooperative choices. Comparative cross-algorithm experiments reveal PPO-ACT's superior critical behavioral characteristics: the minimum enhancement factor $r$ required to trigger cooperative emergence is significantly reduced, while the $r$ threshold for sustaining stable cooperation also decreases accordingly. Our PPO-ACT framework overcomes the inherent limitation of standard PPO where randomly initialized actor-critic networks fail to discover cooperative strategies under low enhancement factors ($r<5.1$). Through curriculum learning that implements phased reward shaping (Phase 1: $r=5.0$ → Phase 2: $r=4.0$), the algorithm progressively guides policy networks to develop cooperation-enabling representations. This stands in sharp contrast to conventional PPO's behavior at $r=4.0$, where the system inevitably converges to all-defector outcomes. In such cases, random initialization of policy parameters traps agents in defection-dominated Nash equilibria, completely preventing cooperation emergence. 
	
	PPO-ACT and PPO exhibit typical bistable convergence patterns - systems exclusively reach all-cooperator or all-defector states, with extremely low probabilities of mixed cooperator-defector coexistence. This convergence behavior demonstrates nonlinear amplification effects inherent in the policy update mechanism, where minor initial strategy deviations become exponentially amplified through reinforcement learning, ultimately driving the system to phase transition into pure-strategy steady states. Q-learning exhibits a gradual improvement pattern in cooperation levels. However, even with high enhancement factors ($r=6.0$), the algorithm fails to achieve cooperation fractions exceeding $60\%$. This performance ceiling directly reflects the fundamental constraints of discrete Q-tables, which cannot effectively model spatial correlations between agents. Although the Fermi update rule can generate cooperative behavior when $r\geq3.7$, it requires $r\geq5.1$ to achieve stable full cooperation, demonstrating the inefficiency of local update rules. The experimental results show that PPO-ACT enables cooperators to persist under more stringent gain conditions. Our model offers crucial design principles for developing spatially adaptive multi-agent architectures.

	\subsection{Hyperparameter sensitivity analysis of PPO-ACT}
	\label{sec:hyperparam}
	
	This section analyzes the impact of four key hyperparameters on PPO-ACT performance: the learning rate $\alpha$, discount factor $\gamma$, value function loss weight $\delta$, and entropy regularization weight $\rho$. We only study the parameters of the Phase 2. Comparative experiments conducted across varying enhancement factors $r$ reveal the optimal value for each hyperparameter. The horizontal axis is enhancement factor $r$ and the vertical axis is fraction of collaborators.
	
\begin{figure}[htbp!]
		\centering
		\includegraphics[width=0.7\linewidth]{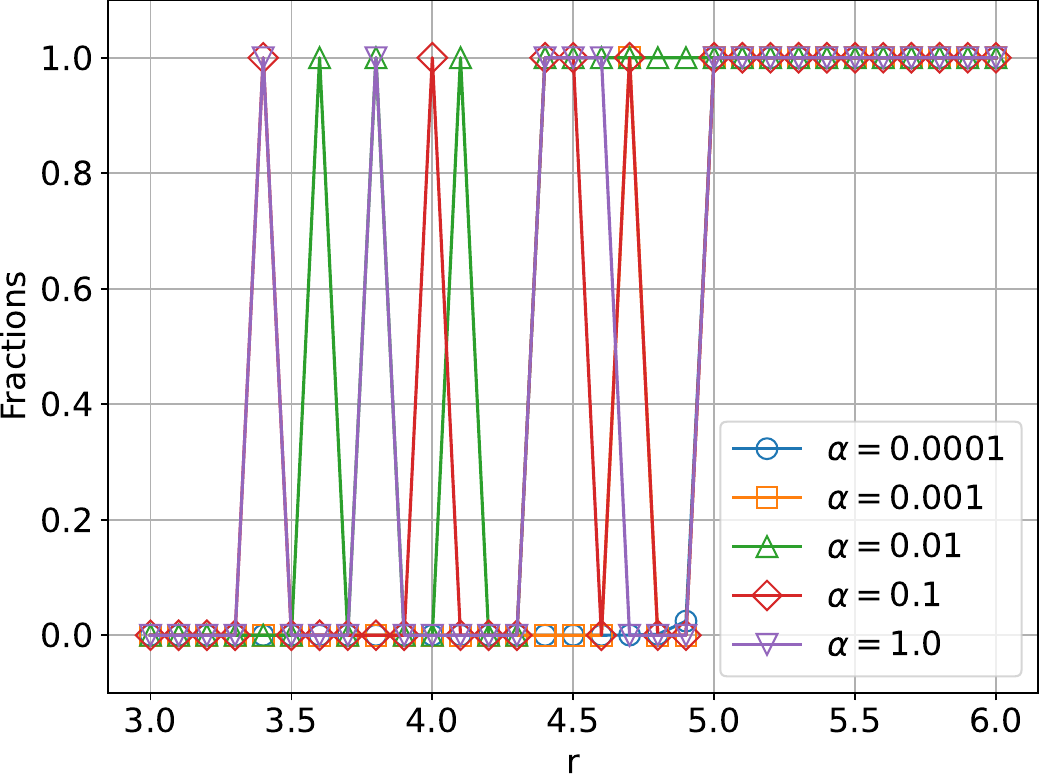}
		\caption{Impact of initial learning rate $\alpha$ on results. In this experiment, the model performed the best when $\alpha=0.001$.}
		\label{fig:cdrcombinedalpha}
	\end{figure}

	In PPO-ACT, the learning rate $\alpha$ controls the step size for updating parameters in both the policy network (Actor) and value network (Critic). Selecting an appropriate learning rate is crucial for training stability and convergence speed. In this experiment, $\rho=0.01$. As shown in Fig.~\ref{fig:cdrcombinedalpha}, when $\alpha<0.001$, the model converges slowly due to insufficient update steps, requiring higher $r$ values to achieve full cooperation. When $\alpha>0.001$, excessively large learning rates lead to unstable policy updates, making it difficult to reach the optimal solution. At $\alpha=0.001$, PPO-ACT demonstrates optimal convergence characteristics, achieving stable full cooperation at relatively low $r$ values. Therefore, we select $\alpha=0.001$ as the default step size parameter for PPO-ACT.
	
\begin{figure}[htbp!]
		\centering
		\includegraphics[width=0.7\linewidth]{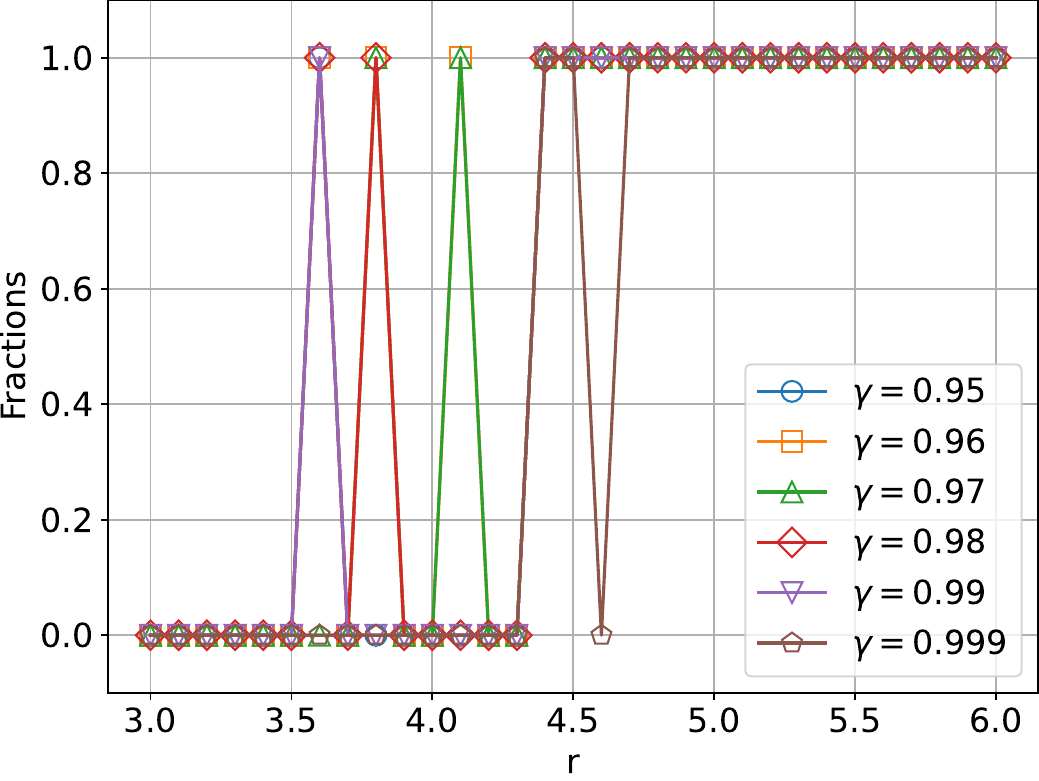}
		\caption{Impact of discount factor $\gamma$ on results. Experimental data indicate $\gamma=0.96$ as the parameter value yielding maximum model efficacy.}
		\label{fig:cdrcombinedgamma}
	\end{figure}

	The discount factor $\gamma$ is a crucial hyperparameter in PPO-ACT that determines the agent's consideration of future rewards.  In this experiment, $\rho=0.01$. As shown in Fig.~\ref{fig:cdrcombinedgamma}, experimental results demonstrate that at $\gamma=0.96$, PPO-ACT exhibits strongest cooperation promotions. Therefore, the $\gamma$ of Phase 2 is set to 0.96.
	
\begin{figure}[htbp!]
		\centering
		\includegraphics[width=0.7\linewidth]{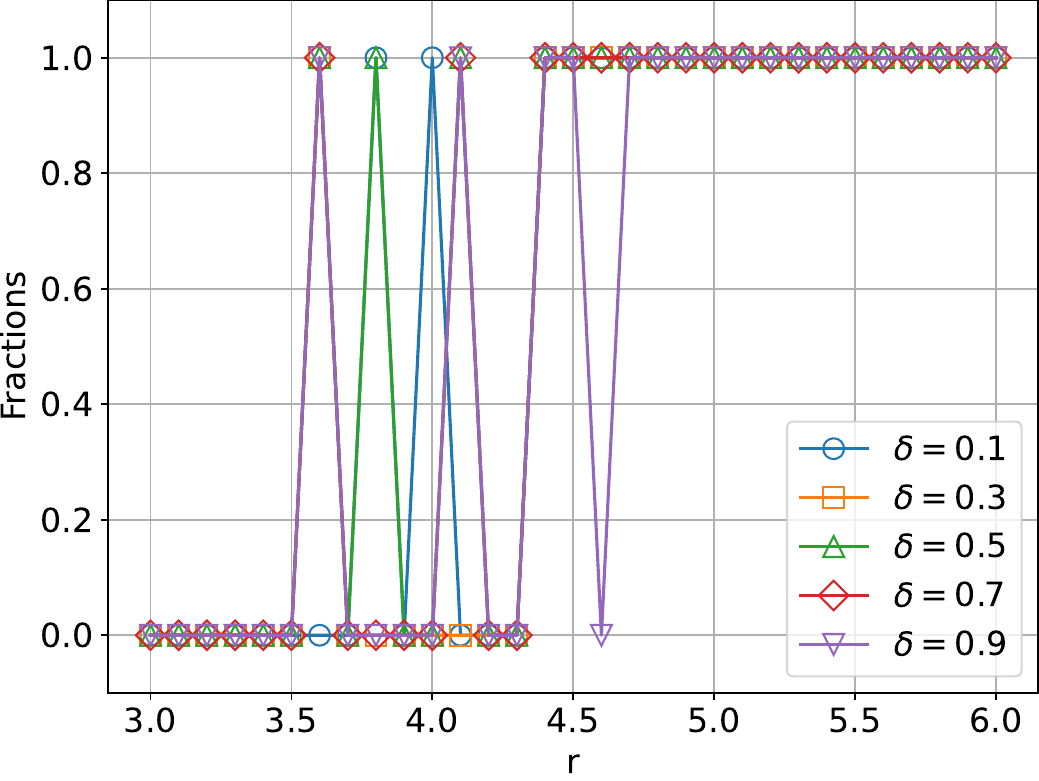}
		\caption{Impact of discount factor $\delta$ on results. The model reaches its performance peak when $\delta$ is set to 0.96.}
		\label{fig:cdrcombineddelta}
	\end{figure}

	The value function loss weight $\delta$ balances the contribution of value function errors relative to the policy gradient updates. As shown in Fig.~\ref{fig:cdrcombineddelta}, experimental results reveal a non-monotonic relationship between $\delta$ and cooperation levels. When $\delta < 0.5$, insufficient emphasis on value estimation leads to inaccurate state-value predictions, requiring higher r values to establish cooperation. At $\delta = 0.5$, PPO-ACT achieves optimal performance, accurately evaluating both immediate and long-term rewards while maintaining stable policy updates. Excessive values ($\delta > 0.5$) overweight the value function at the expense of policy optimization, resulting in slower adaptation to changing game conditions. This demonstrates the importance of balanced optimization between policy and value networks in PPO-ACT's dual-network architecture.
	
\begin{figure}[htbp!]
		\centering
		\includegraphics[width=0.7\linewidth]{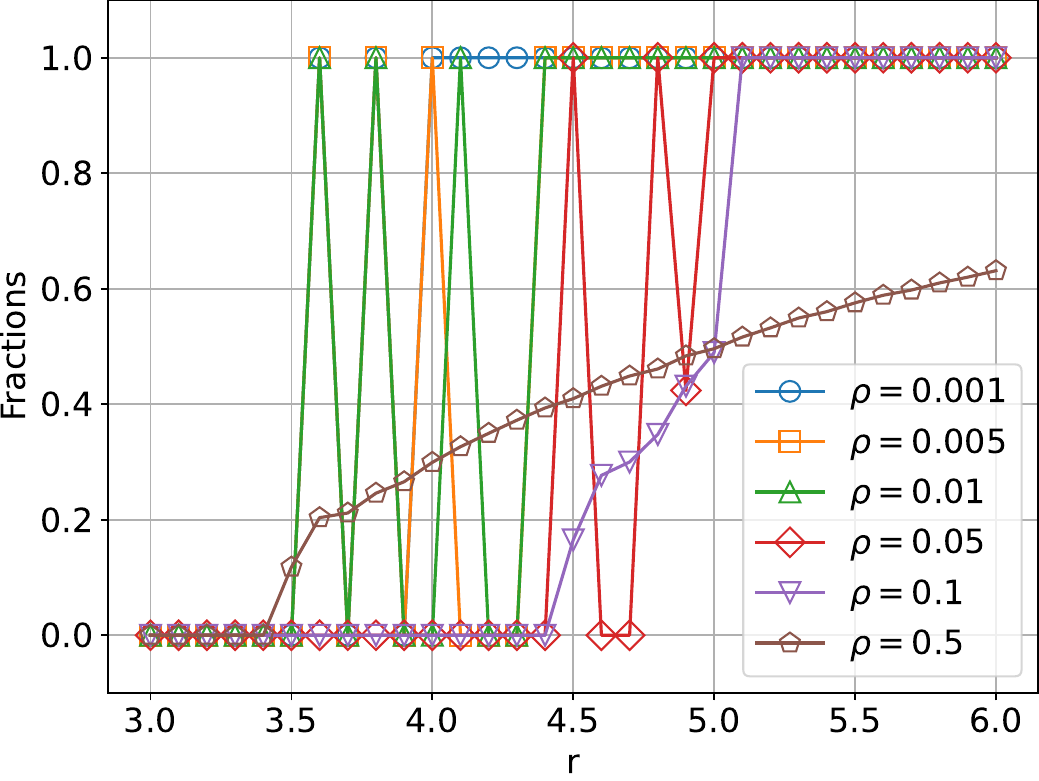}
		\caption{Impact of discount factor $\rho$ on results. The optimal model performance was achieved at $\rho=0.001$}
		\label{fig:cdrcombinedrho}
	\end{figure}

	The entropy weight $\rho=0.001$ ensures stable policy transfer by balancing exploration and exploitation (see  Fig.~\ref{fig:cdrcombinedrho}). Higher $\rho$ values hinder convergence to cooperation as excessive exploration disrupts learned strategies during Phase 2. This carefully tuned low entropy promotes reliable cooperation at challenging reward levels while maintaining necessary adaptability. 
	
	PPO-ACT achieves an optimal balance between training speed, stability, and final performance with $\alpha=0.01$,  $\gamma=0.96$, $\delta=0.5$, and $\rho=0.001$ (see Table~\ref{table:para}). These findings provide important references for parameter tuning in different scenarios. Notably, the clipping parameter $\varepsilon$ and advantage estimation weight $\lambda$ show no impact within common ranges, warranting further investigation.
	
	\section{Conclusions}
	\label{sec:con}
	
	The PPO-ACT framework developed in this study achieves an innovative integration of proximal policy optimization with adversarial curriculum transfer in evolutionary game theory. This synthesis offers an original methodological approach for addressing cooperation evolution in spatial PGG. Through systematic theoretical analysis and experimental validation, we demonstrate that PPO-ACT's two-stage training paradigm exhibits significant advantages in algorithmic performance. Comparative experimental data demonstrate PPO-ACT's dual advantage over conventional Q-learning and Fermi update rules. The algorithm sustains cooperative equilibria under substantially reduced benefit conditions while exhibiting markedly accelerated convergence rates. Particularly in challenging scenarios like all-defector initial conditions, PPO-ACT's curriculum learning approach demonstrates exceptional adaptability, verifying the effectiveness of our algorithmic design.
	
	From a theoretical perspective, this study reveals the unique advantages of PPO-ACT's policy optimization framework in cooperation evolution. The Actor-Critic architecture enables spatiotemporal optimization of long-term cooperative benefits. These technical innovations address the limitations of traditional imitation dynamics in global optimization. PPO-ACT agents demonstrate learning capabilities that extend beyond basic cooperation strategies. Through the autonomous exploration and adversarial curriculum transfer process, these agents develop advanced skills for enhancing strategic effectiveness via spatial configuration optimization. The experimental results confirm the validity of PPO-ACT's training methodology. These findings additionally demonstrate the distinct advantages of the combined PPO and ACT approach in facilitating cooperative evolution.
	
	Regarding practical applications, the PPO-ACT framework offers new insights for multi-agent system design. Its core concepts can be extended to related research in economics, sociology, and other fields, providing methodological guidance for solving various social dilemma problems. The framework's extensibility also establishes a foundation for studying more complex evolutionary game scenarios.
	
	In terms of spatial dynamics, our research finds that PPO-ACT spontaneously forms stable cooperative-defector phase separation patterns. Cooperators minimize edge exposure through tight clustering, while defectors create ``exploitation frontiers" surrounding cooperative clusters. This self-organizing phenomenon provides algorithmic-level micro-explanations for network reciprocity theory, revealing how spatial structures promote cooperation evolution.
	
	Despite these achievements, several limitations remain. The computational complexity and sensitivity to hyperparameter settings require further improvement. Additionally, scalability in large-scale systems needs verification. Future studies will extend PPO-ACT applications to more complex network structures. Additional work should examine cooperative behaviors in populations with heterogeneous agents. 
	Furthermore, incorporating language-based utility functions \cite{capraro2024language} could enhance the model's ability to simulate human-AI interaction scenarios, where linguistic framing plays a critical role in decision-making.
	These efforts will expand both the theoretical depth and practical applications of this work.
	
	In summary, the PPO-ACT framework provides new theoretical tools for understanding cooperative behavior in complex systems. Its innovative design and empirical results not only advance evolutionary game theory but also offer important references for related research in economics and sociology. Particularly, the framework's curriculum-driven policy adaptation and spatial self-organization capabilities provide novel insights for designing adaptive multi-agent systems.
	
	\section*{CRediT authorship contribution statement}
	
	\textbf{Zhaoqilin Yang}: Writing – original draft, Writing – review and editing, Validation, Methodology, Conceptualization.
	\textbf{Chanchan Li}: Conceptualization, Investigation, Writing – review and editing.
	\textbf{Xin Wang}: Writing – review and editing, Visualization, Software.
	\textbf{Youliang Tian}: Funding acquisition, Resources, Supervision.
	
	\section*{Declaration of competing interest }
	
	The authors declare that they have no known competing financial interests or personal relationships that could have appeared to influence the work reported in this paper.
	
	\section*{Data availability}
	
	No data was used for the research described in the article.
	
	\section*{Acknowledgments}
	This work was supported by the Natural Science Special Project (Special Post) Research Foundation of Guizhou University (No.[2024] 39); National Key Research and Development Program of China under Grant 2021YFB3101100; National Natural Science Foundation of China under Grant 62272123; Project of High-level Innovative Talents of Guizhou Province under Grant [2020]6008; Science and Technology Program of Guizhou Province under Grant [2020]5017, [2022]065; Science and Technology Program of Guiyang under Grant [2022]2-4.
	
	\bibliographystyle{elsarticle-num-names}
	\bibliography{cas-refs}
\end{document}